# Phosphine on Venus Cannot be Explained by Conventional Processes

*William Bains[1,9, #, *], Janusz J. Petkowski[1, #, *], Sara Seager[1,2,3], Sukrit Ranjan[1a], Clara Sousa-Silva[1,2], Paul B. Rimmer[4,5,6], Zhuchang Zhan[1], Jane S. Greaves[7], Anita M. S. Richards[8]*

[1]Dept. of Earth, Atmospheric, and Planetary Sciences, Massachusetts Institute of Technology, 77 Mass. Ave., Cambridge, MA, 02139, USA.

[2]Dept. of Physics, Massachusetts Institute of Technology, 77 Mass. Ave., Cambridge, MA, 02139, USA.

[3]Dept. of Aeronautics and Astronautics, Massachusetts Institute of Technology, 77 Mass. Ave., Cambridge, MA, 02139, USA.

[4]Department of Earth Sciences, University of Cambridge, Downing Street, Cambridge CB2 3EQ, UK.

[5]Cavendish Laboratory, University of Cambridge, JJ Thomson Ave, Cambridge CB3 0HE, United Kingdom

[6]MRC Laboratory of Molecular Biology, Francis Crick Ave, Cambridge CB2 0QH, United Kingdom

[7]School of Physics and Astronomy, Cardiff University, Cardiff CF24 3AA, UK.

[8]Jodrell Bank Centre for Astrophysics, Department of Physics and Astronomy, The University of Manchester, Alan Turing Building, Oxford Road, Manchester, M13 9PL, UK.

[9]School of Physics & Astronomy, Cardiff University, 4 The Parade, Cardiff CF24 3AA, UK

[a] SCOL Postdoctoral Fellow

# These authors contributed equally to this work, and are listed alphabetically.

* Correspondence to: bains@mit.edu, jjpetkow@mit.edu.

# Abstract

The recent candidate detection of ~1 ppb of phosphine in the middle atmosphere of Venus is so unexpected that it requires an exhaustive search for explanations of its origin. Phosphorus-containing species have not been modelled for Venus' atmosphere before and our work represents the first attempt to model phosphorus species in the Venusian atmosphere. We thoroughly explore the potential pathways of formation of phosphine in a Venusian environment, including in the planet's atmosphere, cloud and haze layers, surface, and subsurface. We investigate gas reactions, geochemical reactions, photochemistry, and other non-equilibrium processes. None of these potential phosphine production pathways are sufficient to explain the presence of ppb phosphine levels on Venus. If $PH_3$'s presence in Venus' atmosphere is confirmed, it therefore is highly likely to be the result of a process not previously considered plausible for Venusian conditions. The process could be unknown geochemistry, photochemistry, or even aerial microbial life, given that on Earth phosphine is exclusively associated with anthropogenic and biological sources. The detection of phosphine adds to the complexity of chemical processes in the Venusian environment and motivates *in situ* follow up sampling missions to Venus. Our analysis provides a template for investigation of phosphine as a biosignature on other worlds.

# Introduction

A biosignature is a feature of a planet that provides evidence for the presence of life on that planet (Catling *et al.* 2018; Schwieterman *et al.* 2018). Few, if any, remotely detectable feature of a planet is unambiguous evidence for life, and so any feature must be interpreted in the context of other knowledge about the planet. Atmospheric trace gases are favored biosignatures both for solar system bodies and for exoplanets, and a wide range have been suggested (Seager and Bains 2015; Seager *et al.* 2012). However, a detailed analysis of how a biosignature gas could be generated abiotically has only been carried out for molecular oxygen (Meadows 2017; Meadows *et al.* 2018), Other work in general does not discuss potential abiological routes to candidate biosignature gases, only noting that on Earth the only (or major) source of the gas is biology. But as the case of oxygen illustrates, a solely biological source on Earth does not preclude abiological sources on other planets. In this paper we provide a detailed analysis of abiological routes to phosphine, specifically in the context of Venus' atmospheric and geological chemistry. The analysis is motivated in part by recent claims that phosphine is present in Venus' atmosphere, but also more generally to provide a template for the analysis of this gas as a biosignature in any planetary context.

Venus is about the same size and mass as Earth, and is sometimes called Earth's sister planet. Venus' atmospheric chemistry and surface conditions, however, are quite different from Earth's. The interior chemical composition of Venus is poorly known. It is assumed to be similar in chemical composition to the Earth's crust and mantle, mainly because of the similarity between Earth's and Venus' size and overall bulk density (Smrekar *et al.* 2014).

Unlike the bulk planet composition, the atmospheres of Earth and Venus are very different. Our understanding of the chemistry of the Venusian atmosphere and clouds is incomplete, especially when it comes to the experimentally-derived concentrations of chemical species, like phosphoric acid, that are central to the calculations presented in this paper. Nevertheless, the Venusian clouds and hazes are known to have a complex vertical atmospheric profile with several distinct layers. The main cloud layer (~48 km – ~70 km) is composed of droplets, which are believed to be made primarily of photochemically-produced sulfuric acid (Oschlisniok *et al.* 2012). Haze extends from below the clouds through the cloud layer to at least 100 km, and may be composed of elemental sulfur as well as sulfuric acid (Taylor *et al.* 2018; Titov *et al.* 2018) (Fig. 1). The main sulfuric acid cloud decks also contain an unidentified UV-absorbing species. The UV absorber is very dynamic, with variable distribution in space and time within clouds (Haus *et al.* 2016; Lee *et al.* 2019; Lee *et al.* 2021) (reviewed in (Marcq *et al.* 2018; Taylor *et al.* 2018; Titov *et al.* 2018)). The complexity of the Venusian environment could, *a priori*, provide unexpected chemistry that could lead to the formation of phosphine.

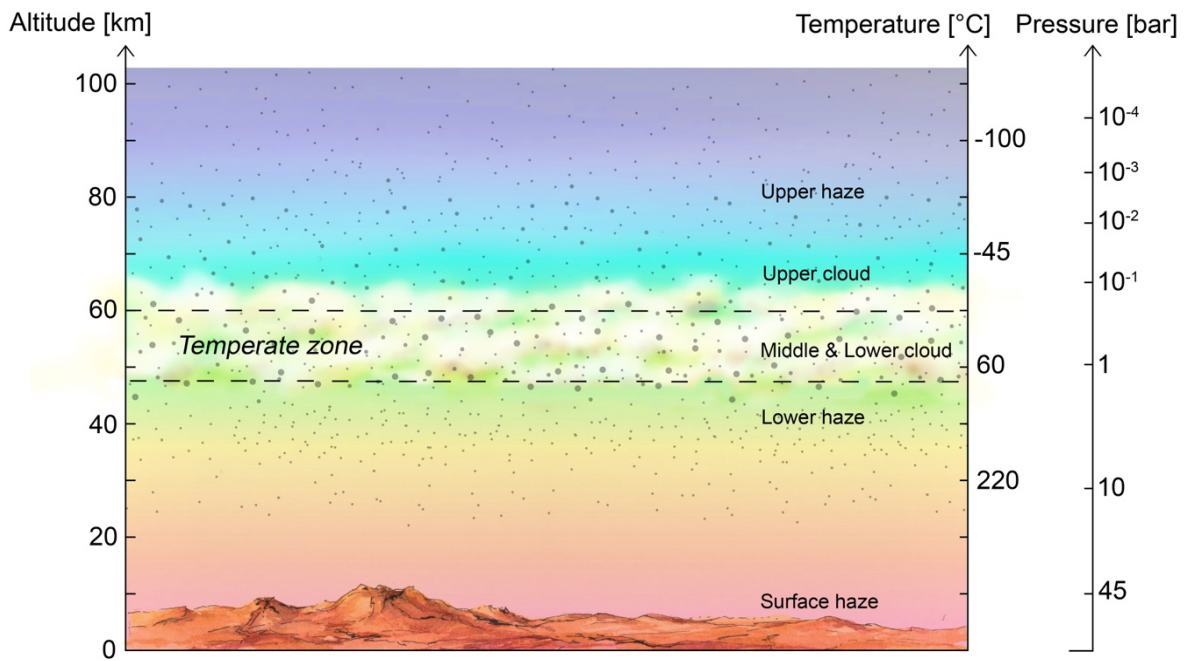

**Fig. 1.** A simplified schematic representation of the vertical structure of the main atmospheric layers on Venus (figure modified from (Seager *et al.* 2021)).

The recent candidate phosphine detection in the Venus' cloud decks adds further questions to the already complex picture of the chemical composition of the atmosphere of Venus (Greaves *et al.* 2020c). The detection was based on a single millimeter wavelength absorption line, and requires confirmation by the detection of additional phosphine spectral features. The detection has been contested (as discussed in detail below), but re-analysis of the data by several methods, and the tentative detection of a phosphine signal in the Pioneer MS data support the detection. This paper does not address the presence of phosphine in the atmosphere of Venus. For our purposes of using Venus as a test case for phosphine as a biosignature on a rocky planet, we will assume that the detection is valid.

If the detection of phosphine in Venus' clouds at 55 – 60 km altitude is correct, the presence of phosphine in Venus' atmosphere is highly unexpected, and requires explanation, as would the detection of phosphine on any other solar system body, or on any exoplanet. For some Solar System planets such as Jupiter, Saturn and Earth the explanation of phosphine production is well-known, as we discuss in section 1.1 below. This paper is the first step in providing such an explanation for the case of Venus. We start the introduction with a short summary of the recent detection of phosphine in the atmosphere of Venus and put it in the context of similar detections on other Solar System planets. Next, we review the chemistry and biology of phosphine gas, focusing on its unique production by life here on Earth (Section 1.1). We conclude the introduction with the motivation for the work presented in this paper (Section 1.2), the overall approach (Section 1.3), and the outline of the employed methods and the obtained results (Section 1.4).

## 1. 1. Phosphine in Solar Systems Bodies

### 1. 1. 1. Detection of Phosphine on Venus and Other Planets

The recent candidate detection of ppb amounts of phosphine in the atmosphere of Venus is a highly unexpected discovery. Millimetre-waveband spectra of Venus from both ALMA (Atacama Large Millimeter/submillimeter Array) and the JCMT (James Clerk Maxwell Telescope) telescopes at 266.9445 GHz show a $PH_3$ absorption-line profile against the thermal background from deeper, hotter layers of the atmosphere. The initial detection using the JCMT telescope in 2017 suggested an abundance of ~20 ppb, the initial follow-up detection using ALMA in 2019 suggested an abundance of ~7 ppb (Greaves *et al.* 2020c). Both detections have been disputed; we discuss this debate in more detail below (Section 2.2.1.). Mogul at al (Mogul *et al.* 2020a; Mogul *et al.* 2020b) have also claimed detection of phosphine in historical Pioneer spacecraft LNMS (Pioneer Venus Large Probe Neutral Mass Spectrometer) mass spectrometry data. If the Greaves et al (Greaves *et al.* 2020a; Greaves *et al.* 2020b; Greaves *et al.* 2020c) and Pioneer LNMS (Mogul *et al.* 2020a; Mogul *et al.* 2020b) detections are valid, then there is at least 1 ppb phosphine present in the atmosphere of Venus, and there may periodically be substantially more. Throughout this paper we will describe the predicted abundance as 1 ppb unless otherwise stated, as a conservative minimum. The thermal emission has a peak emission at 56 km with the FWHM (full-width at half-minimum) spans approximately 53 – 61 km (Greaves *et al.* 2020c). Phosphine is therefore presumed to be present above ~55 km: whether it is present below this altitude, and if present, what its abundance is, is not determined by these observations. The upper limit on phosphine occurrence is not defined by the observations, but is set by the half-life of phosphine at <80 km, as discussed below.

Phosphine has previously been detected in the atmospheres of three solar system planets: Jupiter, Saturn, and Earth. Phosphine is present in the giant planet atmospheres of Jupiter and Saturn, as identified by ground-based telescope observations at submillimeter and infrared wavelengths (Bregman *et al.* 1975; Larson *et al.* 1977; Tarrago *et al.* 1992; Weisstein and Serabyn 1996). In giant planets, $PH_3$ is expected to contain the entirety of the atmospheres' phosphorus in the deep atmosphere layers (Visscher *et al.* 2006), where the pressure, temperature and the concentration of $H_2$ are sufficiently high for $PH_3$ formation to be thermodynamically favored. In the upper atmosphere, phosphine is present at concentrations several orders of magnitude higher than predicted by thermodynamic equilibrium (Fletcher *et al.* 2009). Phosphine in the upper layers is dredged up by convection after its formation deeper in the atmosphere, at depths greater than 600 km (Noll and Marley 1997).

An analogous process of forming phosphine under high $H_2$ pressure and high temperature followed by dredge-up to the observable atmosphere cannot happen on worlds like Venus or Earth for two reasons. First, molecular hydrogen is a trace species in the atmospheres of rocky planets like Earth or Venus, so the formation of phosphine is not favored as it is in the deep atmospheres of the $H_2$-dominated giant planets. On Earth $H_2$ reaches 0.55 ppm levels (Novelli *et al.* 1999), on Venus it is much lower at ~4 ppb (Gruchola *et al.* 2019; Krasnopolsky 2010). Second, rocky planet atmospheres do not extend to a depth where, even if their atmosphere were composed primarily of hydrogen, phosphine formation would be favored (the possibility that phosphine could be formed below the surface and then being erupted out of volcanoes is addressed separately in Section 3.2.3 and Section 3.2.4, but is also unlikely).

Despite such unfavorable conditions for phosphine production, Earth is known to have $PH_3$ in its atmosphere at ppq to ppt levels (see e.g. (Gassmann *et al.* 1996; Glindemann *et al.* 2003; Pasek *et al.* 2014) and reviewed in (Sousa-Silva *et al.* 2020)). $PH_3$'s persistence in the Earth atmosphere is a result of the presence of microbial life on the Earth's surface (as discussed in Section 1.1.2 below), and of human industrial activity.

Neither the deep formation of phosphine and subsequent dredging to the surface nor its biological synthesis has hitherto been considered a plausible process to occur on Venus.

### 1. 1. 2. Phosphine is Exclusively Associated with Life on Earth

On Earth phosphine is a gas exclusively associated with life and is not made by any other natural atmospheric or geological chemical process (see e.g. (Gassmann and Glindemann 1993; Glindemann *et al.* 2003; Glindemann *et al.* 2005a; Glindemann *et al.* 1996)) and reviewed in (Bains *et al.* 2019a; Bains *et al.* 2019b; Sousa-Silva *et al.* 2020)). Terrestrial phosphine fulfils the criteria for being a biosignature gas, a gas whose detection indicates the presence of life (Catling *et al.* 2018; Seager and Bains 2015; Seager *et al.* 2016; Sousa-Silva *et al.* 2020; Walker *et al.* 2018). Previous work predicted that, if detected on a temperate rocky planet, phosphine would be a robust biosignature gas due to spectroscopic potential and limited false positives in such environments, although detection with near-future telescope will only be possible for a few planetary scenarios (Sousa-Silva *et al.* 2020). Since phosphine is mostly studied in the context of industrial chemistry, agriculture and laboratory chemical synthesis, its biology is not widely known. This warrants a brief introduction on the chemistry and biology of phosphine in the context of its biosignature potential on rocky planets.

On Earth, biological $PH_3$ production is associated with microbial activity in environments that are strictly anoxic (lacking oxygen) and highly reduced. The majority of reports of biological $PH_3$ production come from the studies of environments with anaerobic niches such as wetlands, sewage and animal intestinal tracts, flatus, and feces (reviewed in (Sousa-Silva *et al.* 2020)). Several studies have also reported the production of $PH_3$ from mixed bacterial cultures in the lab (Jenkins *et al.* 2000; Rutishauser and Bachofen 1999). Despite the fact that the exact metabolic pathway leading to $PH_3$ production in anaerobic bacteria is still unknown, it is clear that phosphine is a biosignature gas on Earth, albeit strictly associated with the anaerobic biosphere. On Earth phosphine could be made directly by microbial reduction of more oxidized phosphorus species or indirectly by microbial production of reduced phosphorus compounds, such as hypophosphite, and their subsequent disproportionation to $PH_3$ (Gassmann and Glindemann 1993; Glindemann *et al.* 1999; Glindemann *et al.* 2005a;

Glindemann *et al.* 1996). In either case however the presence of phosphine is an indicator of the presence of life. For more information on phosphine in the context of terrestrial biology see recent studies by (Bains *et al.* 2019a; Bains *et al.* 2019b; Sousa-Silva *et al.* 2020).

### 1. 2. Motivation

As discussed, we wish to extend the analysis of phosphine as a candidate biosignature gas by analyzing possible abiotic sources of phosphine. As a case study, and a base for future research, we model the production of phosphine in the Venusian environment. The presence of phosphine in the atmosphere of Venus would be unexpected, and so its detection, if confirmed, requires explanation. If the detection is confirmed by further observations, the presence of phosphine in Venus' atmosphere suggests that our understanding of Venusian atmospheric chemistry is at least incomplete, and that the source of that phosphine needs to be identified. In light of the exclusively biological production of phosphine on Earth, the only rocky planet hitherto known to have phosphine in its atmosphere, the question arises whether the detection of phosphine on Venus could indicate the presence of life. For such a claim to even be entertained, all other possible sources of phosphine should be identified and eliminated. Regardless of whether phosphine is present on Venus, this investigation serves as a starting point for futures detections of phosphine features in observational data of temperate exoplanets. We emphasize that, even if the detection of phosphine is confirmed in the atmosphere of Venus, this can only be considered as evidence of the presence of life if *all* other sources of phosphine can be ruled out (Catling *et al.* 2018). This paper is a first step in that undertaking, considering possible non-biological mechanisms for making phosphine in the atmosphere, surface or subsurface of Venus.

### 1. 3. Approach: Photochemistry, Kinetics, and Thermodynamics

The ideal approach to identify the possible source of any gas in a planet's atmosphere would be to exhaustively model the rate of all possible reactions that could create and destroy that gas. Presently this is impossible. Exhaustive modelling requires knowledge of all the components of the atmosphere, surface, and subsurface of the planet. While some components of Venus' atmosphere are well known, many, including gases relevant to phosphine reactivity, remain unknown. In addition, exhaustive modelling requires accurate knowledge of the rates of all possible reactions between component molecules under all relevant conditions. Many reaction rates for known species in the Venusian environment have not been measured.

We therefore break the modelling problem into two parts. 1] We construct a photochemical model accounting for the formation and destruction of phosphine based on previous photochemical models of Venus' atmosphere. 2] We separately and complementarily use a thermodynamic approach to model formation pathways for phosphine. While the thermodynamic modelling is not intended to substitute for the full kinetic modelling of chemical reactions, it plays a useful and necessary role to *rule out* chemical reactions that could spontaneously produce phosphine.

Together the two modelling units provide upper bounds on Venusian phosphine production.

### 1. 4.     Paper Outline

In this paper, we apply chemical modelling to attempt to explain the production of the highly unexpected discovery of the trace gas phosphine in the atmosphere of Venus (Greaves *et al.* 2020c).

The main body of the paper is divided into two sections, modelling the photochemistry and kinetics of phosphine in the atmosphere (Section 2) and thermodynamics in the atmosphere, surface, and subsurface (Section 3). Detailed methods for these sections are provided in online supplementary material (Supplementary Section 1.1, Supplementary Section 1.2 and Supplementary Section 1.3).

In Section 4 we summarize other processes, including lightning and exotic physical and chemical phenomena that could in principle lead to the formation of phosphine on Venus.

In the Discussion Section (Section 5) we explore several unconventional explanations for the phosphine on Venus, including exotic geochemistry, photochemistry and biologically-driven formation of phosphine. A range of chemical reactions can produce phosphine under Venus conditions, but all of these require reactants that are themselves extremely unlikely to form on Venus, a problem we term "displaced improbability". We conclude the paper by arguing that the source of phosphine on Venus, if the presence of $PH_3$ is confirmed, cannot be explained by our current knowledge of the planet. All potential sources fall short by many orders of magnitude. We argue that further aggressive observations of Venus and its atmosphere, as well as the development of astrobiology-focused space missions, should get the highest priority and would be crucial for an unambiguous explanation for the source of phosphine in the Venusian atmosphere.

## 2.    Photochemistry and Kinetics of Phosphine in the Atmosphere of Venus

The overall goal of our photochemical calculations is to determine if photochemically-driven mechanisms can maintain the detected 1 ppb of $PH_3$ at any altitude. This is not yet possible within a self-consistent model because synthesis rates of $PH_3$ from oxidized species are largely unknown. To account for the limitations caused by missing $PH_3$ kinetics, we make the complex chemistry of phosphine in the Venusian atmosphere tractable by modelling phosphine photochemical destruction and synthesis networks separately.

We proceed by first calculating the destruction rates for $PH_3$, for which reaction kinetics are relatively well known. We do so by (1) using a photochemical model to estimate the vertical radical concentration profiles in the Venusian atmosphere, and (2) using the radical profiles to estimate $PH_3$ lifetimes (and hence destruction rates) throughout the atmosphere. Separating the photochemical model calculations and lifetime estimates enables us to repeat our lifetime calculations with radical profiles derived from a different model (Bierson and Zhang 2019), permitting us to test the sensitivity of our conclusions to the choice of photochemical model (Ranjan *et al.* 2020). Second, we explore the photochemical pathways for the synthesis of $PH_3$ and determine whether the $PH_3$ synthesis network can compensate for the known $PH_3$ destruction mechanisms and sustain a ~1 ppb concentration of phosphine at any altitude in the Venusian atmosphere.

We show that photochemical synthesis of $PH_3$ is unable to explain the observed $PH_3$ concentration. Although the major source of uncertainty in this calculation is the extremely poor knowledge of the $PH_3$ synthesis pathways, our approach is conservative such that these uncertainties do not affect our main conclusions

## 2. 1. Introduction to Photochemistry and Kinetics Analysis

In Section 2.2., we summarize the photochemical models used in this work (Section 2.2.1., Section 2.2.2. and Section 2.2.3.), including the addition of $PH_3$ to the photochemical network, and estimate the lifetime of phosphine in the Venusian atmosphere (see Supplementary Section 1.1. and its subsections in Supplementary Information). We discuss in detail all the known processes that affect the lifetime of phosphine, including destruction of phosphine by atmospheric radicals, direct UV photolysis and vertical transport in the atmosphere of Venus. We also discuss significant limitations and uncertainties of phosphine lifetime calculations.

The estimation of the lifetime of phosphine on Venus is key for determining production rates that are required to maintain the detected ~1 ppb concentration in the Venusian atmosphere. We compare the photochemical destruction rates from our photochemical model with the predicted maximum possible photochemical production rate of phosphine, to assess the possibility of its photochemically-driven formation (Section 2.2.2. and 2.2.3.). We explain why our predicted phosphine photochemical production is many orders of magnitude lower than that needed to explain the observed abundance of phosphine.

(Greaves *et al.* 2020c) provided a preliminary description of a photochemistry model for the Venusian atmosphere that includes phosphorus species. Here we provide a more complete description of that model, and apply it to phosphine chemistry on Venus. The model uses the ARGO 1D photochemistry-diffusion code (Rimmer and Helling 2016) to solve the atmospheric transport equation for the steady-state vertical composition profile. ARGO is a Lagrangian photochemistry/diffusion code. The code follows a single parcel of gas as it moves from the bottom to the top of the atmosphere, as determined by a prescribed temperature profile. The code updates temperature, pressure, and actinic ultraviolet flux at each height in the atmosphere. In this reference frame, bulk diffusion terms are accounted for by time-dependence of the chemical production, $P_i$ (cm$^3$ s$^{-1}$), and loss, $L_i$ (s$^{-1}$), and so below the homopause, the chemical equation being solved is effectively:

$$\frac{\partial n_i}{\partial t} = P_i[t(z, v_v)] - L_i[t(z, v_z)]n_i, \qquad (1)$$

where $n_i$ (cm$^{-3}$) is the number density of species $i$, $t$ (s) is time, $z$ (cm) is atmospheric height, and $v_z = K_{zz}/H_0$ (cm/s) is the effective vertical velocity due to Eddy diffusion, from the Eddy diffusion coefficient $K_{zz}$ (cm$^2$ s$^{-1}$). The model is run until the abundance of every major and significant minor species (any with $n_i > 10^5$ cm$^{-3}$) does not change by more than 1% between two global iterations.

The handful of known reactions of PH$_3$ with the major reactive Venusian species O, Cl, OH, and H were combined with previously published Venus atmospheric networks of (Krasnopolsky 2012; Krasnopolsky 2013) and (Zhang *et al.* 2012), and the network of STAND2019 of (Rimmer and Rugheimer 2019), which includes H/C/N/O species. This model and its results are the same as those presented in (Greaves *et al.* 2020c). Details of the reaction networks, initial conditions and modelling are provided in Supplementary Section 1.1. and its subsections in Supplementary Information. See also Supplementary Figs. S1, S2 and Supplementary Tables S1, S2.

This whole-atmosphere model allows us to assess the lifetime of PH$_3$ throughout the atmosphere self-consistently. The model accounts for photochemistry, thermochemistry and chemical diffusion. UV transport calculation was modified in two ways. First, we ignore the UV absorption of SO$_2$ for the first three global iterations, and include it afterwards. This seems to help the model to converge. After the first three global iterations, we include UV absorption by SO$_2$ and absorption by the 'mysterious absorber' with properties described by (Krasnopolsky 2007) (see Supplementary Section 1.1.1. in Supplementary Information).

With these conditions, using the photochemical network described below, convergence required 33 global iterations of the model.

The counterbalance of photochemical destruction of phosphine is the possibility that phosphine is photochemically generated in gas or droplet phases. The possibility of gas phase production was considered as follows. A network of reactions that could generate PH$_3$ from H$_3$PO$_4$ was constructed; H$_3$PO$_4$ was selected as the starting molecule because H$_3$PO$_4$ is predicted to be the most abundant phosphorus species in Venus' atmosphere at cloud level and above, and because H$_3$PO$_4$ is the only phosphorus species for which gas phase kinetic data is available. The maximum possible rate of phosphine production was calculated as the flux through this network assuming no back reactions. More detail on the network, its construction and estimation of the reaction rates is provided in Supplementary Information, Supplementary Section 1.2., Supplementary Figs. S4, S5, S6, S7. The possibility of photochemical production of phosphine in cloud droplets is discussed briefly in Section 5.2.

## 2. 2. Results of the Photochemistry and Kinetics Analysis

### 2. 2. 1. Abundance of Phosphine

We chose to model processes that can maintain a stable abundance of 1 ppb phosphine in the atmosphere of the planet, for the following reason.

The initial detection by Greaves et al using the JCMT telescope (Greaves *et al.* 2020c) in 2017 was interpreted as an abundance of ~20 ppb, derived from three different analytical methods (Greaves *et al.* 2020a). The re-processed follow-up detection using ALMA in 2019 (Greaves *et al.* 2020b) was interpreted as an abundance of 1-4 ppb, when planet averaged, variable across the planet, an interpretation that is robust to a variety of analytical methods (Greaves *et al.* 2020a; Greaves *et al.* 2020b). The original detections have both been challenged both on the statistical grounds (Snellen *et al.* 2020; Thompson 2021) as well as interpretation of the data, suggesting that the detection is $SO_2$ instead of $PH_3$ (Akins *et al.* 2021; Lincowski *et al.* 2021; Villanueva *et al.* 2020). The TEXES/NASA-IRTF NIR spectrometry data has been interpreted as showing an upper phosphine abundance limit of 5 ppb on March 2015 (Encrenaz *et al.* 2020). The upper limits for phosphine above the cloud top assessed from the SOIR/VEx spectra from localized region of the Venus terminator collected between August 2006 and January 2010 suggest $PH_3$ abundances of less than 1 ppb (Trompet *et al.* 2020). The Pioneer LNMS data has not been used directly to infer an abundance, but suggests a similar order of magnitude to $H_2S$ in 1978, which is modelled to be present at ~1 ppb at the cloud level (see Supplementary Information, Figure S3 (Krasnopolsky 2008)). The differences in the estimates of the abundance of $PH_3$ in Venus' atmosphere could therefore be due to variability of phosphine abundance in space and time, due to differences in detection methods or analyses, or both. If phosphine is present, however, the detections are all consistent with an abundance of at least 1 ppb in the cloud decks. As the purpose of this paper is to explore potential abiotic routes to this biosignature gas, the presence of the gas is taken as a starting point. To be conservative, we assume an abundance of 1 ppb. If a process cannot produce enough phosphine to explain 1 ppb, then it also cannot produce enough phosphine to explain 20 ppb. If any process (biotic or abiotic) *can* explain the presence of phosphine at 1 ppb, this does not mean that it can produce enough to maintain the 20 ppb that was potentially seen in 2017 using the JCMT telescope. So as a criterion to rule out potential sources of phosphine, assuming a lower abundance is a conservative assumption.

### 2. 2. 2. Lifetime and Necessary Production Rate of $PH_3$ in the Venusian Atmosphere

The abundance of phosphine on Venus is a result of a balance between its production and destruction. Estimating Venusian $PH_3$ destruction rate (and hence its lifetime) as a function of altitude is key for understanding the $PH_3$ production rates required to maintain a ~1 ppb atmospheric concentration. Figure 2 presents our estimates of $PH_3$ destruction rate and lifetime as a function of altitude, broken down by specific destruction mechanisms.

We begin by commenting broadly on $PH_3$ photochemical destruction rates in the Venusian atmosphere. Attack by O is the main loss mechanism in the high atmosphere (>60 – 80 km), attack by Cl the main loss mechanism in the middle atmosphere, and thermolysis the main loss mechanism at the planet surface; this is consistent with calculations performed with radical profiles derived from other models of Venus, albeit ones that do not consider $PH_3$ (Bierson and Zhang 2019). Direct photolysis is included, but is found not to be the dominant loss mechanism at any height in the atmosphere for any of the models considered. The presence of $PH_3$ suppresses radical concentrations in the lower atmosphere. The concentrations of radicals are low in the lower atmosphere, and so even in small abundances, $PH_3$ becomes a significant scavenger; consequently models that exclude $PH_3$ (e.g., (Bierson and Zhang 2019)) may overestimate photochemical destruction rates in the deep atmosphere.

We next discuss the chemistry of atomic chlorine, which determines the profile of $PH_3$ in the mid atmosphere. Atomic Cl is predicted to occur well below the limit of detection, with mixing ratios of $<10^{-17}$ beneath the clouds according to all the atmospheric models we consider. Even at these mixing ratios, Cl significantly affects the lifetime of $PH_3$ below the clouds of Venus. In our model the vertical profile of Cl atoms is complex. In brief, $ClS_2$ is produced by thermal reactions between sulfur species, CO and HCl below 5 km, and is efficiently broken down to Cl atoms by 327 – 485 nm photons that penetrate below 35 km. Above 30 km Cl is removed by reaction with chemical products of $SO_3$ which itself is produced by thermal dissociation of $H_2SO_4$. Cl abundance is predicted to be $<1$ $cm^{-3}$ near the surface (the Cl is produced thermochemically near the surface, and then locked into $ClS_2$), $>100$ $cm^{-3}$ at 25 - 35 km (from $ClS_2$ photolysis), and above 50 km, $<1$ $cm^{-3}$ between 42 and 54 km (due to reactions with chemical products of $SO_3$), and then increases from 1 $cm^{-3}$ to $10^8$ $cm^{-3}$ between 58 and 100 km due to HCl photolysis (See Supplementary Information, Section 1.1.5.3 for further details on Cl chemistry in our model).

However other models using different networks show different Cl atom abundances. The atomic and radical profiles from Bierson (Bierson and Zhang 2019), Krasnopolsky (Krasnopolsky 2007) and our profiles disagree with each other by approximately five orders of magnitude, which means that the predicted chemical lifetimes for $PH_3$ due to destruction by these atoms and radicals differs by several orders of magnitude.

If destruction by atoms and radicals was the only way to remove $PH_3$, then the lifetime of $PH_3$ would be very poorly constrained. It would depend on abundances of species that cannot be measured, and which can vary over almost five orders of magnitude between models. However, the thermal decomposition, diffusion timescale and photochemical destruction of $PH_3$ are robust to differences in chemical networks and provide us with a confident upper limit to the lifetime of $PH_3$ in the atmosphere of Venus. We therefore move on to the role of transport.

$PH_3$ has a lifetime of < 1 second in the high atmosphere (>78 - 98 km) due to high levels of UV radiation and its concomitant radicals. In the deep atmosphere (<50 km), which is UV-shielded, $PH_3$ lifetime to photochemical destruction may be much longer (up to $10^{11}$ seconds). Vertical transport of $PH_3$ to high altitudes ultimately limits the $PH_3$ lifetime in much of the lower atmosphere. However, transport in the lower atmosphere of Venus is slow: consequently, $PH_3$ lifetimes may be as high as ~400 years in parts of the lower atmosphere. If we instead estimate the lifetime using the radical concentration profiles of (Bierson and Zhang 2019), we predict lifetimes of $\leq 700$ years in the deep atmosphere, because the $PH_3$ must diffuse to a higher $z_0$[1] (up to 98 km), compared to our model (78 km).

The comparatively long lifetime of $PH_3$ predicted for parts of the deep atmosphere (~100s of years) motivates us to consider the possibility that low photochemical or abiotic production of $PH_3$ could result in accumulation of phosphine over time and diffuse upwards to explain 1 ppb $PH_3$ abundance at the level probed by (Greaves *et al.* 2020c). This scenario requires an efficient unknown phosphine formation mechanism deep in the atmosphere, and/or efficient transport to the detection altitudes of 53-61 km but not to the destruction altitude (>78-98 km). Our calculations suggest there is no such transport pattern for the Venusian atmosphere.

---

[1] Altitude at which the photochemical lifetime of $PH_3$ becomes short ($\leq 10^4$ s), i.e. where the radical population become high; see Supplementary Section 1.1.1.

The rate of destruction of $PH_3$ (at the cloud level or below) is much slower than on Earth, because of the much lower concentration of OH radicals in the Venusian atmosphere. A much smaller production rate is therefore needed to generate a 1 ppb concentration in the atmosphere than would be true on Earth. We calculated the total, planet-wide outgassing flux necessary to maintain an atmospheric concentration of 1 ppb in the atmosphere of Venus at the detection altitudes of 53-61 km. We find that a flux of $\sim 10^8$ phosphine molecules $cm^{-2} s^{-1}$ (averaged across the whole planet) is needed to reproduce the observed phosphine mixing ratio of 1 ppb above 55 km (Greaves *et al.* 2020c). This is equivalent to ~26 kg/second or $\sim 8 \times 10^5$ tonnes $year^{-1}$. For comparison, methane is produced at a rate of $\sim 340 \times 10^6$ tonnes $year^{-1}$ from non-anthropogenic sources on Earth, $\sim 14 \times 10^6$ tonnes of which are geological (i.e. not dependent on life) (Saunois *et al.* 2016).

In the remainder of this paper, we explore the possibility of an efficient abiotic phosphine formation mechanism in the Venusian atmosphere.

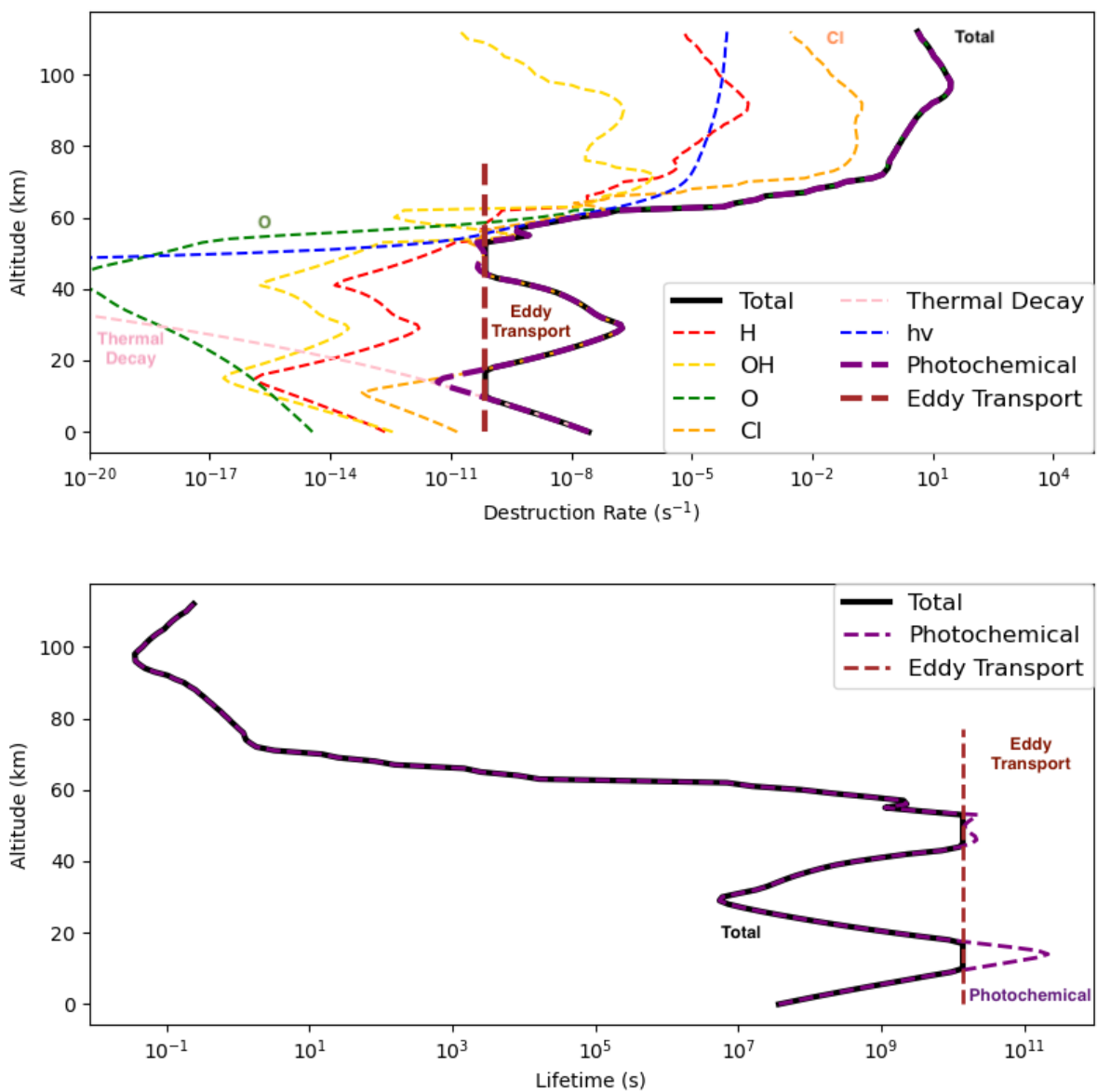

**Fig. 2.** The lifetime of phosphine in the Venusian atmosphere. *Top panel*: Removal rates for PH$_3$ in the Venusian atmosphere, as a function of altitude. x axis: Destruction rate (s$^{-1}$), y axis: Altitude (km). Individual photochemical loss processes are shown in thin dashed lines. Also shown is the loss rate due to diffusion to the upper atmosphere, calculated by inverting the diffusion timescale. Thick black line presents overall loss rate, which is the minimum of the photochemical and diffusion loss rates. *Bottom panel*: Photochemical, diffusion, and overall lifetimes of PH$_3$ in the Venusian atmosphere, calculated by inverting the corresponding loss rates. x axis: Lifetime (s), y axis: Altitude (km). Overall, the photochemical lifetime of PH$_3$ is long in the lower atmosphere but short in the upper atmosphere, meaning that transport to the upper atmosphere ultimately limits PH$_3$ lifetime in much of the lower atmosphere. Even so, PH$_3$ lifetimes of order centuries are possible in the lower atmosphere.

### 2. 2. 3. Photochemical Synthesis of Phosphine Cannot Explain the Observed PH$_3$ Abundance in the Atmosphere of Venus

Photochemical synthesis of phosphine, by reduction of oxidized phosphorus species by atmospheric radicals, could in principle lead to the formation of phosphine. We argue however that photochemically driven reactions in Venus' atmosphere cannot produce PH$_3$ in sufficient amounts to explain the detection of ~1 ppb. We find that the reactions involving atmospheric radicals capable of reducing oxidized phosphorus species (e.g., hydrogen radicals) are too slow, and the required forward reaction rates are too low, by factors of $10^5$ or more (see Figure 3). We present our reasoning in detail below.

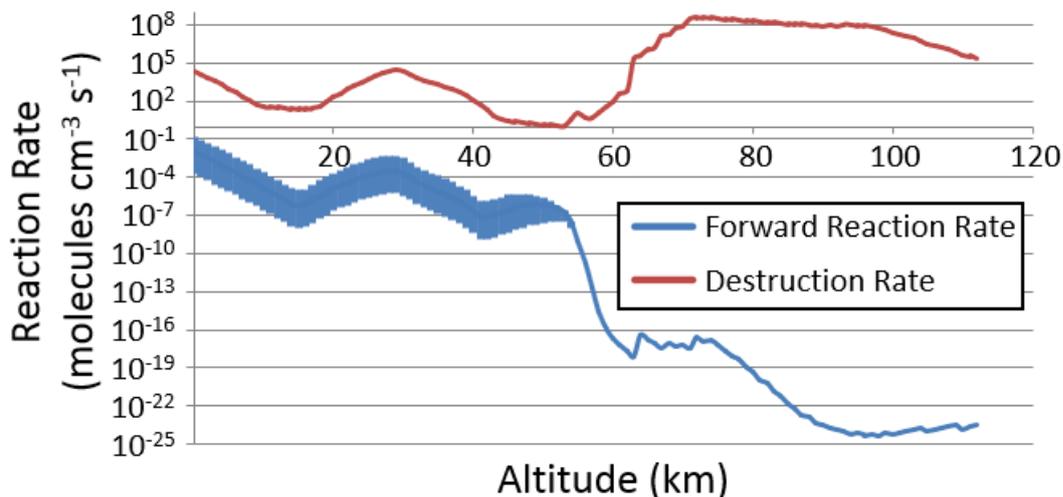

**Fig. 3.** The photochemical production and destruction rates of phosphine. x axis: Altitude (km), y axis: Reaction rate (molecules cm$^{-3}$ s$^{-1}$). Maximum rate of forward reaction through the kinetic network as a function of altitude (blue line) compared with the photochemical destruction rate (red line). The base of the clouds is assumed to be at any altitude between 45 km and 55 km, which gives a range of forward rates reflecting a range of phosphorus species concentrations, themselves depending on the lower boundary of the cloud layer as described in Supplementary Information Section 1.3.2.2. Under no conditions the rate of the photochemical formation of phosphine is sufficient to balance the photochemical destruction rate, therefore making the photochemical production of phosphine unlikely.

Figure 3 shows that there is no altitude at which the maximum possible forward reaction rate is sufficient to counter the destruction rate: the minimum ratio of destruction/synthesis rates is 8.46x10$^{-6}$. Figure 4 analyses which reactions in the network are responsible for the slow production of phosphine. The main 'blockage' in the network (Figure 4) for PH$_3$ synthesis is the series of reactions that can lead from P=O to PH or PH$_2$. The conversion of phosphoric acid (H$_3$PO$_4$) to the P$^{(+3)}$ radical H$_2$PO$_3$ is also a rate-limiting process, supporting the idea that the spontaneous production of phosphite or phosphorous acid is not favored (discussed further below in Section 3.2.1.2); note that phosphorous acid itself – H$_3$PO$_3$ – is not stable in gas phase. We discuss potential chemistry of H$_3$PO$_3$ in the droplet phase, as well as its potential role as a transient intermediate, in Section 3.2.1.2 and Section 3.2.1.3 below.

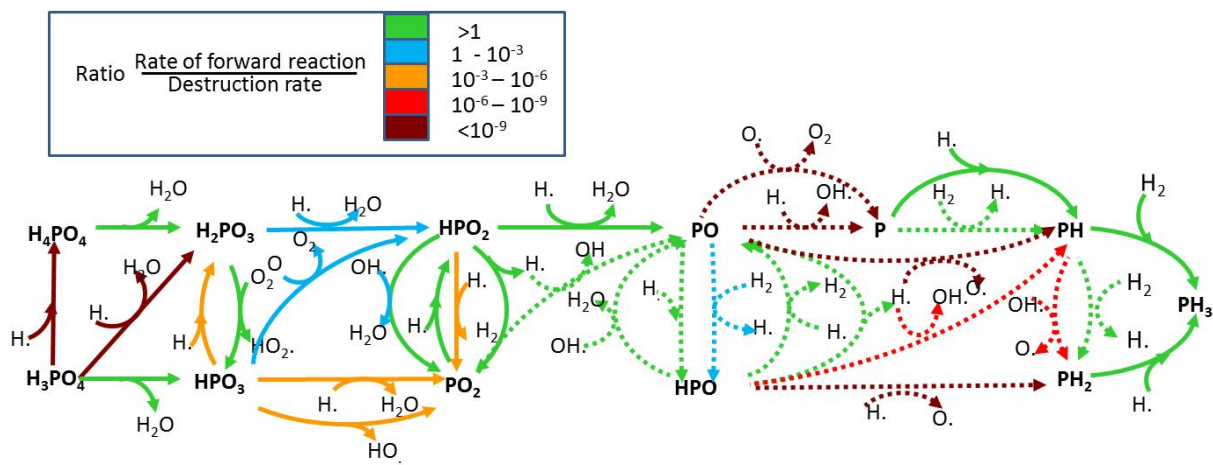

**Fig. 4.** Exploration of the potential photochemical pathways for the synthesis of PH$_3$. The reaction network was constructed as described in Supplementary Information, Supplementary Section 1.2. The destruction rate of phosphine was calculated from the photochemical model (Supplementary Information, Supplementary Section 1.1.1). Maximum possible forward reaction rates were calculated as described in Supplementary Information, Supplementary Section 1.2. For each altitude, the ratio $R$ = reaction rate/destruction rate was calculated for each reaction. The reactions are colored by the *maximum R* for any altitude for that reaction. There is no path to PH$_3$ synthesis through the network that does not cross at least one reaction that has an $R<10^{-6}$, i.e. is at least nine orders of magnitude too slow to account for the observed levels of phosphine. Therefore, there is no reaction path that can efficiently produce phosphine photochemically. The transformation of P=O to PH or PH$_2$ is the main bottleneck of the network. The forward kinetic network is constructed as a function of altitude. Reactions are colored for the assumption that the cloud base occurs at 48 km. Figure modified from (Greaves *et al.* 2020c).

We note in summary that that our analysis is very conservative because it is purposely highly biased towards predicting the production of phosphine, for two reasons:

1. We assume that *all* of the atmospheric phosphorus is concentrated into one species, the species that is reacting in each reaction. Such scenario is highly improbable. In reality phosphorus species would predominantly be present as H$_3$PO$_4$ or P$_4$O$_{10}$ (see Section 3.2.1.1), and all other species would be trace gases.
2. We assume that only forward (reducing) reactions occur. If back (oxidizing) reactions were also considered, they would reduce the calculated net rate of reduction, and lower the overall production rate of phosphine.

Therefore, our network provides the maximum possible phosphine production rate from known photochemical processes. The maximum rate predicted is more than four orders of magnitude too low to account for the presence of ~1 ppb PH$_3$ in Venus' atmosphere. In reality back reactions would significantly lower the efficiency of the formation of PH$_3$. Several such back reactions could occur, the net result of forward and back reactions occurring at the same time is the phosphorus-catalyzed recombination of H, O and OH into H$_2$O instead of the production of reduced phosphorus species. The precedent for such phosphorus-catalyzed recombination chemistry is known in terrestrial flame chemistry (Twarowski 1993; Twarowski 1995; Twarowski 1996). We note however that this hypothesis needs more detailed modelling and experimental studies to be confirmed.

Our forward PH$_3$ production reaction network contains no provision for reactions of oxidized phosphorus species with sulfur or oxidized chlorine species like ClO, which play a substantial role in Venusian atmospheric chemistry (Marcq *et al.* 2018; Sandor and Clancy 2018; Taylor and Hunten 2014). No reaction kinetics are reported for reaction of oxidized phosphorus species with reactive, oxidizing S or Cl species. It is unknown if such hypothetical photochemical processes involving sulfur or chlorine species can lead to the reduction of oxidized phosphorus species and, as a result, to the production of phosphine. We discuss such unknown chemical processes as a potential source of phosphine on Venus in Section 5.2.

Our approach suggests that phosphorus monoxide (PO) could be a significant component of the reaction chemistry of phosphorus in Venus' atmosphere. PO has not been observed or modelled as an atmospheric species on Venus to date. PO's presence could be confirmed by directed observation, as it was done in the past for PO (Tenenbaum *et al.* 2007) and phosphorus oxoacids (Turner *et al.* 2018) in the interstellar medium. We emphasize however that we are postulating the existence of PO as a transient intermediate species, not a major component of the Venusian atmosphere.

### 2. 3. Summary and Conclusion of the Photochemistry and Kinetic Analysis

We have carried out a detailed analysis of photochemical and other endergonic chemistry that could produce phosphine under Venus conditions. Our models provide the destruction rate and lifetime for phosphine in Venus' atmosphere, and hence a flux rate necessary to maintain ~1 ppb phosphine stably in the atmosphere. Our analysis confirms that none of the modelled kinetic pathways can explain the levels of phosphine observed, falling short by many orders of magnitude, even using the most conservative assessments available.

We note that these are all calculations of gas phase photochemistry. Solid phase photochemistry is not relevant, as no significant UV penetrates to the ground on Venus. We address the question of the UV photochemistry of the cloud droplets in Section 5.2.

## 3. Thermodynamic Analysis of Potential Phosphine-Producing Reactions

### 3. 1. Introduction to Thermodynamics of Phosphine Production

In the absence of the kinetic data for chemical reactions that could lead to phosphine formation we employ a thermodynamic approach to investigate the plausibility of phosphine production on Venus.

Thermodynamics predicts the free energy to be gained from allowing a system to relax to equilibrium. For example, a gas mixture of hydrogen and oxygen will be predicted to be of higher energy than the same gas mixture in which some hydrogen has been reacted with some oxygen to produce water, and so we predict that if the system reacts then water will be produced. The *amount* of each of the reagents is an important component of this analysis (see Supplementary Section 1.3. for more detailed exposition and Supplementary Section 2.1. for an example thermodynamics calculation). We can ask whether the reaction

$P_4O_{10} + H_2 \rightarrow PH_3 + H_2O$

will have a net positive or a net negative free energy of reaction *only if* we know the concentrations of all four reactants. If the energy of reaction is positive given the concentration of these species under Venusian conditions, then thermodynamics predicts that the forward reaction will require energy, the back reaction will release energy, and so the reaction will proceed *from* $PH_3$ and $H_2O$ *to* $P_4O_{10}$ and $H_2$. Thus if this reaction is predicted to have a positive free energy, given the abundance of atmospheric gases and assuming 1 ppb $PH_3$, then we can say robustly that the production of 1 ppb $PH_3$ by this reaction is not consistent with our knowledge of Venus. This is not to say that *no* $PH_3$ could be produced. For example, if the abundance of phosphine was $4.3 \times 10^{-34}$ ppb, and all the other reagents were at the temperature, pressure and concentration expected at 60 km on Venus, then the reaction above would be at equilibrium. What thermodynamic analysis shows is that, at 60 km, 1 ppb $PH_3$ cannot be explained by formation from this reaction. For every 10 kJ/mol free energy of reaction calculated under the assumption of Venus conditions at the base of the clouds, the abundance of phosphine has to be reduced 23-fold from 1 ppb to bring the reaction to equilibrium.

If no combination of conditions (different temperatures, pressures, reducing agents and concentrations), from any observation or model, yields a negative free energy for this reaction assuming 1 ppb of phosphine, then this reaction can be confidently *ruled out* as a source of 1 ppb phosphine on Venus. A thermodynamic analysis cannot substitute for the full kinetic modelling of chemical reactions. A rapidly reacting system will approach thermodynamic equilibrium. If the reaction is slow compared to the timescale of the system (e.g. transport or observational timescales), then the reaction will not reach equilibrium and phosphine will not be produced regardless of the thermodynamics (as is illustrated by the case of the reduction of calcium phosphate in the high atmosphere of Venus, discussed in Section 4.2). If a reaction is fast but it is thermodynamically disfavored, 1 ppb of phosphine cannot be produced by this reaction under Venus conditions. This is true of catalyzed and uncatalyzed reactions; the reason that the reaction is fast does not matter. Thus, thermodynamics does not predict when a reaction can occur, but predicts when one cannot explain the presence of 1 ppb phosphine. It is therefore a useful tool to rule out possible chemical pathways for phosphine production, if the kinetic data is not available.

We approach the calculation of the thermodynamics of chemical reactions in the Venusian environment by calculating the free energy ($\Delta G$) of any reaction involving stable chemical species detected or modelled in Venus' atmosphere that could generate phosphine, both in the atmosphere and on the surface. We tested hundreds of partial pressure and cloud altitude combinations, for a total of thousands of conditions for each of the dozens of reactions.

We also explore the thermodynamics of the subsurface formation of phosphine by employing the concept of oxygen fugacity of crustal and mantle rocks.

Calculation of the free energy of reaction was performed using standard methods (see Supplementary Information, Supplementary Section 1.3.1., See also Supplementary Tables S3, S4.). Nonideality of gases was calculated using Berthelot's equation (Rock 1969). Solids were assumed to be in their ideal state, i.e. as pure materials. Reactions were chosen as follows. To produce phosphine, a reaction must have 1) a source of phosphorus, 2) a source of hydrogen and 3) a reducing agent. The relative abundance of the sources of phosphorus in the atmosphere were calculated as described below (see Supplementary Information, Supplementary Section 1.3.2.). All reducing gases, potential reducing solids, and gaseous sources of hydrogen that have been measured or modelled were used to construct all possible hypothetical reducing reactions with all sources of phosphorus. The vertical concentration profiles of gases were taken from the photochemical model described above in Section 2 and in Supplementary Information, Supplementary Section 1.1. The thermodynamics of the production of phosphine and of phosphorous acid (which could disproportionate to form phosphine) were also modelled (see Supplementary Information, Supplementary Section 1.3.2. for further details). Detailed modelling of the Venusian subsurface chemistry is not practical, as the rock compositions are not known, and a very large number of different minerals could be present. We therefore modelled the oxygen fugacity ($fO_2$), for a range of temperatures (700 – 1600 K), of subsurface rocks needed to generate phosphine in the subsurface Venusian environment, as described in more detail below in Section 3.2.3. and in Supplementary Information, Supplementary Section 1.3.3. See also Supplementary Tables S7, S8.

### 3. 2. Results of the Thermodynamic Analysis of Potential Phosphine-Producing Reactions

#### 3. 2. 1. Surface and Atmospheric Thermodynamics of Phosphine Production

##### 3. 2. 1. 1. Identification of Dominant Atmospheric Phosphorus Species

Phosphine, a reduced form of phosphorus, is not a dominant species in the oxidized Venusian environment. The oxidized Venusian conditions favor the formation of oxidized phosphorus compounds. To identify the dominant atmospheric phosphorus species, we have modelled the relative abundance of oxidized phosphorus species under Venus' atmosphere conditions.

Both P(+3) and P(+5) oxidized phosphorus species can be present as oxyacids or as acid anhydrides. The thermodynamic model shows that $P_4O_6$ is thermodynamically preferred over $P_4O_{10}$ in Venus' lower atmosphere (<35 km) (Fig. 5). In the lower atmosphere, dehydrated forms of phosphorus dominate over hydrated forms, due to the combination of high temperature and low water concentration.

$P_4O_6$ as a dominant phosphorus species on Venus may be surprising, but it is in agreement with previous theoretical studies on brown dwarfs and gas giants done by (Visscher *et al.* 2006). At temperature and pressure regimes of higher altitudes we find $H_3PO_4$ dominates. Visscher et al find the most stable form of phosphorus in analogous regimes in brown dwarfs is $NH_4H_2PO_4$ (i.e. ammonium dihydrogenphosphate). This species would not form on Venus, where the gas phase concentration of ammonia is essentially zero. Its free acid analogue, which would be formed by incubating $NH_4H_2PO_4$ in acid, is $H_3PO_4$.

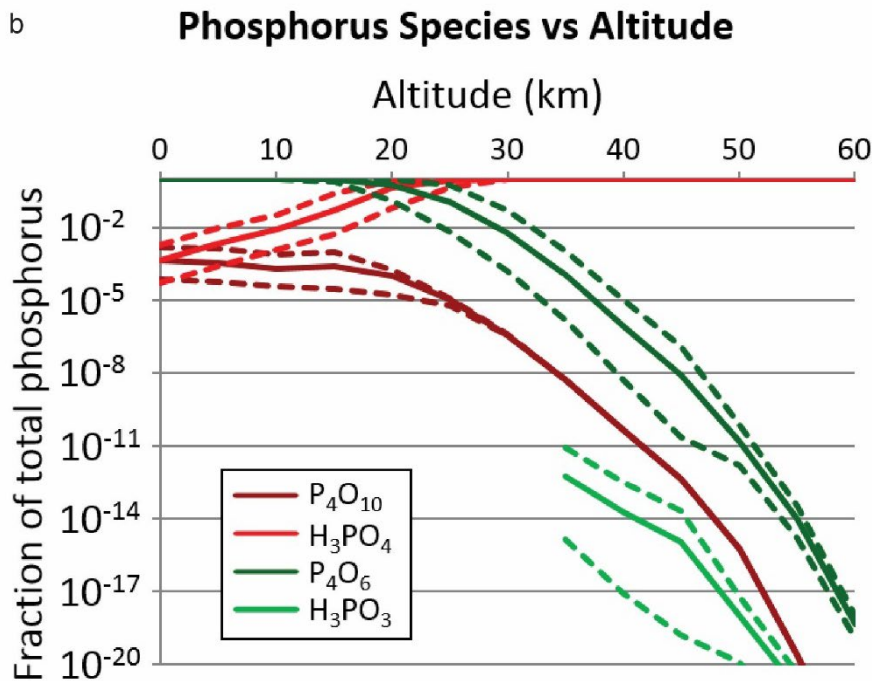

**Fig. 5.** Model of the relative abundance of phosphorus oxyacid species under Venus atmosphere conditions, as a function of altitude. x axis: Altitude (km), y axis: Fraction of total phosphorus. Solid lines show the dominat phosphorus species. Dashed lines show upper and lower limits for the relative fractions of each species, as modelled in different chemical environments (Supplementary Information, Supplementary Section 1.3.2.2.). $P_4O_6$ and $H_3PO_4$ are the thermodynamically dominat phosphorus species in the lower (<35 km) and the upper (>35 km) atmosphere of Venus, respectivley. Note that phosphorous acid ($H_3PO_3$) cannot exist in gas phase, and so only exists in the cloud droplets.

The model predicts that by far the dominant species in the cloud layer is phosphoric acid ($H_3PO_4$). The principle uncertainties in the model are the abundance of water in the atmosphere (which influences the ratio of oxide to oxyacid) and the abundance of reducing agents. We discuss the abundance of reducing agents in the next section.

We note that our model is incomplete. In reality highly concentrated $H_3PO_4$ consists of a mixture of 'pure' $H_3PO_4$, $H_3PO_4 \cdot H_2O$ complexes, and many dehydration products (e.g., $H_4P_2O_7$, $H_5P_3O_{10}$ etc.). However detailed thermodynamic data for such minor phosphorus species under Venus conditions is not available, therefore our model serves as a best possible approximation.

### 3. 2. 1. 2. Formation of Phosphine by Reduction of P species in the Venusian Atmosphere-Surface Environment Cannot Proceed Spontaneously

Our calculations show that formation of phosphine in the Venusian atmosphere and on the surface is very unlikely to proceed spontaneously. None of the tested reactions in thousands of considered conditions that make phosphine or phosphorous acid are thermodynamically favorable. All chemical reactions that can produce phosphine in the Venusian environment are on average 100 kJ/mol too energetically costly (10 - 400 kJ/mol) to proceed spontaneously (see Figure 6, Supplementary Figures S9-S11).

We divide our analysis into separate analyses of reduction reactions and of disproportionation reactions. In principle P(V) or P(III) species present in Venus' atmosphere could be reduced by gases in the atmosphere to form phosphine. P(III) species could also disproportionate to P(V) species and phosphine; specifically, the disproportionation of phosphorous acid ($H_3PO_3$) to phosphine is a well-known laboratory preparation method for phosphine, and it could be that analogous reactions are forming phosphine in Venus' atmosphere. In this section we consider reduction reactions, in the next section (Section 3.2.1.3) we consider disproportionation reactions.

Reduction of P(V) or P(III) species by atmospheric gases is highly unlikely to be a net producer of phosphine. Reduced gases known or modelled to be present in Venus' atmosphere include $H_2$, OCS, CO, $H_2S$, and elemental sulfur (as gas or haze). In the case of OCS, CO and elemental sulfur a third component is needed to convert $P_4O_{10}$ or $P_4O_6$ to $PH_3$ to provide hydrogen atoms. None of the reactions have a negative free energy of reaction under Venus atmospheric conditions.

The reduction of oxidized phosphorus species by surface minerals is ruled out. The only common reduced surface minerals are likely to be iron minerals. Iron(II) sulfide and iron(II) chloride are not stable under Venus surface conditions (Fegley 1997)(Supplementary Figure S8) and reduced iron oxides cannot reduce $P_4O_6$ to $PH_3$ (Supplementary Figure S11).

As noted above, surface mineral phosphorus (if present) is likely to be present as phosphate (Zolotov and Garvin 2020). The reduction of mineral phosphate by reduced atmospheric species to produce $PH_3$ is also highly unlikely thermodynamically (Supplementary Figure S10). We considered four model minerals, calcium phosphate (whitlockite) $Ca_3(PO_4)_2$, calcium fluorophosphate (fluorapatite) $Ca_5(PO_4)_3F$, magnesium phosphate $Mg_3(PO_4)_2$, potassium phosphate $K_3PO_4$, and their reduction by the reducing atmosphere species: $H_2$, OCS, $H_2S$, CO, elemental sulfur ($S_8$ or $S_2$). We note that, although chemical reactions occurring below 30 km are unlikely to be the source of the observed phosphine, there remains the possibility that surface minerals could be transported above 30 km as dust, and so we considered mineral reduction as a source of phosphorus at all altitudes up to 60 km.

We summarize the thermodynamics of reduction of atmospheric and surface phosphorus species in Figure 6, where we show the distribution of number of reduction reactions that make phosphine as a function of their free energy and as a function of altitude (Figure 6).

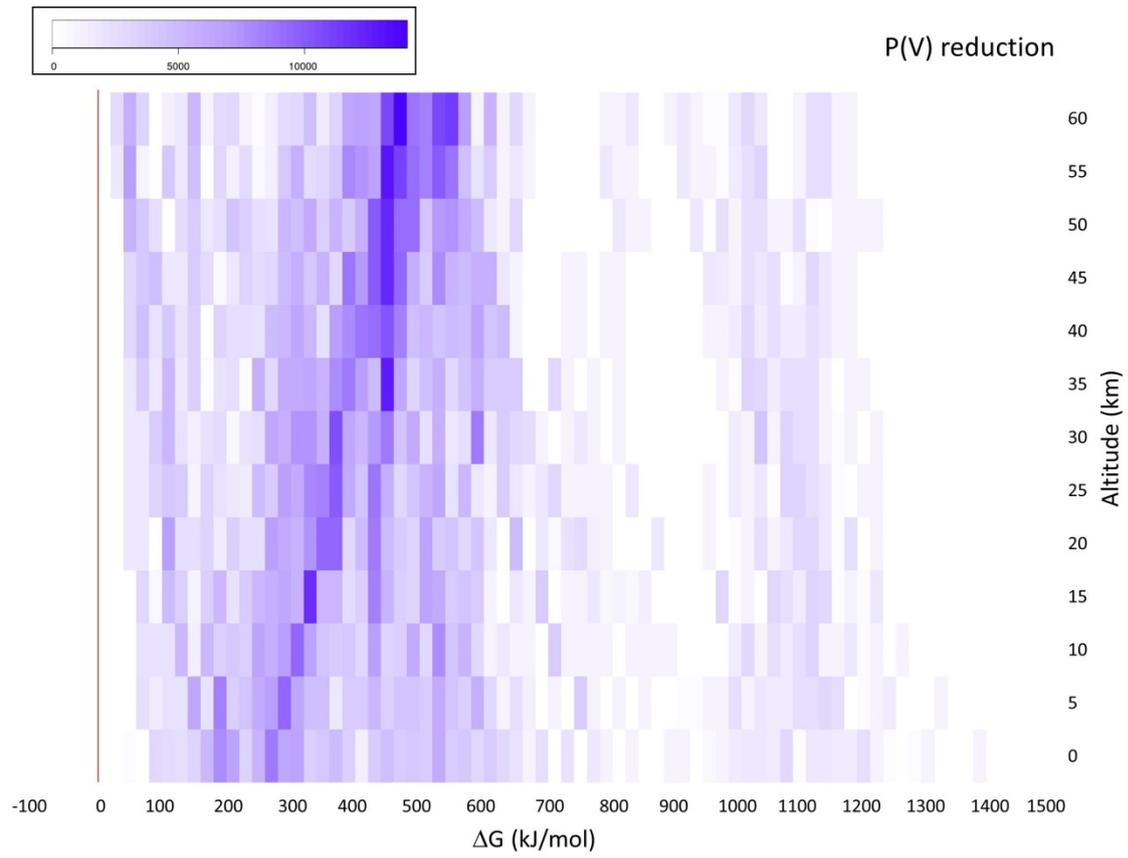

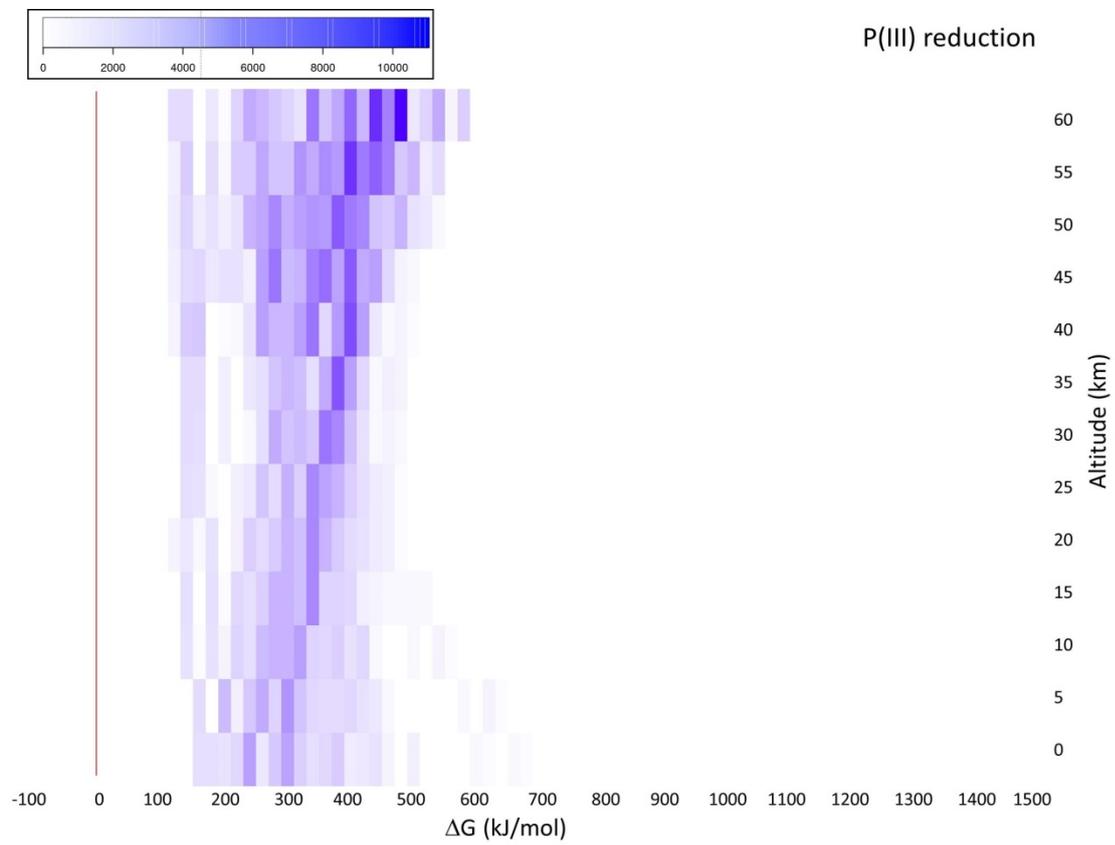

**Fig. 6.** The infeasibility of phosphine production in the Venusian atmosphere and surface by reduction. The y axis shows altitude above the surface and each column (x axis) is a bin of data in a range of Gibbs Free Energy (ΔG). Red vertical line shows ΔG=0. The darker the color of a cell the more reactions/conditions fall within a given ΔG range. The Gibbs Free Energies are from reactions of surface and atmospheric phosphorus species with gaseous or solid reducing agents. Reactions with gases were calculated with a high or a low gas concentration, derived from published data (Table S5), in all combinations, assuming a concentration of 1 ppb $PH_3$. Reduction reactions of $P_4O_6$, $P_4O_{10}$, $H_3PO_4$ and $H_3PO_3$ were considered (the last of these only in solution phase in the clouds), as well as surface reduction of phosphate minerals (see also Table S6). None of the conditions give a negative free energy, which would indicate a reaction that spontaneously produced phosphine. Thermodynamics was only followed to the altitude of the cloud tops, after which phosphorus species and water are expected to freeze out making reactions of stable phosphorus compounds kinetically implausible. Phosphine production by reduction is not thermodynamically favored under the conditions of the Venusian atmosphere, surface and subsurface conditions. Heatmaps were generated using Heatmapper (Babicki *et al.* 2016).

The minimum free energy of reduction reactions under any Venus conditions was found to be 47 kJ/mol, which implies that the maximum phosphine abundance that these reactions can explain is $\sim 2.8 \times 10^{-6}$ ppb.

We note here that reduced phosphorus species are reported to be formed from phosphate under hydrothermal environments on Earth (Herschy *et al.* 2018). However, formation of reduced phosphorus species (e.g. phosphite) under the hydrothermal conditions on Earth is not a model for the formation of reduced phosphorus species formation on Venus. The mechanism of formation of reduced phosphorous species in hydrothermal systems relies on the abundant presence of liquid water. Such conditions cannot occur on the surface of Venus, where water is a trace species and is present as a supercritical gas.

We could argue that the complex atmosphere of Venus is not fully characterized, and specifically that the clouds may be more reduced than we think, and that the more reduced character of the Venusian atmosphere might explain the presence of phosphine. While such a statement is formally true, it is not supported by the current observational evidence of the atmosphere of Venus and therefore it is unlikely.

### 3. 2. 1. 3. *Formation of Phosphine from Disproportionation of P(III) Species in the Venusian Atmosphere-Surface Environment Cannot Proceed Spontaneously*

The disproportionation of the P(III) compound phosphorous acid ($H_3PO_3$) to form phosphine via the reaction:

$4H_3PO_3 \rightarrow PH_3 + 3H_3PO_4$

is a well-known laboratory preparative route for phosphine. We therefore ask whether similar chemistry could be a source of phosphine on Venus. $H_3PO_3$ cannot exist in gas phase, where P(III) oxides are present solely as $P_4O_6$; however $H_3PO_3$ can exist in solution in water and in concentrated sulfuric acid (Sheldrick 1966). Thus $H_3PO_3$ could be formed in the cloud decks by reduction of $H_3PO_4$, or by solution of $P_4O_6$ in the liquid phase of the clouds. It could also be formed as a transient intermediate in a reaction of $P_4O_6$ with an H-bearing species at any altitude, and then rapidly disproportionate to phosphine and $H_3PO_4$ or $P_4O_{10}$.

There are therefore two classes of reactions that could lead from P(III) species to phosphine; the conversion of $P_4O_6$ to a notional intermediate $H_3PO_3$ in gas phase, where it can be considered an intermediate that immediately disproportionates to $PH_3$ and $H_3PO_4$, and the formation of $H_3PO_3$ in liquid phase in cloud droplets that subsequently rain out to the lower regions of the atmosphere where $H_3PO_3$ disproportionates to $PH_3$. Below we show that both are highly unlikely as sources of phosphine.

For $P_4O_6$ to be converted to $H_3PO_4$ and $PH_3$ (via the notional formation of $H_3PO_3$ as a reaction intermediate, as recently discussed by (Schulze-Makuch 2021)), a source of hydrogen atoms is required. $H_2O$, HCl and $H_2S$ could potentially be such a source. The reaction with water is well known on Earth, where $P_4O_6$ dissolves in clean, cold liquid water to form phosphorous acid. Notional reactions of $P_4O_6$ with HCl yield $PCl_3$ as a co-product, and with $H_2S$ yield a phosphorus sulfide. The free energy of $PCl_3$ is known. $P_4S_3$ is the only phosphorus sulfide, of many, for which gas phase free energy of formation is known. We therefore use the following reactions as models for reaction of $P_4O_6$ with $H_2O$, HCl, and $H_2S$ (where the species in square brackets are transient intermediates) (see also Figure S12):

1. $P_4O_6 + 6H_2O \rightarrow$ *[$4H_3PO_3$]* $\rightarrow 3H_3PO_4 + PH_3$
2. $P_4O_6 + 1½H_2O \rightarrow$ *[$4H_3PO_3$]* $\rightarrow ¾ P_4O_{10} + PH_3$
3. $P_4O_6 + 3HCl + 3H_2O \rightarrow$ *[$3H_3PO_3 + PCl_3$]* $\rightarrow 2¼ H_3PO_4 + ¾ PH_3 + PCl_3$
4. $P_4O_6 + 6HCl \rightarrow$ *[$2H_3PO_3 + 2PCl_3$]* $\rightarrow 1½ H_3PO_4 + ½ PH_3 + 2PCl_3$
5. $P_4O_6 + 9HCl \rightarrow$ *[$H_3PO_3 + 3PCl_3 + 3H_2O$]* $\rightarrow ¾H_3PO_4 + ¼PH_3 + 3PCl_3$
6. $P_4O_6 + 3H_2S \rightarrow$ *[$2H_3PO_3 + ½P_4S_3 + 1½S$]* $\rightarrow ½P_4S_3 + 1½S + ½PH_3 + 1½H_3PO_4$

There is no realistic way to estimate the concentration of $P_4S_3$, $PCl_3$ or $POCl_3$, so this was assumed for the sake of exposition to be $10^{-13}$ ppt (i.e. 1 part in $10^{22}$). This corresponds to around 1 molecule per liter at 1 bar, and seems a plausible lower limit (lower concentrations favor the forward reaction producing phosphine, so this is a conservative assumption).

The free energy of reaction of the reactions above, as well as the disproportionation of $H_3PO_3$ in the clouds, was calculated for all combinations of plausible Venus conditions. The result is summarized in Figure 7. No set of conditions favor the production of phosphine.

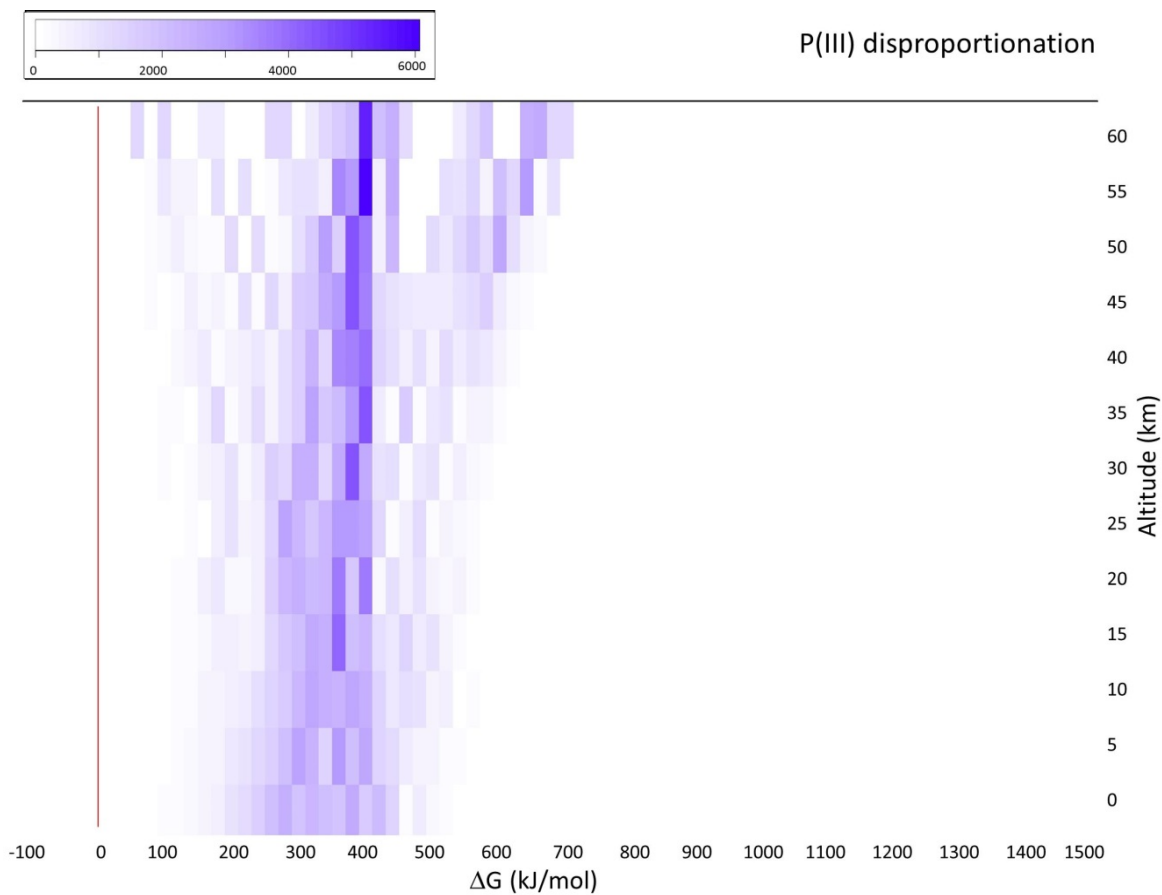

**Fig. 7.** The infeasibility of phosphine production in the Venusian atmosphere and surface by disproportionation. The y axis shows altitude above the surface and each column (x axis) is a bin of data in a range of Gibbs Free Energy (ΔG). Red vertical line shows ΔG=0. The darker the color of a cell the more reactions/conditions fall within a given ΔG range. The Gibbs Free Energies are from disproportionation reactions of P(III) species, either $P_4O_6$ via $H_3PO_3$ as an intermediate or from $H_3PO_3$ in the cloud deck (see also Table S6). None of the conditions give a negative free energy, which would indicate a reaction that spontaneously produced phosphine. Thermodynamics was only followed to the altitude of the cloud tops, after which phosphorus species and water are expected to freeze out making reactions of stable phosphorus compounds kinetically implausible. Phosphine production by disproportionation is not thermodynamically favored under the conditions of the Venusian atmosphere, surface and subsurface conditions. Heatmaps were generated using Heatmapper (Babicki *et al.* 2016).

The minimum free energy of disproportionation reactions under any Venus conditions was found to be 22 kJ/mol, which implies that the maximum phosphine abundance that these reactions can explain is ~$1.1 \times 10^{-3}$ ppb.

Why does reaction $P_4O_6 + 6H_2O \rightarrow$ *[$4H_3PO_3$]* $\rightarrow 3H_3PO_4 + PH_3$ presented above not produce phosphine on Venus when analogous chemistry does in the lab on Earth? There are three reasons. First, the terrestrial lab reaction is done in liquid water. The activity of water in liquid water is much higher than the activity of gaseous water in Venus' very dry atmosphere; low water activity disfavors the reaction. Secondly, the reaction is strongly disfavored at high temperatures (where $P_4O_6$ is abundant on Venus); at low temperatures (such as at cloud level) $P_4O_6$ is a rare, trace species, disfavoring the forward reaction. Lastly, reaction $P_4O_6 + 6H_2O \rightarrow$ *[$4H_3PO_3$]* $\rightarrow 3H_3PO_4 + PH_3$ is *not* thermodynamically favored under terrestrial lab conditions. The reaction that is favored is the reaction of $P_4O_6$ with liquid water to form $H_3PO_3$ solution. If that solution is then dried (which requires input of energy) to yield pure $H_3PO_3$, and if that $H_3PO_3$ is then heated (which requires input of energy) then the state of the system is changed such that the reaction to form $PH_3$ is favored. Adding $P_4O_6$ to cold water on its own does not produce $PH_3$, even on Earth, because solutions of $H_3PO_3$ do not spontaneously disproportionate under cold water temperature conditions.

Finally, $H_3PO_3$ could be formed either by solution of $P_4O_6$ or by reduction of $H_3PO_4$ in the cloud droplets, where $H_3PO_3$ is stable. If droplets then fall to lower, hotter regions of the atmosphere, that $H_3PO_3$ could disproportionate to form $PH_3$. We evaluate the amount of phosphine that this process could produce below.

Figure 5 shows the equilibrium amounts of phosphorus species as a function of altitude in the atmosphere, which illustrates that $H_3PO_3$ is a small fraction of phosphorus species at cloud level on Venus. This calculation takes into account the equilibrium between $P_4O_6$ in gas and droplet phase and $H_3PO_3$ in droplet phase, and so accounts for the equilibrium

$$P_4O_6 + 6H_2O \leftrightarrow 4H_3PO_3$$

We calculate the amount of $H_3PO_3$ in the clouds as follows. We can calculate from the volume of cloud particles (Esposito *et al.* 1983), the fraction of P that is present as $H_3PO_3$, and assuming that P is present in all cloud particles at an average of 1 molar, that there would be 0.25 millimoles, or 20 milligrams in the *entire* cloud deck of Venus (See Supplementary Section 2.5.2; Table S12). This assumes that the cloud layer extends down to 40 km, which is probably below its actual extent.

$H_3PO_3$ would not disproportionate appreciably in the temperature regime of the clouds. The kinetics of the disproportionation of phosphorous acid have not been studied. We can however use the disproportionation of hypophosphorous acid ($H_3PO_2$) as a proxy:

$$3H_3PO_2 \rightarrow 2H_3PO_3 + PH_3$$

As $H_3PO_3$ accumulates as a product of this reaction, $H_3PO_3$ cannot disproportionate substantially faster than $H_3PO_2$ under the same conditions. The kinetics of $H_3PO_2$ disproportionation have been studied (Shechkov *et al.* 2003). The first order rate of disproportionation at 420 K (approximately the cloud base temperature) is $\sim 2 \times 10^{-4}$ sec$^{-1}$. If this rate applies to $H_3PO_3$ disproportionation, then a constant concentration of 22 mg/planet of $H_3PO_3$ disproportionating at this rate would produce a flux of phosphine of $\sim$ 130 grams/year/planet (assuming that the solution of $P_4O_6$ into droplets and its subsequent conversion to $H_3PO_3$ was not rate limiting). The disproportionation of the $H_3PO_3$ would be faster if droplets containing $H_3PO_3$ fell below the cloud layer and evaporated (Seager *et al.* 2021), and the $H_3PO_3$ disproportionated in the higher temperature regions of the lower atmosphere. However, to generate 800,000 tonnes/year of $PH_3$ necessary to maintain a constant 1 ppb $PH_3$ through this mechanism the entire cloud deck would have to 'rain out' every microsecond, which is ridiculous. These calculations discount the fact that $P_4O_6$ is oxidized in sulfuric acid to $H_3PO_4$ (Krasnopolsky 1989), which will further reduce the concentration of *all* P(III) species in the droplets.

### 3. 2. 2. Sensitivities in Thermodynamics Analysis

We note that the sensitivity analysis to the concentrations of gases in the Venusian atmosphere shows that only very substantial systematic errors (at least $10^4$-fold difference) in gas abundance measurements or modelling could account for the production of phosphine. Such dramatic differences from current expectation are therefore highly unlikely (see Supplementary Section 2.3. and Figure S13).

If an unknown, non-volatile material that was a less powerful reducing agent than hydrogen was present in the clouds, could it reduce phosphoric acid to phosphorus acid? (If it were more powerful than hydrogen then it would split water and generate hydrogen, as discussed below.) This cannot be definitively ruled out in the absence of specifics, but if hydrogen cannot reduce phosphoric acid to phosphorous acid under Venus conditions, then a less powerful reducing agent is unlikely to be able to do so.

### 3. 2. 3. Subsurface Thermodynamics of Phosphine Production

#### 3. 2. 3. 1. *Phosphorus Abundance in Venusian Rocks*

The abundance of phosphorus in Venusian rocks is not known. The only direct measurement of the composition of Venus is X-ray fluorescence data from the Vega landers. These did not detect phosphorus, although it detected abundant silicon (Smrekar *et al.* 2014). As phosphorus and silicon X-ray fluorescence signals are very close (Leake *et al.* 1969), all the Vega result can tell us is that at the site of the Vega landers phosphorus was not an abundant element compared to silicon. Models of the bulk composition of Venus suggest one similar to the Earth (Smrekar *et al.* 2014). We therefore take the Earth as our model.

A survey of igneous terrestrial rock shows a wide range of phosphorus content, but an average of ~0.2% P by weight. Notably, isotopic markers of lower mantle rocks (Hart *et al.* 1992) are not associated with increased phosphorus content. This suggests that both surface volcanism and mantle plume volcanism will produce rocks with similar phosphorus content (see Supplementary Section 2.4.4. for details).

#### 3. 2. 3. 2. *Formation of Phosphine in the Venusian Subsurface Environment Cannot Proceed Spontaneously*

Volcanism could contribute phosphine to the atmosphere through two mechanisms. The first is if the equilibrium thermodynamics of rocks near the surface (i.e. in the upper mantle or crust) favored phosphine production. The second is if rocks from the lower mantle, under different conditions of temperature and pressure, could be brought to the surface through plume volcanism and react to generate phosphine. In this section we address the first, surface chemistry source. In Section 3.2.4. below we address mantle plume volcanism.

We note that the rate of volcanism on Venus is not known. Studies of surface topology and cratering suggest that Venus is volcanically active. Volcanism is believed to be primarily through hotspot volcanism driven by mantle plumes, and not plate tectonics. Smrekar et al identify 9 volcanic hotspots (Smrekar *et al.* 2010) analogous to the Hawaii island chain on Earth (for comparison, Earth has 6-8 such currently active plume volcanic regions, depending on definition and scale). (Gülcher *et al.* 2020) identify 37 potentially currently active volcanic areas (compared to an average of 32 terrestrial volcanoes that have erupted in any one of the past 50 years (Siebert 2013)). The volume of flood volcanism on Venus is more than five times the combined area of flood volcanic basalts on Earth (Ivanov and Head 2013), Byrne (Byrne 2020) reviewed evidence of volcanism on Venus, and concluded that the planet was probably 2-3 times as volcanically active as Earth. However (Mikhail and Heap 2017) postulated that the overall volcanic flux is much lower than that on Earth. We must conclude therefore that the rate of volcanism on Venus is unknown, but is unlikely to be more than five times that on Earth.

We use thermodynamics to estimate the potential production of phosphine by crustal volcanism. It is impractical to perform calculations of the thermodynamics of specific reactions in the subsurface of Venus, because the composition of the rocks is not known and the thermodynamics of individual reactions are not known. We therefore simplify the problem of calculating whether subsurface chemistry could generate phosphine by using the concept of oxygen fugacity ($fO_2$). Oxygen fugacity is the notional concentration of free oxygen in a mineral at thermodynamic equilibrium; the higher the concentration, the more oxidizing the rock is. (See (Frost 1991) and Supplementary Section 1.3.3. for more details on $fO_2$ and its calculation. See also Supplementary Tables S7, S8.). A higher oxygen fugacity (concentration of free oxygen in the crustal rocks) means a more oxidized rock and a lower probability of reduction of phosphates. We find that the oxygen fugacity of plausible crust and mantle rocks is 8 - 15 orders of magnitude too high to support reduction of phosphate. It is therefore extremely unlikely that subsurface activity on Venus, including volcanism, would produce substantial amounts of phosphine.

We present our reasoning as follows. We compared the fugacity of the phosphate/phosphine equilibrium to the fugacity of standard mineral buffers representative of terrestrial rocks. The results are shown in Figure 8.

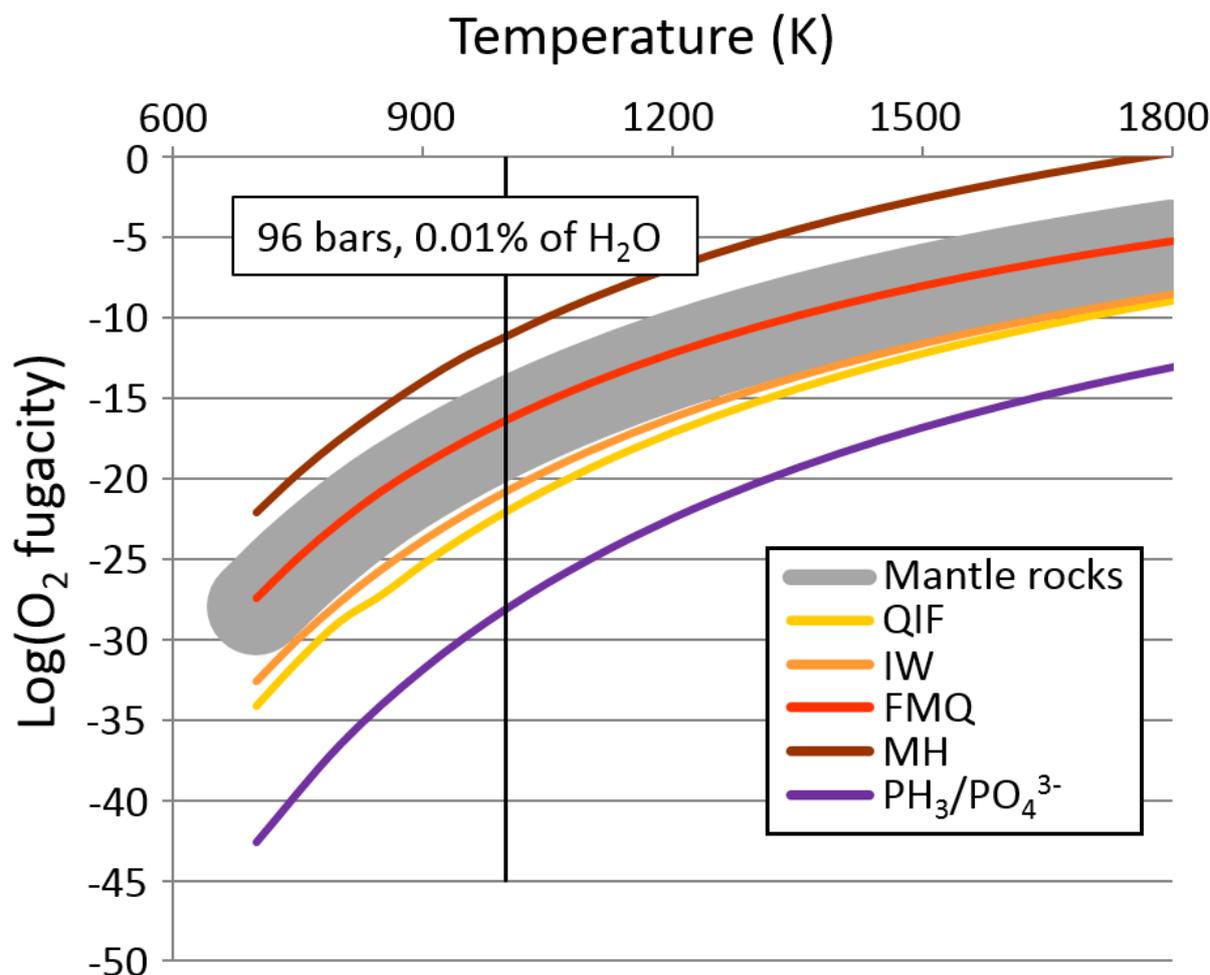

**Fig. 8.** Comparison of the fugacity of the phosphate/phosphine equilibrium to the fugacity of the standard mineral buffers of terrestrial rocks. x axis: log $O_2$ fugacity, y axis: Temperature (K). Fugacity of the production of phosphine from phosphate minerals is calculated for 96 bars and 0.01 % water in the rocks. The fugacity of the phosphate/phosphine equilibrium is shown as a purple line. The other curves are $O_2$ fugacities of standard rock buffers. The phosphate/phosphine $fO_2$ curve lies below the QIF buffer line (the most reduced rock of the buffers shown) which falls below the typical $fO_2$ of terrestrial mantle or crustal rocks (grey band region). Therefore, typical terrestrial rocks are too oxidized to produce $PH_3$ from phosphates and the formation of phosphine is highly unlikely under Venusian subsurface conditions.

To interpret any $fO_2$ curve, any point above a fugacity line will mean that the oxidized member of a reaction will be favored, anything below a fugacity line means that the reduced member is favored.

The phosphate/phosphine $fO_2$ curves lie substantially below the QIF buffer line, which itself falls well below the typical $fO_2$ of mantle or crustal rocks. Rare cases of very reduced rocks are found in some locations, e.g. (Ulff-Møller 1985), with an $fO_2$ of ~QIF-1. However, such rocks are unlikely to contain any water, because it would react with the metallic iron in the rock. The $fO_2$ of Lunar and asteroidal olivines and plagioclase is usually around IW-2 to IW+2 (Karner *et al.* 2004). All of them are too oxidized to produce $PH_3$ from phosphate. This means that in crustal and mantle rocks, phosphorus will overwhelmingly be present as phosphate.

The results of our fugacity calculations are also supported by observations that $PH_3$ is not known to be made by volcanoes on Earth, although in principle reduced phosphorus species could be produced in ocean-floor hydrothermal systems through serpentenization reactions (Pasek *et al.* 2020) (an environment with no analogue on Venus). Estimation of the production of $PH_3$ through volcanism on a simulated anoxic early Earth concluded that only trace amounts of volcanic phosphine can be produced through this process. The predicted maximum production rate of phosphine on the early Earth is only ~100 tonnes per year (Holland 1984), even assuming a highly reduced planet with abundant water. The volcanic production of phosphine in more oxidized, dehydrated planetary scenarios is even more unlikely.

The redox state of the crustal rocks on Venus is unknown. The relatively reduced QIF buffer is an Fe(II)/Fe(0) buffer: to have a substantially more reducing rock, a more electropositive metal than iron would need to be present in significant amounts as elemental metal, which itself would imply that all the iron (and nickel) in the rock would have to be reduced to elemental metal as well. This is a possible but implausible scenario.

We validate our approach by calculation of the fugacity of the terrestrial $H_2S/SO_2$ equilibrium. The results from the computed $SO_2/H_2S$ line (Figure S14) are qualitatively consistent with field observations on Earth and modelling on Mars (see Supplementary Section 2.4.1.). Another way to demonstrate that subsurface chemistry cannot generate atmospheric phosphine is to consider the amount of volcanism that would be necessary to generate the observed amount of phosphine in the atmosphere. We find that to maintain ~1 ppb of $PH_3$ on Venus a volcanic flux many orders of magnitude greater than that on Earth is required. We modelled volcanic outgassing as follows.

The thermodynamics inherent in Figure 8 does not state that phosphine cannot be made by geochemistry, just that the ratio of phosphine to phosphate would be extremely small. We estimate the amount of volcanism that would be needed to maintain an atmospheric abundance of ~1 ppb as follows. We calculated the ratio of phosphate to phosphine (formally of P(+5):P(-3)) that would be produced by volcanic rocks using the $f(O_2)$ approach described above, based on the $f(O_2)$ values of six redox buffers with redox states between IW (Iron/Wustite: Fe/FeO) and MH (Magnetite/Haematite: $Fe_3O_4/Fe_2O_3$) buffers, including the IW and MH buffers themselves, and for a range of temperatures, pressures and rock water content that reflect the extreme ranges plausible for Venus' crust. From this, the amount of phosphorus that would have to be erupted to provide the flux of 25.96 kg/sec (needed to maintain an abundance of ~1 ppb in the atmosphere) can be calculated. These fluxes are shown in Figure 9. (See Supplementary Information Section 2.4.3 (Figure S16) and Supplementary Information Section 2.4.4 (Figure S17) for details of the data sources and calculations).

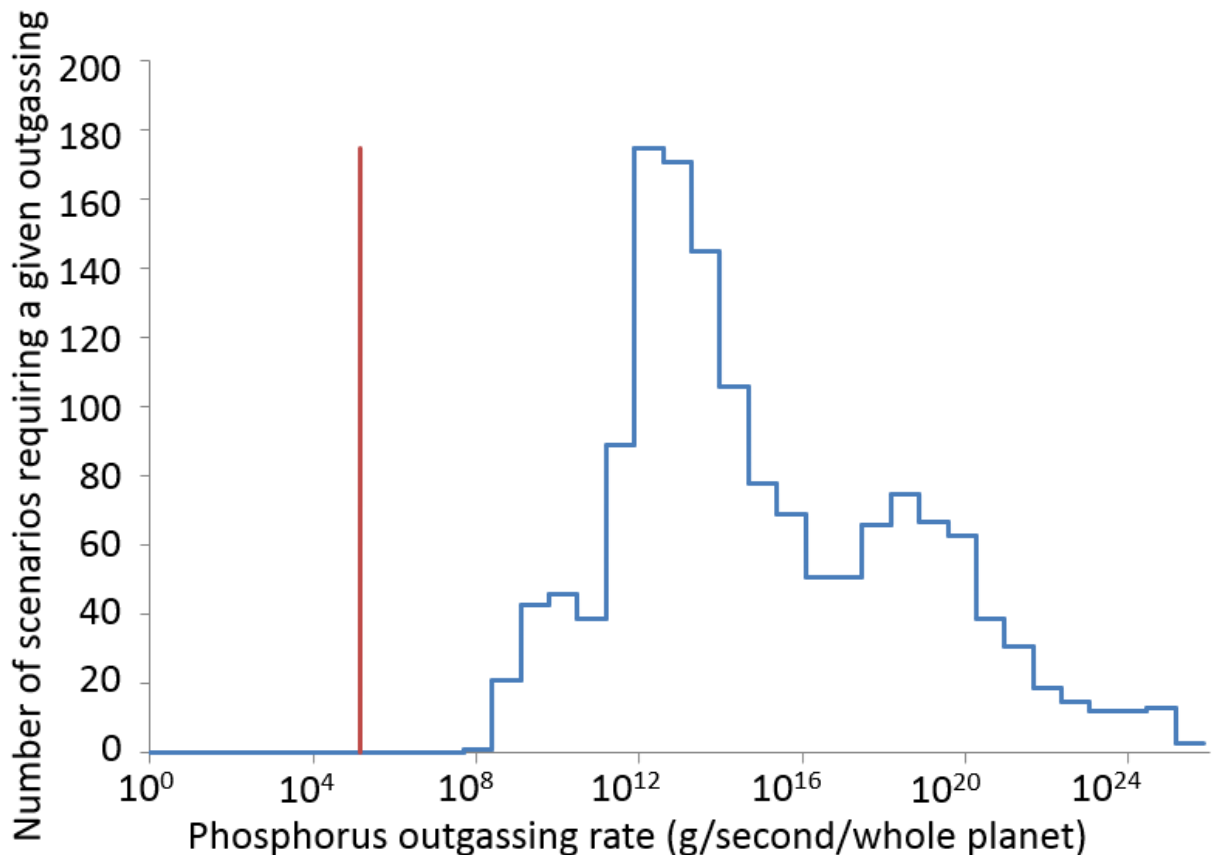

**Fig. 9.** The assessment of the volcanic production of phosphine. x axis: total phosphorus outgassing rate in grams of phosphorus per second across the whole planet, binned in log(5) bins. y axis: number of combinations of $f(O_2)$ buffer, temperature, pressure and water content for which that outgassing rate provided 25.96 kg/sec phosphine flux. Blue line – model output. Red line – estimated terrestrial phosphorus outgassing flux (See Supplementary Section 2.2.3 for details). To explain the observed abundance of phosphine at least many hundred times more volcanism on Venus than on Earth is required.

Few conditions require a total flux of less than $10^9$ grams of phosphorus per second. For comparison, the flux of phosphorus from modern day Earth volcanism (of all sorts) is ~ 143 kg/second (shown as a vertical red line of Figure 9 – see Supplementary Section 2.4.3 for details). This is 1390 times lower than the most extreme rate predicted for Venus, representing outgassing at 90 bar and 1600 K, with a fugacity of the Iron/Wustite buffer (at the bottom range of plausibility for mantle rocks), and from rocks containing 1.5 % water. 1.5 % water is a high value even for terrestrial mid-ocean ridge and mantle plume melts (Moore 1970; Saal *et al.* 2002; Weis *et al.* 2015). Arc volcano magmas can have up to 5 % water (Anderson 1973), but as these are directly derived from subducted crustal ocean floor their high water levels cannot be replicated on a planet without oceans. More realistic values of $f(O_2)$, water content, temperature and pressure require tens of thousands of times more volcanism on Venus than on Earth to produce the amount of phosphine required. We note that the Venusian crust (and by inference the upper mantle, due to the resurfacing event) may be more oxidized than Earth (Wordsworth 2016), making the lower outgassing rates even less probable. We consider it highly unlikely that Venus has more than >1000 times the volcanic activity of Earth needed to explain the presence of phosphine in its atmosphere.

Fugacity is dependent on pressure, temperature and water concentration. We probed the sensitivity of our conclusions to variation in all three parameters (see Supplementary Section 2.4.2. and Figure S15). No realistic values of pressure (up to 10,000 bar), water content (up to 5%) or temperature (up to 1800 K) can support phosphine production (Figure S15). We note that phosphorous acid and phosphites cannot be produced by volcanoes, as they break down at temperatures >~450 K. We discuss other reduced phosphorus species in Section 3.2.1.2. and Section 3.2.1.3.

### 3. 2. 4. Mantle Phosphides as a Source of Phosphine

Our argument above suggests that crustal and upper mantle phosphorus is overwhelmingly present as oxidized P(V), and that these are unlikely sources of $PH_3$. By contrast, Earth's lower mantle contains at least some regions that are highly reduced (Smith *et al.* 2016) and in which phosphorus is likely to be present as phosphides rather than phosphate. Phosphides are stable to extremely high temperatures and pressures (Japel *et al.* 2002), and so could be formed deep in the mantle and brought to the surface through plume volcanism, if such volcanism occurs on Venus. Mineral phosphides are hydrolyzed by acid solutions in water to form phosphine (Bumbrah *et al.* 2012), or phosphite or hypophosphite which could subsequently disproportionate to phosphine (Pasek and Lauretta 2005). Therefore, if phosphide were erupted from the lower mantle to the surface it could be converted to phosphine. We note that Truong and Lunine have recently suggested that plume volcanism could be a source of 1 ppb phosphine (Truong and Lunine 2021). We find this scenario unlikely for four, independent reasons.

Firstly, it is not clear that lower mantle phosphides are commonly erupted to the surface unchanged. On Earth, mantle plume magma is estimated to rise on a timescale of $10^6 - 10^7$ years at temperatures in excess of 3000 K (Condie 2001), during which time phosphorus species would reach thermodynamic equilibrium relevant to the temperature and pressure of the upper mantle and then the base of the crust, i.e. P(V) phosphate or P(III) $P_4O_6$-related anions. Thus, although mantle plume volcanism *originates* in the lower mantle, its chemistry will not be *lower mantle chemistry* by the time it erupts.

Secondly, it is not clear that solvolysis of mineral phosphides in concentrated sulfuric acid will generate phosphine. While hydrolysis of industrial-grade iron in dilute (0.5 M) sulfuric acid efficiently generates phosphine (Geng *et al.* 2010), and hydrolysis of phosphide chemicals (Bumbrah *et al.* 2012) and minerals by water or dilute acid generates reduced phosphorus species (Pasek *et al.* 2014), there have been no studies on the reaction of phosphides with *concentrated* sulfuric acid. The attack of concentrated sulfuric acid on materials follows different chemistry than the attack on those same materials by solutions of sulfuric acid in water. Concentrated sulfuric acid is an oxidizing agent, and is known to rapidly oxidize phosphine at low temperatures (see Supplementary Section 2.5.3.1.); hot concentrated sulfuric acid will also oxidize metals such as iron and copper, producing $SO_2$ as a gaseous product (rather than reacting with iron to produce hydrogen, as is the case with dilute acid). A likely outcome of reacting phosphides with *concentrated* sulfuric acid would be an oxidation reaction, such as

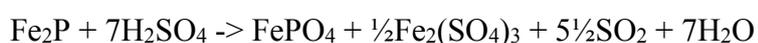

$Fe_2P + 7H_2SO_4 \rightarrow FePO_4 + \frac{1}{2}Fe_2(SO_4)_3 + 5\frac{1}{2}SO_2 + 7H_2O$

In the lower atmosphere where $H_2SO_4$ is likely to be dissociated into $SO_3$ and $H_2O$, direct oxidation of phosphides by $SO_3$ is likely. This is speculative, but is more in line with the known chemistry of sulfuric acid (Liler 1971) than the generation of a highly reducing gas ($PH_3$) using a strong oxidizing agent ($H_2SO_4$). This chemistry could be experimentally tested.

Thirdly, even if we assume both rapid and efficient delivery of phosphides to the surface and rapid and efficient conversion of phosphides to phosphine, the amount of phosphides released into the atmosphere, and scale and frequency of such volcanic eruptions needed for this scenario of phosphine production to be possible, makes it seem unlikely. We argue as follows. If mineral phosphides were efficiently converted to phosphine, then at 800,000 tons of phosphorus would need to be erupted from the lower mantle every year, as noted above, to explain the presence of ~1 ppb phosphine in the atmosphere. An average abundance of ~0.2% of P in terrestrial rocks (by weight) implies a mass of ~$4 \times 10^8$ tons of lower mantle rock would have to be erupted per year to account for the observed phosphine level, or ~0.2 km$^3$ of lower mantle basalt. This would be a quite substantial eruption, and it would be unlikely for it to be happening just when Greaves et al (Greaves *et al.* 2020c) were making their observations. For context, the Siberian and Deccan traps, vast volcanic flood plains that represent the most extensive volcanism in the Phanerozoic on Earth, were probably created by massive plume eruptions that at their peak produced 1.3 km$^3$ of basalt/year, which would deliver ~$10^{12}$ g of phosphorus to the surface, per year, the large majority as phosphate (Renne and Basu 1991; Sen 2001). As discussed above, no evidence for such recent catastrophic volcanism exists on Venus, although more observations would be needed to rule out such active flood volcanism on modern Venus.

Lastly, we note that if solvolysis of phosphide minerals in rocks is generating phosphine, then it would also be expected to generate other reduced gases. Any rock sufficiently reduced to contain phosphorus as phosphide would contain carbon as carbide and sulfur as sulfide, and possibly nitrogen as nitride. Hydrolysis of that rock would produce hydrogen sulfide quantitatively, and possibly methane, acetylene or ammonia as well. The precedent of terrestrial rock abundance suggests an S:P ratio of ~2:1 as noted in SI section 2.4.3. There is no evidence for anomalous hydrogen sulfide levels in the Venusian atmosphere.

We conclude that four lines of argument suggest that lower mantle volcanism is not a source of phosphine on Venus. However, if lower mantle plume volcanism is generating significant atmospheric gases, there should be other evidence such as extensive highly reduced flood basalts on the surface. This prediction can be tested by *in situ* sampling.

### *3. 2. 4. 1. Possible Scenarios for a Volcanic origin of Venusian Phosphine*

What set of assumptions could allow volcanism to explain the presence of 1 ppb phosphine in the atmosphere? Photochemical models are complex, and the estimates of phosphine production rates derived thereof are not easy to derive from first principles. Here we outline a simple analytical calculation to place an extreme lower bound on phosphine production rate required to explain the reported signal. The advantage of this approach is that it is easy to understand and validate. In this way, it completements the full photochemical calculation also presented in this paper.

Photochemical models agree that phosphine will be efficiently destroyed at high altitudes by photolysis and reactions with photolytically-generated radicals; phosphine that diffuses to this altitude must be destroyed, and its destruction must be counterbalanced by production in steady-state. We can calculate an extreme lower bound on the minimum production flux of phosphine required to compensate solely for destruction by transport to the upper atmosphere and explain the reported 1 ppb phosphine detected at 61 km altitude as follows:

For reasons discussed above, we assume a phosphine column equivalent to 1 ppb phosphine at altitudes ≥61 km. 1 ppb of phospine at 61 km corresponds to a phosphine column density ($N_{PH3}$; cm$^{-2}$) at $z$ ≥61 km only if

$$N_{PH3} = r_{PH3} * p(61 \text{ km}) / (g_{Venus} * u_{Venus})$$

Where $r_{PH3}=10^{-9}$ is the phosphine mixing ratio, $p(61 \text{ km})=1.94 \times 10^5$ Barye is the atmospheric pressure at 61 km, $g_{Venus}=887$ cm s$^{-2}$ is the acceleration due to gravity on Venus, and $u_{Venus}=43.45$ amu is the mean molecular mass of the Venusian atmosphere[2]. Then, $N_{PH3} = 3 \times 10^{15}$ cm$^{-2}$. In reality due to diffusion there must also be also be substantial phosphine below 61 km, but, in the interest of conservatism, let us ignore that: at least this much $PH_3$ must be present to explain the reported signal (if $PH_3$ is present at all).

Now, let us calculate the timescale for this $PH_3$ to circulate to the upper atmosphere, where it is destroyed. We consider this altitude to be $z_1$=98 km, corresponding to the altitude by which $PH_3$ lifetime is <1 s in all three of the models of the Venusian atmosphere we consider here. This vertical mixing timescale is (Jacob 1999, Equation 4.23):

$$T = \frac{(\delta z)^2}{2 \cdot K_z}$$

Where $\delta z$ is the height difference and $K_z$ is the eddy diffusion coefficient. The longer the timescale then the slower the transport of $PH_3$ to the destruction altitude and hence the lower the flux needed to explain a given abundance. In the interest of conservatism, we choose extremal $\delta z$ and $K_z$ to maximize the timescale and minimize the required $PH_3$ production flux. $\delta z = z_1 - z_0$, where $z_0$ is the altitude where $PH_3$ is transported from. A reasonable choice for $z_0$ would be 61 km, the $PH_3$ detection altitude; we instead adopt $z_0$=0, i.e. assume it takes as long for phosphine to diffuse to 98 km from the surface as it does from 61 km. We make this unphysical choice in the name of conservatism. So, $\delta z$= 98 km. Finally, we adopt $K_z$ = 2200 cm$^2$ s$^{-1}$, i.e. the minimum eddy diffusion coefficient in the atmosphere (realized near the surface). Then:

T = (98 km)$^2$/(2*2200 cm$^2$ s$^{-1}$) = 2.2x10$^{10}$ seconds

---
[2] https://nssdc.gsfc.nasa.gov/planetary/factsheet/venusfact.html, accessed 2/1/2021

Inverting, i.e. assuming that destruction is solely by transport from surface to 97 km, the corresponding "destruction rate" is: $4.5 \times 10^{-11}$ s$^{-1}$

By multiplying these two numbers, we arrive at our very conservative lower bound on required production flux:

$1.3 \times 10^5$ cm$^{-2}$ s$^{-1}$

Which is $\sim 10^{-3}$ times the rate estimated from a full photochemical model. This is the flux that could be produced by surface volcanoes under the most extreme of the scenarios modelled above, and is within what might be expected from mantle plume delivery of phosphides to the surface assuming that those phosphides were efficiently converted to phosphine.

To summarise, we can explain the phosphine on Venus as being the result of volcanism only if we assume that *both* [1] *and* [2] below are true

[1]: volcanism can produce phosphine at 800 tonnes/year or more:-
   EITHER plume mantle volcanism deliver phosphine efficiently to the surface (which requires:-
   - Deep mantle phosphides traverse the mantle over a period of >1 million years essentially unchanged, AND
   - phosphides are efficiently converted to a highly reducing gas in an oxidizing atmosphere, AND
   - Other reduced volatile elements, notably sulfur, are not converted to reduced gases at the same time
   
   OR Upper mantle / crust volcanism delivers phosphine to the surface, which requires:-
   - The mantle of Venus is substantially more reduced than that of Earth AND
   - The upper mantle contains as much water as MORBs with the highest water content on Earth AND
   - Eruptions occur at 1500 K or above

[2]: phosphine's lifetime is $10^3$ times longer than the photochemical model suggests
   - There is a mechanism that efficiently transfers PH$_3$ from the surface to 61 km, so efficiently that no PH$_3$ remains at low altitudes and PH$_3$ experiences no photochemical loss, OR equivalently the majority of erupted phosphide is transported to 61 km without loss, AND
   - PH$_3$ transport from 61 km to its destruction altitude of ≤98 km is slow, occurring on timescales comparable to the whole-atmosphere circulation timescale. PH$_3$ is not destroyed during this slow transport

The assumptions made in the atmospheric chemistry are not physically plausible, or even self-consistent. Nevertheless, it is necessary to make them to arrive at a PH$_3$ production flux that can match volcanic production in even the most extreme scenario. More physical assumptions regarding photochemistry and transport are represented by the full photochemical model, which produces a required flux three orders of magnitude higher.

The combination of unlikely volcanic chemistry fluxes, unlikely volcanic chemistry and physically unrealistic atmospheric chemistry assumptions appears to us to be unlikely to be applicable to Venus.

### 3. 2. 4. 2.  Other Sources of Phosphide-Containing Materials as a Source of Phosphine

We also exclude other phosphide sources as likely sources of phosphine on Venus.

Mineral phosphides are known on Earth, where they are rare but widely distributed. A mineral fulgurite - a glass resulting from lightning strikes was proposed as a potential source that could in principle contain reduced phosphorus species (Pasek and Block 2009). It is estimated that fulgurites probably contain < 0.5% phosphorus (Gailliot 1980), and are widely stated as being 'rare' (e.g. (Glover 1979; Petty 1936; Pye 1982)). Phosphides can also originate in pyrometamorphic rocks. Pyrometamorphic rocks form as a result of fossil fuel fires, a process that is probably not relevant to Venus (Britvin *et al.* 2019; Kruszewski *et al.* 2020).

We also consider that meteoritic delivery of phosphides to Venus is unlikely as a potential source of observed amounts of phosphine.

Iron-nickel meteorites are known to contain reduced species of phosphorus, mostly as phosphides (Geist *et al.* 2005). Such metal-rich meteorites could also be a source of phosphide and hence, upon its hydrolysis, of phosphine. For example, reduced phosphorus species can be found in the meteoritic mineral schreibersite $(Fe,Ni)_3P$, the most common mineral containing reduced phosphorus (Pech *et al.* 2011), and in other minerals (Buseck 1969; Ma *et al.* 2014; Pratesi *et al.* 2006; Zolensky *et al.* 2008). It has been suggested that schreibersite was a source of reduced phosphorus species on early Earth (Baross *et al.* 2007), and could in principle continue to be a trace source of reduced phosphorus species today.

The accretion rate of meteoritic material to the Earth today is of the order of 20-70 kilotons/year (Peucker-Ehrenbrink 1996). ~6% of this material is iron/nickel meteorites (Emiliani 1992) which contain phosphides at a level of an average of 0.25% phosphorus by weight (Geist *et al.* 2005). If we rely on the extremely conservative assumption that hydrolysis of $(Fe,Ni)_3P$ phosphides to phosphine is 100% efficient, that would deliver a maximum of ~10 tonnes of phosphine to the Earth every year, or about 110 milligrams/second, which is a negligible amount globally (Greaves *et al.* 2020c; Sousa-Silva *et al.* 2020). This estimated maximal yearly meteoritic delivery of phosphine on Venus is ~5 orders of magnitude too low to explain detected amounts.

Our calculations are also in agreement with previous estimates of the phosphine production through meteoritic delivery, which were also found to be negligible (Holland 1984) and with very recent work by Carrillo-Sánchez who show that the great majority of meteoritic phosphorus species is oxidized (even though the severe conditions of atmospheric entry do create trace amounts of elemental P, this elemental P gets readily oxidized as well) (Carrillo-Sánchez *et al.* 2020).

### 3. 3. Conclusions of the Thermodynamic Analysis of Potential Phosphine-Producing Reactions

We show with our thermodynamic analyses that none of the known possible routes for production of $PH_3$ on Venus can explain the presence of ~1 ppb phosphine. All fall short, often by many orders of magnitude.

The thermodynamics of known reactions between chemical species in the atmosphere and on the surface of Venus are too energetically costly and cannot be responsible for the spontaneous formation of phosphine.

Similarly, the formation of phosphine in the subsurface is not favored. Oxygen fugacity of the crustal and mantle rocks is many orders of magnitude too high to reduce mineral phosphates to phosphine.

Finally, we show that the hydrolysis of phosphide minerals, both from crustal and mantle rocks, as well as delivered by meteorites, cannot provide sufficient amounts of phosphine.

## 4. Other Potential Processes of Phosphine Formation

### 4. 1. Potential Endergonic Processes of Phosphine Formation

Several potential sources of energy that could drive the formation of $PH_3$ should be mentioned briefly for completeness, although we argue that none of them could be responsible for the observed abundance of phosphine on Venus.

Lightning strikes cannot create sufficient amounts of phosphine to explain the observed ~1 ppb amounts of phosphine in the atmosphere of Venus. Lightning may be capable of producing a plethora of molecules that are thermodynamically disfavored. However, our calculations suggest that lightning's production of $PH_3$ is at most ~5 orders of magnitude too low to explain detected amounts (Sousa-Silva *et al.* 2020). We estimate that the maximum amount of phosphine produced by lightning in one Venusian year under some very optimistic assumptions is 3.5 tonnes, which is 5 orders of magnitude lower than that necessary to explain ~1 ppb in the atmosphere (see Supplementary Section 2.5.1. (Figure S18; Table S9-S11) for details on the estimation of phosphine production by lightning).

We note that our predicted value of phosphine production through lightning is an upper bound and, in reality, the lightning-induced production of reduced phosphorus species in Venusian atmosphere is likely to be much less efficient. The well-studied formation of analogous N species by lightning strikes on Earth favors formation of nitrates and nitrites, and not the thermally less stable reduced forms of N like ammonia (Ardaseva *et al.* 2017; Mancinelli and McKay 1988; Rakov and Uman 2003).

Moreover, the above calculations agree with several studies on the formation of reduced phosphorus species, including $PH_3$, by laboratory-simulated lightning. Such experiments can produce traces of phosphine from discharges onto phosphate salt solutions, but at very low efficiency (Glindemann *et al.* 1999; Glindemann *et al.* 2004).

Mechanochemically-driven reduction of phosphate to phosphine in rocks, by tribochemical weathering at quartz and calcite or marble inclusions, was postulated as a potential abiotic source of phosphine (Glindemann *et al.* 2005b). However scaling the results presented in (Glindemann *et al.* 2005b) to plausible global earthquake activity (even under very optimistic assumptions that all the rock moved during an Earthquake-induced landslide can be the substrate for this chemistry) suggests that the flux of phosphine produced would be at least two orders of magnitude too small to account for the observed abundance of phosphine in Venus' atmosphere. In addition, tribochemical production of phosphine in crustal rocks requires a local fluid to provide hydrogen atoms, which is very unlikely to be present in Venus' crust. The crustal rocks are above the critical temperature of water and under an atmosphere with $\sim 3 \times 10^{-5}$ partial pressure of water; they are therefore expected to be extremely desiccated with no local hydrogen source. (see Supplementary Section 2.5.5 for more details on tribochemical production of phosphine).

A very large comet or asteroid impact could theoretically generate a highly reduced atmosphere for millions of years that could lead to formation of conditions that are more favorable for phosphine production (Kasting 1990). We note however that a scale of such impact has to be comparable to the hypothetical impact that is postulated to have created a transient $H_2$-rich atmosphere on early Earth ~4.48 billion years ago (Benner *et al.* 2019; Service 2019). Even the Chicxulub impactor, which resulted in a crater 150 km wide and contributed to the extinction of the dinosaurs did not manage to significantly change the redox state of Earth's atmosphere (although it had dramatic effects on radiative balance, and hence climate (Brugger *et al.* 2017; Toon *et al.* 2016)). An impact as large as Chicxulub occurs every 50-100 million years. It is statistically highly unlikely that an even larger cataclysm of this sort happened in recent Venusian history. The radar mapping of the surface of Venus does not show sufficiently large recent craters on the surface of Venus and therefore does not support the recent large impact scenario (Ivanov and Head 2011; Kreslavsky *et al.* 2015). Smaller impacts could only generate phosphine through delivery of meteoritic phosphide, which is insufficient to account for phosphine production as discussed above in Section 3.2.4.2. and in (Greaves *et al.* 2020c).

Lastly, solar X-rays and solar wind protons carry substantial energy, but are absorbed at high altitudes, and so could not penetrate to the clouds where phosphorus species might be found and where phosphine is detected, and hence cannot drive the formation of phosphine.

### 4. 2.     Other Potential Exergonic Processes as Sources of Phosphine

In principle some exotic chemistry on Venus, not considered before, could be responsible for the formation of phosphine. In this section we address a few potential examples, including formation of phosphine from elemental phosphorus or production of phosphine with reducing agents more powerful than molecular hydrogen. We argue that all such scenarios just replace the implausibility of making phosphine with another, equally implausible set of conditions which could then produce phosphine (i.e. a "displaced improbability").

For example, if elemental phosphorus could be erupted from Venusian volcanoes, it could be reduced by atmospheric gases to phosphine. However, the production of elemental phosphorus from phosphate rock chemistry under Venus' conditions is itself extremely improbable on thermodynamic grounds (see Supplementary Section 2.5.3.2. (Figure S19) and Supplementary Section 2.5.3.3. (Figure S21) for details on the possibility of formation of elemental phosphorus on Venus). In principle, elemental phosphorus could be generated from phosphoric acid by reaction with elemental carbon. Graphite has been suggested as the 'unknown UV absorber' (Shimizu 1977). However, the thermodynamics of this reaction do not favor phosphorus production under cloud conditions (see Supplementary Section 2.5.3.2. (Figure S20)). Thus, invoking elemental phosphorus as a source of phosphine by any route replaces the implausibility of making phosphine with the implausibility of making elemental phosphorus.

Other minerals could be suggested as being present on the surface, such as highly reduced lower mantle minerals as suggested above. As another example, we consider Berlinite (aluminum phosphate $AlPO_4$). If Berlinite were present, then there is a possibility that Berlinite be reduced by atmospheric gases at the surface to produce phosphine in a reaction with a negative free energy under Venus conditions. Specifically, the reaction

$4H_2S + AlPO_4 \rightarrow PH_3 + \frac{1}{2}Al_2O_3 + 2\frac{1}{2}H_2O + 4S$

has a negative free energy at below 5 km altitude if $PH_3$ is present at 1 ppb, $H_2S$ is at its highest modelled level and $H_2O$ is simultaneously at its lowest predicted level. This is an improbable but not impossible series of assumptions. However, Berlinite itself is thermodynamically unlikely to be present on Venus' surface (see Supplementary Section 2.5.4. (Figure S23, Table S13)). Similarly, the thermodynamics of calcium phosphate reduction by carbon monoxide to phosphine are favorable at 120 km altitude: however this requires the reaction to occur at 170 K (where any reaction will be extremely slow), and that the mineral to be lofted to this altitude, both of which are extremely unlikely (see Supplementary Section 2.5.3.4. (Figure S22)). The presence of unexpected minerals, or expected minerals at unexpected locations, on Venus is a testable hypothesis that could be the subject of remote or *in situ* observation.

Other reducing agents could exist on the surface of Venus, and be more powerful reducing agents than hydrogen. Previous suggestions for rare Venusian surface minerals include lead or bismuth sulfide, elemental metals or other materials (Schaefer and Fegley Jr 2004; Treiman *et al.* 2016). Some Venusian mountaintops show 'snowcaps' of a highly radar-reflective material. The chemical composition of these deposits is unknown (Taylor *et al.* 2018), and could conceivably be a source of exotic chemistry. However, we know that water is present (as gas) in Venus' atmosphere. If a more powerful reducing agent than hydrogen is present on the surface, then the reaction:

$$X + H_2O \rightarrow H_2 + XO$$

would happen spontaneously, oxidizing that reducing agent and reducing water to hydrogen. To invoke a more powerful reducing agent than hydrogen one therefore has to explain both what it is *and* why it does not react with water present in the atmosphere. However again their presence could be testable by observation.

## 5. Summary and Discussion

### 5.1. Summary

Phosphorus-containing species have not been modelled for Venus' atmosphere prior to Greaves et al (2020), other than overall thermodynamic calculation of the dominant phosphorus oxidation and hydration states (Krasnopolsky 1989). This work represents the first full description of a model of phosphorus species on Venus.

We have modelled processes that might produce phosphine under Venus conditions. This does not address whether phosphine is present, which is still a matter of controversy as noted above. While we do not wish to distance ourselves from the controversy, the purpose of this paper is to explore where phosphine might come from on a rocky planet, using Venus as a specific example, if it is present. We have assumed here that it is present, at ~1 ppb, and that presence requires an explanation.

We have explored every plausible chemical and physical process (and a number of implausible but possible ones) that could lead to the formation of phosphine on Venus, making conservative estimates where exact values were not known. We have shown that all conventional explanations of phosphine production that can explain the recent tentative detection ~1 ppb of phosphine in the Venusian atmosphere (Greaves *et al.* 2020c) are highly unlikely. Specifically, we have explored photochemical production (at least 5 orders of magnitude below the rate required to explain the observed ~1 ppb levels), atmospheric equilibrium thermodynamics (on average ~100 kJ/mol too energetically costly), surface and subsurface chemistry (8 – 15 orders of magnitude too low), and a range of other processes. We conclude that phosphine on Venus is produced by a physical or chemical process that is not expected to occur on terrestrial rocky planets.

### 5.2. Unknown Chemistry as an Explanation for the Presence of $PH_3$

If no conventional chemical processes can produce phosphine, is there a not yet considered process or set of processes that could be responsible for its formation?

One of the possibilities is that chemical species exist in the crust, or in the atmosphere of Venus, that we have not considered. Perhaps an unknown atmospheric chemical drives phosphine formation, especially considering that the photochemistry of Venus' atmosphere is not fully understood. Such a mechanism would have to be compatible with what we do know about Venus; for example, a powerful reductant would have to be compatible with the observed presence of water in Venus' atmosphere, as discussed in Section 4.2.

A specific example of such a mechanism would be photochemistry in the cloud droplets. The photochemistry of phosphorus species in sulfuric acid droplets is completely unknown, and so in principle phosphine could be produced photochemically in the sulfuric acid droplets of the cloud layer. However, we consider this unlikely, not least because it is known that phosphine is rapidly oxidized by sulfuric acid to phosphoric acid. Even if a photochemical process did produce phosphine in sulfuric acid, it seems unlikely that it would escape oxidation back to phosphoric acid. In fact, we expect the sulfuric acid cloud layer to be a sink for phosphine (one which we have not incorporated into the models above for lack of kinetic data). See Supplementary Section 2.5.3.1. for more on cloud droplet chemistry, and the chemistry of phosphine in sulfuric acid.

Other, completely unknown chemistry could be a source of phosphine, but in the absence of suggestions as to what that chemistry might be, such speculation cannot be considered a hypothesis to be tested.

### 5. 3. Phosphine as a Venus Cloud Biosignature Gas

Could living organisms in the temperate clouds of Venus produce phosphine? For decades many have speculated that the Venusian clouds are a suitable habitat for life (Cockell 1999; Grinspoon 1997; Grinspoon and Bullock 2007; Morowitz and Sagan 1967; Schulze-Makuch *et al.* 2004; Schulze-Makuch and Irwin 2002; Schulze-Makuch and Irwin 2006). The anomalous UV absorber in Venus' atmosphere has been proposed as a biosignature (Limaye *et al.* 2018; Seager *et al.* 2021), though chemical processes may be the source (Frandsen *et al.* 2020; Pérez-Hoyos *et al.* 2018; Wu *et al.* 2018). Unknown chemical species in the clouds absorb more than half of the UV flux that the planet receives, an absorption which is not constant across the planet but has unexplained temporal and spatial differences and constraints (Jessup *et al.* 2020; Lee *et al.* 2019; Marcq *et al.* 2019). Recent work has developed the case for phosphine as a biosignature gas in anoxic environments (Bains *et al.* 2019a; Bains *et al.* 2019b; Sousa-Silva *et al.* 2020). We emphasize that a biosignature is a sign that life is present. It may or may not be produced directly by life. While we do not know whether life on Earth produces phosphine itself, or rather if life produces reduced phosphorus species such as phosphite or hypophosphite that subsequently disproportionate to phosphine, the association of phosphine with biology (and in recent centuries with human technology) is clear (Bains *et al.* 2019a; Gassmann and Glindemann 1993; Glindemann *et al.* 1999; Glindemann *et al.* 2005a; Glindemann *et al.* 1996; Sousa-Silva *et al.* 2020). Specifically, we previously proposed that $PH_3$ production on Earth is associated with strictly oxygen-free, highly reduced, hot, moderately acid ecosystems (pH<5, 80 ºC) or cooler, very acid conditions (pH <2, 20 ºC) (Bains *et al.* 2019a; Bains *et al.* 2019b; Sousa-Silva *et al.* 2020). The Venusian clouds have some apparent parallels to these environments on Earth where life produces $PH_3$, although obviously the Venusian clouds are not reduced. We therefore explored the possibility that the Venusian $PH_3$ is produced by life.

We emphasize that the presence of phosphine in Venus' atmosphere does not prove the presence of life. Any explanation for the unexpected finding of $PH_3$ in Venus' atmosphere has to be tested, and to be tested it has to be articulated. Here we apply the same thermodynamic methods used above to test the hypothesis that life could explain the presence of phosphine on Venus. The reader should understand that this leaves *unexplored* the many other problems with the concept of life on Venus, such as the extremely low water activity, and the presence of concentrated sulfuric acid which is a powerful oxidizing agent and rapidly destroys the large majority of terrestrial biochemicals.

Could $PH_3$ on Venus also be associated with biological activity? We have argued above that producing phosphine in the Venusian atmosphere requires energy. A unique feature of life is that it captures chemical energy and uses it to drive chemical reactions that would not happen spontaneously in the environment (such as production of $O_2$ via photosynthesis on Earth). One widely accepted criterion for a biosignature is a gas completely out of equilibrium with its environment (Krissansen-Totton *et al.* 2016; Lovelock 1975), as phosphine is on Venus.

To make phosphine from phosphate, an organism would have to use a reducing agent. Here we ask whether such a reducing agent is within the scope of the reducing power of terrestrial biochemicals. The redox reactions involving phosphorus species that could be of biochemical origin are of the general form of:

1) $XH + H^+ + H_2PO_4^- \rightarrow H_2PO_3^- + H_2O + X^+$
2) $4XH + 4H^+ + H_2PO_4^- \rightarrow PH_3 + 3H_2O + OH^- + 4X^+$
3) $3XH + 3H^+ + H_2PO_3^- \rightarrow PH_3 + 2H_2O + OH^- + 3X^+$

Where $XH$ and $X^+$ are the reduced and oxidized form of a biological reducing agent respectively. Reactions are assumed to occur at pH=7. We test if under Venus conditions, would an unreasonably strong reducing agent be required to produce phosphine (i.e. we assess if we can we rule out biological production of phosphine on biochemical thermodynamic grounds). To estimate the thermodynamics of biological phosphine production, we assume that a cell living in a cloud droplet is composed mainly of water that takes in phosphorus and reduces it to phosphine. (Figure 10).

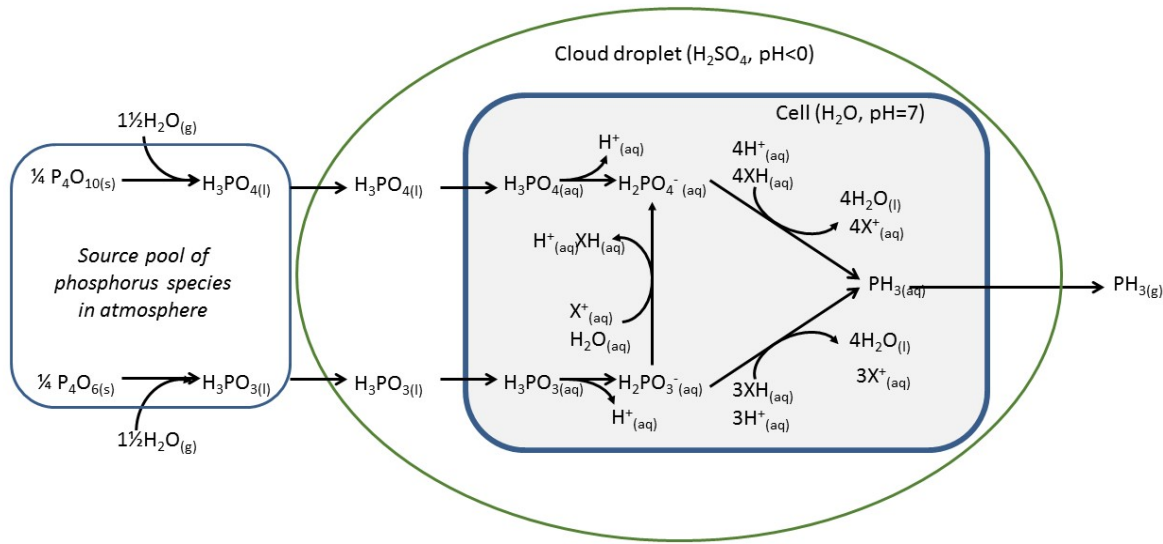

**Fig. 10.** A model for biological production of phosphine on Venus. The favored path for reduction of atmospheric phosphorus species to phosphine is reduction of phosphoric acid to phosphine (upper reaction pathway on the schematic above). Correspondingly, the reduction of phosphite to phosphine is disfavored, because of the low concentration of the phosphite reactant (lower reaction pathway on the schematic above). If the concentration of phosphite is allowed to rise in the cell, then reduction of phosphate to phosphite becomes less energetically favorable, and reduction of phosphite to phosphine correspondingly more favorable. It is plausible to suggest, though it is speculative, that phosphite would accumulate in cells to a level where its reduction to phosphine was thermodynamically neutral, allowing a multi-step reduction pathway for phosphate. HX: biological reducing agent, such as NADH.

Phosphorus species were assumed to be present in the extracellular droplet phase as oxyacids at 1 molal concentration (see Supplementary Section 1.3.2.2). We have assumed that, like terrestrial acidophiles, the putative Venusian organisms keep their interior at pH>5, as do Earth organisms, even those living at environments of pH=0 or pH=12 (Baker-Austin and Dopson 2007; Horikoshi 2016). An internal pH=7 was assumed here. The energy implicit in converting phosphate from the external pH (pH=0) to the intracellular pH (pH=7) was calculated as discussed extensively in (Bains *et al.* 2019a), and Supplementary Section 2.6. If the free energy needed to convert extracellular phosphorus to intracellular singly ionized forms at 1 mM was calculated as positive, it was assumed that the cell could not import phosphorus and no phosphine production could occur (i.e. the phosphorus was assumed to enter the cell by passive diffusion). The ratio of $H_2PO_3^-/H_2PO_4^-$ inside the cell immediately after transport was assumed to be the same as the ratio of $H_3PO_3/H_3PO_4$ outside the cell (but see below).

We estimated the thermodynamics of reduction of phosphorus species to phosphine using NADH, $FADH_2$, ubiquinone and two iron-sulfur proteins as model agents to illustrate the range of reducing power of different biological reducing agents in terrestrial biochemistry (see Supplementary Section 2.6.; Figure S24). We do not expect these specific chemicals to be present in putative Venusian life; we use them solely for illustration. Our result shows that NADH or the two iron-sulfur (Fe-S) proteins, but not $FADH_2$ or ubiquinone, can power the production of phosphine (see Supplementary Section 2.6.; Figure S25) Others have also suggested life based on an iron-sulfur based redox metabolism in the clouds of Venus (Limaye *et al.* 2018).

We emphasize that the overall process of reducing phosphate in a Venusian environment remains energy-consuming. The putative organisms must gain energy to generate the reducing agents that can then make phosphine. However, we note that life on Earth produces many compounds from common chemicals in the environment, sometimes in large amounts, that require substantial energy investment to make (Seager *et al.* 2012). In itself the expenditure of energy for the biosynthesis of $PH_3$ is not a criterion for ruling out a biological source for phosphine.

We conclude that the energy needed to reduce phosphate to phosphine is not beyond that deployed by terrestrial biochemistry in redox reactions. However, there remain major problems with the concept of life in the clouds of Venus. The clouds are often described as being 'habitable' because of their moderate pressure (~1 bar) and temperature (~60 °C). However, moderate temperature and pressure do not necessarily make the clouds habitable (Seager *et al.* 2021) (and in any case pressure is irrelevant - terrestrial life can grow at any pressure from >1000 bar (Nunoura *et al.* 2018) to <1 millibar (Pavlov *et al.* 2010)). To survive in the clouds, organisms would have to survive in an extremely chemically aggressive environment, one that is highly acidic and with an extremely low concentration of water (highly dehydrating and very low water activity). Sulfuric acid is a notoriously aggressive reagent towards sugars and aldehydes, reducing dry sucrose to charcoal in seconds. In principle life could exist in an aqueous droplet inside the sulfuric acid cloud drop (as drawn in Figure 10 above), but this poses formidable problems in itself. No known biological membrane could remain intact against such a chemical gradient, and the energy required to counteract leakage of water out of the cell (or sulfuric acid into it) could be orders of magnitude greater than the energy used by terrestrial halophiles to maintain their internal environment.

We conclude that, while we cannot rule out life as a source of the phosphine on Venus, the hypothesis that the phosphine is produced by life cannot *a priori* be favored over the hypothesis of unknown photochemistry or unknown atmospheric chemistry. All seem unlikely, and hence all call for further investigation. We note, after (Catling *et al.* 2018), that the extraordinary claim of life should be the hypothesis of last resort only after all conceivable abiotic alternatives are exhausted.

### 5. 4.    Future Work on Identifying a Source for Phosphine on Venus

Our analysis argues that no conventional source can explain the presence of ~1 ppb phosphine on Venus (and hence no source could explain higher abundances either). All the explanations modelled in Sections 2, 3 and 4 above, suggest that if phosphine is present on Venus then it requires a significant change in our understanding of the chemistry of the planet. While one such change could be to postulate the presence of life, geological sources of phosphine are both more accessible to modelling and easier to test using remote observations or relatively simple *in situ* measurements. The same is true of other unexplained aspects of the chemistry of Venus, such as the 'unknown UV absorber', the depletion of $SO_2$ in the cloud layer, the anomalous destruction of carbonyl sulfide below the clouds and others (summarized in (Bierson and Zhang 2019)). Such an investigation would likely require a combination of remote observation campaigns combined with orbiter and lander missions, supported by laboratory work on Earth.

Phosphine may be made by an unknown abiotic surface or cloud chemical processes. Knowledge of those processes will rely at least in part on more detailed knowledge of the Venusian atmosphere and geology. Neither the detailed chemistry nor the photochemistry of any of the potential phosphorus components of Venus' atmosphere are known, but could be investigated on Earth as a preliminary step for remote measurements and *in situ* observations. Progress towards identification of the source of phosphine on Venus can be made by laboratory experimentation here on Earth, especially regarding the properties of phosphorus species under Venus atmosphere and cloud conditions, including studies of chemical reactivity and solubility of phosphorus species in concentrated sulfuric acid and high $CO_2$.

The first priority for observation should be to confirm the presence of phosphine in the atmosphere of Venus with observations of additional spectral features, in the microwave or infrared region of the spectrum where phosphine is a strong absorber (Sousa-Silva *et al.* 2014; Sousa-Silva *et al.* 2020; Sousa-Silva *et al.* 2013). Subsequently, observations should focus on constraining the distribution and abundances of phosphine throughout the Venusian atmosphere. The photochemical model described above suggests that the lifetime of phosphine in the lower atmosphere could be years to centuries. If this is correct and phosphine diffuses from its source into the lower atmosphere (or originates there) then this would imply substantially higher concentrations in the 20-30 km altitude range than in the cloud decks. If phosphine is made by hydrolysis of mineral phosphides, then we would also expect diphosphine to be generated in the same reaction (Greenwood and Earnshaw 2012), and to be detectable in the lower atmosphere. Future *in situ* observations might probe this.

The data that are especially lacking relate to reliable chemistry measurements and detailed models of Venusian clouds. Such models and measurements should extend their focus beyond sulfur chemistry and focus on phosphorus as well. For example, studies aimed at detection of P-H bonds (strong absorbers around 4.3 and 10 microns (Sousa-Silva *et al.* 2019)) are currently underway. Such studies would require high resolution spectroscopy to distinguish $PH_3$ from overlapping $CO_2$ absorption; the necessary resolution should be within the capabilities of CRIRES+(VLT). Detection of P=O bonds would also be a valuable goal, because our kinetic model suggests that production and reduction of phosphorus monoxide (PO) is a rate-limiting factor in the pathway to atmospheric, abiotic phosphine production.

Missions focused on planetary geology, including landers, could help with *in situ* assessment of the possibility of geochemical production of phosphine on the surface of Venus and confirm or refute our conclusions that the geochemical processes on rocky planets are incapable of efficient phosphine production. Ultimately, long-term *in situ* observations of the clouds of Venus should also be carried out. Such long-term missions capable of detailed studies of clouds, aerosols, hazes and their spectral, physical and chemical properties (including mapping any changes over extended time periods) were proposed before (e.g. EnVision mission (Ghail *et al.* 2016), Aerobot aerial platforms (van den Berg *et al.* 2006) and the Venus Atmospheric Mobile Platform, developed by Northrop Grumman Aerospace (Lee *et al.* 2015)). Simultaneous observation of atmospheric features, such as UV absorber and phosphine distribution, would be more valuable than either alone. Some concepts of the aerial platforms are considered for the upcoming VENERA-D mission by ROSCOSMOS and NASA (Zasova *et al.* 2017). If such missions provided compelling evidence for biological processes, then a sample return missions would be required for any detailed biochemical characterization of a putative Venusian aerial biosphere.

Last but not least, our investigation presented in this paper is a useful template for the future investigations of biosignature gases, when these are detected on an exoplanet. Currently, a major focus in exoplanet astronomy is the near-future detection of the presence of life on exoplanets through detection of gases in exoplanet atmospheres that may be attributed to biological activity (Catling *et al.* 2018; Schwieterman *et al.* 2018; Seager and Bains 2015; Seager *et al.* 2016). A wide range of gases have been suggested, and a smaller number studied, as candidate biosignatures (Seager *et al.* 2012). However, detection is only the first step. Evaluation of the chemical context of the gas in a given planetary scenario is central to ruling life out or supporting the hypothesis that life is a source for that gas. This requires detailed analysis of possible formation and destruction pathways, local geology, atmospheric composition, all with inadequate knowledge as we will know far less about an exoplanet than we do about Venus (Catling *et al.* 2018; Schwieterman *et al.* 2018; Walker *et al.* 2018). We believe that the tentative discovery of the Venusian phosphine and the analysis that is presented in this work can form the basis of a template approach that should be applied to any biosignature gas detection to determine if it is a 'false positive', i.e. a gas that could be produced by abiotic processes. We note that the step of assessing of false positive scenarios for any biosignature gas is highly planet-specific. The task of replicating our approach here with other, less well-characterized worlds will not be easy, but will be essential for the attribution of any gas to a biological origin.

## 6. Conclusions

It was previously predicted that any detectable abundance of $PH_3$ in the atmosphere of a rocky planet would be an indicator of biological activity (Sousa-Silva *et al.* 2020). In this paper we show in detail that no abiotic mechanism based on our current understanding of Venus can explain the presence of ~1 ppb phosphine in Venus' clouds. If the detection is correct and phosphine is present at 1 ppb or more, then this means that our current understanding of Venus is significantly incomplete.

If phosphine is not a biological product, then it must be produced by planetary geo- or atmospheric chemistry. In either case our understanding, not only of Venus but of all terrestrial planets and exoplanets, needs a major paradigm shift. Because the source of phosphine is not known, we call for further aggressive observations of Venus and its atmosphere, laboratory studies of phosphorous chemistry in the context of the Venusian environment, and the development of Venus space missions to study its atmosphere and search for signs of life.

# 7. Acknowledgements

We thank Joanna Petkowska-Hankel for the translation of the original Russian Vega and Venera papers and the preparation of Figure 1. We thank Carver J. Bierson and Xi Zhang for insightful discussions about the atmosphere of Venus, and for sharing a preprint of their article and the vertical radical profiles derived therein. David Grinspoon, an anonymous referee, and Steve Benner for constructive criticism, to John Sutherland for comments on P(III) chemistry, and to many in the online community who indirectly spurred us to refine our arguments. We are grateful to Bob and Anna Damms for Russian translations. We thank the Heising-Simons Foundation and the Change Happens Foundation for funding. SR acknowledges the funding from the Simons Foundation (495062). Clara Sousa-Silva acknowledges the 51 Pegasi b Fellowship and the Heising-Simons Foundation. Paul B. Rimmer acknowledges funding from the Simons Foundation (SCOL awards 599634).

# Phosphine on Venus Cannot be Explained by Conventional Processes: Supplementary Information

## 1. Supplementary Approach and Methods:

### 1. 1. Methods Used in Photochemistry and Kinetics Analysis

#### 1. 1. 1. Photochemical Model of the Venusian Atmosphere

To estimate the vertical radical concentration profiles relevant to $PH_3$ photochemistry, we use the photochemical model of the Venusian atmosphere previously reported in (Greaves *et al.* 2020c). This photochemical model of Venus's atmosphere accounts for photochemistry, thermochemistry, and chemical diffusion.

*1. 1. 1. 1. Modelling Framework*

To solve for a self-consistent set of atmospheric constituent concentrations we employ a 1D photochemistry-diffusion code, called ARGO (Rimmer and Helling 2016), which solves the continuity-transport equation. ARGO is a Lagrangian photochemistry/diffusion code that follows a single parcel as it moves from the bottom to the top of the atmosphere, determined by a prescribed temperature profile. The temperature, pressure, and actinic ultraviolet flux are updated at each height in the atmosphere. In this reference frame, bulk diffusion terms are accounted for by time-dependence of the chemical production, $P_i$ (cm$^3$ s$^{-1}$), and loss, $L_i$ (s$^{-1}$) rates. Below the homopause, molecular diffusion can be neglected, and the equation to be solved is:

$$\frac{\partial n_i}{\partial t} = P_i[t(z, v_v)] - L_i[t(z, v_z)]n_i, \qquad (1)$$

where $n_i$ (cm$^{-3}$) is the number density of species *i*, *t* (s) is time, *z* (cm) is atmospheric height above the surface, and $v_z = K_{zz}/H_0$ (cm/s) is the effective vertical velocity due to Eddy diffusion, from the Eddy diffusion coefficient $K_{zz}$ (cm$^2$ s$^{-1}$). Molecular diffusion into and out of the parcel is accounted for by production and loss 'reactions' that remove specific species as the parcel moves upwards, adding them back as the parcel moves downwards, at a rate determined by molecular diffusion (Rimmer and Helling 2016; Rimmer and Helling 2019). The mixing ratios are saved at a given height before the parcel proceeds to the next height, constructing atmospheric profiles for all species included in the accompanying chemical network.

*1. 1. 1. 2. Photochemistry*

Photochemistry is solved for the depth-dependent actinic flux in the standard way using a 2-stream δ-Eddington approximation (Toon *et al.* 1989), using the atmospheric profiles, and then transport a parcel through the atmosphere again with these updated depth-dependent actinic fluxes. Each time this is accomplished is a single global iteration for the model, and the model is run until every major and significant minor species (any with $n_i > 10^5$ cm$^{-3}$) agrees between two global iterations to within 1%. We modify the standard actinic flux calculation in two ways. First, we ignore the absorption of $SO_2$ for the first three global iterations, and include it afterwards. This seems to help the model to converge. In addition, we have included a 'mysterious absorber' with properties (Krasnopolsky 2012):

$$\frac{d\tau}{dz} = 0.056/km\, e^{-(z-67\text{ km})/3\text{ km}} e^{-(\lambda - 3600\text{ Å})/1000\text{ Å}}, z > 67 \text{ km}; \qquad (2)$$

$$\frac{d\tau}{dz} = 0.056/km\, e^{-(\lambda - 3600\text{ Å})/1000\text{ Å}}, 58\, km \leq z \leq 67 \text{ km}; \qquad (3)$$

$$\frac{d\tau}{dz} = 0, z \leq 58 \text{ km}; \qquad (4)$$

### 1. 1. 1. 3.   Photochemical Network

Chemical networks for Venus are limited and prior work often specializes on one part of the atmosphere over another. A variety of sources were therefore used to assemble a whole atmosphere photochemical network. The chemical network is based on STAND2019 (Rimmer and Rugheimer 2019), which includes H/C/N/O species. (Greaves *et al.* 2020c) extended this network by adding a limited S/Cl/P network relevant for the Venusian atmosphere. This network is a copy of the low altitude atmospheric network of Krasnopolsky (Krasnopolsky 2007; Krasnopolsky 2013) and the middle atmosphere network of Zhang (Zhang *et al.* 2012). The network is further modified by removing any reverse reactions explicitly included in (Krasnopolsky 2007; Krasnopolsky 2013), and instead by self-consistently calculating reverse reactions throughout the atmosphere.

For our chemical network, we use STAND2019 (Rimmer and Rugheimer 2019), which includes H/C/N/O species. We have also added a limited S/Cl/P network relevant for the Venusian atmosphere by copying the low atmospheric network of Krasnopolsky (Krasnopolsky 2007; Krasnopolsky 2013) and the middle atmosphere network of Zhang (Zhang *et al.* 2012). For network reactions that do not involve $PH_3$, we use the networks of Krasnopolsky (Krasnopolsky 2007) and Zhang (Zhang *et al.* 2012), modified as follows. The network of Krasnopolsky include specific reverse reaction rates. We excluded these, and instead used the forward reactions and the thermochemical constants from Burcat (Burcat and Ruscic 2005) for calculating the reverse reactions for those species already included in STAND2019, as well as reactions that include the species S, $S_2$, $S_3$, $S_4$, $S_5$, $S_6$, $S_7$, $S_8$, HS, SO, ClO, ClS, $Cl_2$, $H_2S$, OCS, $SO_2$, $SO_3$, $S_2O$, HOCl, ClCO, $Cl_2S$, $Cl_2S_2$, $HSO_3$, $H_2SO_4$ as described by (Rimmer and Helling 2016; Visscher and Moses 2011). We added reverse reactions for the reactions from the (Zhang *et al.* 2012) middle atmosphere network wherever possible. We supplemented this network with the following reactions:

R1: H + $PH_3$ → $H_2$ + $PH_2$: Arrhenius parameters used were A=7.22×10$^{-11}$ cm$^3$ s$^{-1}$ and E=7.37 kJ mol$^{-1}$ (Arthur and Cooper 1997). These authors state that these parameters are valid over 200-500 K; we confirm that they are consistent with the theoretical calculations of (Yu *et al.* 1999) at higher temperatures to within a factor of 2.

R2: OH + $PH_3$ → $H_2O$ + $PH_2$: Arrhenius parameters used were i.e. A=2.71×10$^{-11}$ cm$^3$ s$^{-1}$ and E=1.29 kJ mol$^{-1}$, from (Fritz *et al.* 1982), based on measurements from 250-450 K.

R3: O + $PH_3$ → $H_2PO$ + H: Rate parameter used was k = $4.75 \times 10^{-11}$ $cm^3$ $s^{-1}$ (Nava and Stief 1989) based on measurement from 208-408 K, with value. The addition reaction probably dominates for T<1000 K; The abstraction reaction O + $PH_3$ → OH + $PH_2$ should become significant above 1000 K. As temperatures never exceed 750 K in the Venusian atmosphere, the abstraction reaction will not be significant on Venus. $H_2PO$ formed will be oxidized further to $H_3PO_4$ as a stable end product.

R4: Cl + $PH_3$ → HCl + $PH_2$: Rate parameter used was k = $2.4 \times 10^{-10}$ $cm^3$ $s^{-1}$, from (Iyer et al. 1983). Iyer et al only study this reaction at 298 K; we are not aware of studies at any other temperature. We therefore adopt this reaction rate throughout the atmosphere, which will formally under-estimate the rate of destruction at higher temperatures.

The rate constant for thermolytic breakdown of $PH_3$ via $PH_3$ + M → $PH_2$ + H + M was calculated as described below.

We include the reverse reactions for these phosphine reactions using the same approach as above, using Burcat polynomials (Burcat and Ruscic 2005).

The reaction of N radical with $PH_3$ has not been observed; an upper limit for this reaction near Earth-ambient conditions is $4 \times 10^{-14}$ $cm^3$ $s^{-1}$, which is 2-3 orders of magnitude lower than attack by O, OH, and H radicals (Hamilton and Murrells 1985). As N radicals are predicted to be present only at very low concentrations in the regions of the Venusian atmosphere relevant to this study (<80 km; (Krasnopolsky 2012)), we therefore neglect loss due to attack by N. Formally this will further under-estimate the $PH_3$ loss rate, but by a trivial amount.

We neglect the possibility that the products of $PH_3$ destruction can recombine to restore $PH_3$, except for the reaction $PH_2$ + H + M → $PH_3$ + M, which is included as the reverse reaction for thermolytic decay of $PH_3$ (see below). This formally overestimates $PH_3$ destruction rates. The products of $PH_3$ destruction are rare even in the $H_2$-rich atmospheres of Jupiter and Saturn, and recombination correspondingly unlikely. In Venus' H-poor atmosphere the products will be even rarer, making the rate of recombination of phosphine from its breakdown products negligible (Sousa-Silva et al. 2020). This assumption can be tested experimentally by searching for $PH_3$ products (e.g. diphosphine ($P_2H_4$) in the Venusian atmosphere.

We also include condensation of $S_n$ species (Lyons 2008) and sulfuric acid $H_2SO_4$ (Kulmala and Laaksonen 1990), as functions of the vapor pressures ($p_{vap}$, bar) which are calculated as follows:

$$log_{10}\, p_{vap}\, (S_2) = 7.024 - \frac{6091\, K}{T} \quad (5)$$

$$log_{10}\, p_{vap}\, (S_3) = 6.343 - \frac{6202\, K}{T} \quad (6)$$

$$log_{10}\, p_{vap}\, (S_4) = 6.003 - \frac{6048\, K}{T} \quad (7)$$

$$log_{10}\, p_{\text{vap}}(S_5) = 5.061 - \frac{4715\,\text{K}}{T} \qquad (8)$$

$$log_{10}\, p_{\text{vap}}(S_6) = 4.804 - \frac{3814\,\text{K}}{T} \qquad (9)$$

$$log_{10}\, p_{\text{vap}}(S_7) = 5.213 - \frac{4114\,\text{K}}{T} \qquad (10)$$

$$log_{10}\, p_{\text{vap}}(S_8) = 4.188 - \frac{3269\,\text{K}}{T} \qquad (11)$$

$$log_{10}\, p_{\text{vap}}(H_2SO_4) = 4.4753 - \frac{3229\,\text{K}}{T+710920\,\text{K}} - \frac{3.1723\times10^6\,\text{K}^2}{T^2} + \frac{4.0832\times10^8\,\text{K}^3}{T^3} - \frac{2.0321\times10^{10}\,\text{K}^4}{T^4} \qquad (12)$$

$SO_2$ can be photochemically converted to $H_2SO_4$ and $S_8$, which can condense and be removed to the clouds.

We add removal of $SO_2$ into the clouds in order to match the top boundary conditions from the lower atmosphere models (Krasnopolsky 2007) to the bottom boundary conditions for the middle atmosphere models. The former is orders of magnitude lower than the latter, implying strong depletion across the cloud layer (Zhang *et al.* 2012). Bierson et al. accomplished this by depleting $SO_2$ via oxidation to $SO_3$ and removal by reaction with $H_2O$ to form sulfuric acid (Bierson and Zhang 2019). To achieve this Bierson et al had to decrease $K_{zz}$ within the cloud layer and fix $H_2O$ to be equal to observed concentrations throughout the atmosphere, which implies an unknown source of $H_2O$ in the clouds and yet undetermined chemical cycle that both provides sufficient $H_2O$ reacting with $SO_3$ to deplete $SO_2$ by orders of magnitude and that maintains $H_2O$ mixing ratios at more than an order of magnitude lower than $SO_2$ mixing ratios. We do not fix the $H_2O$ concentrations, and so instead have depleted $SO_2$ by including rainout with a Henry's Law approximation modified as described by Sander (ref. (Sander 2015), their Section 2.7). Incorporating this loss term brings the $SO_2$ curve into better agreement with observation, and may instead be interpreted as approximating photochemical loss of $SO_2$ via a different mechanism or series of reactions.

### 1.1.1.4. *Thermolytic Decay of $PH_3$*

The thermal decomposition of phosphine is important near the base of the atmosphere. Concentrations of radicals below the clouds of Venus are very uncertain, but even with the largest published predictions (Krasnopolsky 2007) for radical concentrations in the lower atmosphere of ~1000 cm$^{-3}$, reaction of $PH_3$ with radicals will be extremely slow (order >$10^8$ seconds). In this environment, thermal decomposition dominates $PH_3$ destruction and therefore determines the lifetime of $PH_3$. The thermal decomposition of $PH_3$ has been considered theoretically (Cardelino *et al.* 2003). Theoretical values of $k_{\text{uni}}$ (s$^{-1}$) and $k_\infty$ (s$^{-1}$) are given as (Cardelino *et al.* 2003):

$$k_{\text{uni}} = 3.55 \times 10^{14}\,\text{s}^{-1} e^{-35644\,\text{K}/T} \qquad (13)$$

$$k_\infty = 1.91 \times 10^{18}\,\text{s}^{-1} e^{-40063\,\text{K}/T} \qquad (14)$$

but no value for the rate constant at the low-pressure limit, $k_0$ (cm$^3$ s$^{-1}$), is provided. This rate constant has to be determined to use the Lindemann expression to calculate the rate constant over a wide range of pressures:

$$k = \frac{k_\infty}{1+k_\infty/(k_0[M])} \tag{15}$$

where [M] is the number density of the third body (in our case $[M] = n$, where $n$ (cm$^{-3}$) is the atmospheric number density). The rate constant at the low pressure limit can be estimated by considering that $k_{uni}$ was calculated for 1300 bar and 900 K, so $[M] = 1.07 \times 10^{22}$ cm$^{-3}$, and solving Equation (15) with $k = k_{uni}$. Doing so yields:

$$k_0 = 3.4 \times 10^{-8} \text{ cm}^3 \text{ s}^{-1} e^{-35644 \text{ K}/T} \tag{16}$$

An alternative way to estimate $k_0$ from $k_\infty$ is to perform a simple conversion of units, with $k_0 = kT/(1\ bar)\ k_\infty$, which gives:

$$k_0 = 2.6 \times 10^{-4} \text{ cm}^3 \text{ s}^{-1} \left(\frac{T}{300\ K}\right) e^{-40063 \text{ K}/T} \tag{17}$$

Finally, we can use the decomposition of NH$_3$ as an analogue of the decomposition of PH$_3$. The low-pressure thermal decomposition rate limit for NH$_3$ has been experimentally determined over a temperature range of 1740-3300 K (Davidson *et al.* 1990). We assume that the principle difference between the two gases is the activation energy for bond scission (i.e. the bond strength). The activation energy at the high pressure limit for PH$_3$ is of 40063 K (Cardelino *et al.* 2003), for NH$_3$ 48840 K (Cardelino *et al.* 2003). We assume this ratio is the same at the low pressure limit, where the measured activation energy for NH$_3$ is 39960 K (Davidson *et al.* 1990). We therefore find:

$$k_0 = 7.2 \times 10^{-9} \text{ cm}^3 \text{ s}^{-1}\ e^{-32778 \text{ K}/T} \tag{18}$$

The timescales for thermal decomposition derived from these rate constants, along with the timescale using only $k_\infty$, are shown in (ref. (Greaves *et al.* 2020c), their Figure S8). Since our first estimate, Equation (16), yields the longest timescale, and will therefore be most favorable for abiotic PH$_3$ scenarios, we use that value.

*1. 1. 1. 5.   UV Photolysis of Phosphine*

PH$_3$ photolyzes via PH$_3$ + hv → PH$_2$ + H upon absorption of ≤ 230 nm UV (Kaye and Strobel 1984; Visconti 1981). We estimate the photolysis rate coefficient $J_A$ by (Seager *et al.* 2013):

$$J_A = \int_\lambda q_\lambda I_\lambda e^{-\tau_\lambda} \sigma_\lambda d\lambda, \tag{19}$$

where $I_\lambda$ is the solar intensity at the top of the atmosphere, $\tau_\lambda$ is the optical depth of the overlying atmosphere, $\sigma_\lambda$ is the absorption cross-section of PH$_3$, and $q_\lambda$ is the quantum yield of PH$_3$ photolysis. For $I_\lambda$, we use the solar instellation spectrum aggregated by (Hu *et al.* 2012) by concatenating the quiet-sun emission spectrum (from (Curdt *et al.* 2004); 110-119 nm) to the Air Mass Zero reference spectrum from the American Society for Testing and Materials (http://rredc.nrel.gov/solar/spectra/am0/) (>119.5 nm). We scale the insolation by

cos($z$)=0.5 to match the dayside mean cos($z$) adopted by (Krasnopolsky 2012), by 0.5 to account for diurnal variation, and by $0.72^{-2}$ to account for Venus's closer orbit to the Sun.

We take the absorption cross-sections of $PH_3$ from (Chen *et al.* 1991), reported at 295 K. We follow (Kaye and Strobel 1984) in taking the quantum yield of photolysis to be unity at wavelengths ≤ 230 nm.

In calculating $\tau_\lambda$, we include absorption due to $CO_2$ and $SO_2$, following the insight of (Krasnopolsky 2006) that to first order every UV photon <218 nm is absorbed by one of these gases. For the absorption cross-sections of $SO_2$ and $CO_2$, we use the aggregation of (Ranjan and Sasselov 2017), who in these wavelength ranges draw primarily on (Huestis and Berkowitz 2010; Shemansky 1972), and (Manatt and Lane 1993). The UV profile of the Venusian atmosphere was modified to include the 'unknown UV absorber' with properties described by (Krasnopolsky 2007):

$$\frac{d\tau}{dz} = 0.056/km\, e^{-(z-67\text{ km})/3\text{ km}} e^{-(\lambda-3600\text{ Å})/1000\text{ Å}}, \qquad z > 67 \text{ km}; \qquad (20)$$

$$\frac{d\tau}{dz} = 0.056/km\, e^{-(\lambda-3600\text{ Å})/1000\text{ Å}}, \qquad 58\,km \leq z \leq 67 \text{ km}; \qquad (21)$$

$$\frac{d\tau}{dz} = 0, \qquad z \leq 58 \text{ km}; \qquad (22)$$

Our approach neglects scattering; this means UV radiation penetrates deeper into the atmosphere than we model here, meaning we overestimate photolysis rates. Even so, direct photolysis is not a dominant loss process for $PH_3$ in the Venusian atmosphere <100 km (Figure 2, in the main text).

*1. 1. 1. 6. Model Input: Atmospheric Profile of Venus and Initial Boundary Conditions*

We describe the details behind the selection of photochemical model inputs by (Greaves *et al.* 2020c), and reproduce their atmospheric profile of Venus (Figure S1 and Figure S2). In short, we take fixed surface boundary conditions from (Krasnopolsky 2007) for the major atmospheric species, and with initial surface boundary conditions from (Krasnopolsky 2007) for minor species, radicals and atoms (these conditions are reproduced in Table S1). (Greaves *et al.* 2020c) provide validation for this model by comparing with observed concentrations of CO, $H_2O$, HCl, $H_2S$, OCS, $S_3$, SO, $SO_2$ and $PH_3$.

Here and in (Greaves *et al.* 2020c), we are comparing a 1D photochemistry model set up to approximate the global-average height-dependent chemistry of a three-dimensional and dynamic atmosphere. Measurements, on the other hand, often apply to particular latitudinal and longitudinal regions measured at a particular time. As such, the errors shown here estimate not only the error bars of individual measurements, but the variation between measurements taken at the same altitude, wherever possible.

Work of (Krasnopolsky 2013) provides single measurements per height for $S_3$, and we use their error bars as variance. (Krasnopolsky 2013) also gives mixing ratios for CO and OCS, but provides no error bars, but rounds mixing ratios to the nearest multiple of 5, so an ± 5 error was applied to each datapoint. (Marcq *et al.* 2008) measure CO, CSO, $H_2O$ and $SO_2$ mixing ratios at a variety of latitudes at a given altitude. The error bars were averaged for all data points, and the variance was estimated by taking the maximum datapoint plus this error, and the minimum datapoint minus this error. For the measurements of $H_2O$ and HCl, we use the errors from (Bertaux *et al.* 2007) averaged over all datapoints within 10 km of the plotted datapoint. The observations and errors for HCl were plotted on a linear scale, and the 1-$\sigma$ errors reach mixing ratios of zero in the upper atmosphere, which is why the error on log scale is so large. Finally, for the SO and $SO_2$ observations from (Belyaev *et al.* 2012), which show many dozens of datapoints at altitudes between 75 and ~100 km. Since the error bars were far smaller than the difference between datapoints at similar altitudes, the variance is estimated simply by using the maximum and minimum values for the mixing ratios within 10 km regions. The data and errors are shown in Table S2.

### 1. 1. 2. Photochemical Model Input: Atmospheric Profile of Venus

In modeling the Venusian atmosphere, we follow (Krasnopolsky 2007; Krasnopolsky 2012) in taking the temperature-pressure (TP) profile from the Venus International Reference Atmosphere (VIRA). Specifically, we use previously published TP profiles of (Seiff *et al.* 1985): for the deep atmosphere profile (0-32 km) and for the altitudes between 32-100 km, where we use the 45 degrees latitude profile. For the altitudes between 100-112 km we use the VIRA dayside profile from (Keating *et al.* 1985). Figure S1 shows the temperature-pressure profile adopted in this work. We similarly follow (Krasnopolsky 2007; Krasnopolsky 2012) in the Eddy diffusion profile, taking it to be constant at $2.2\times10^3$ cm$^2$ s$^{-1}$ for z<30 km, $1\times10^4$ cm$^2$ s$^{-1}$ for z=47-60 km, $1\times10^7$ cm$^2$ s$^{-1}$ for z>100 km, and connected exponentially at intermediate altitudes. Figure S2 shows the Eddy diffusion profile adopted in this work.

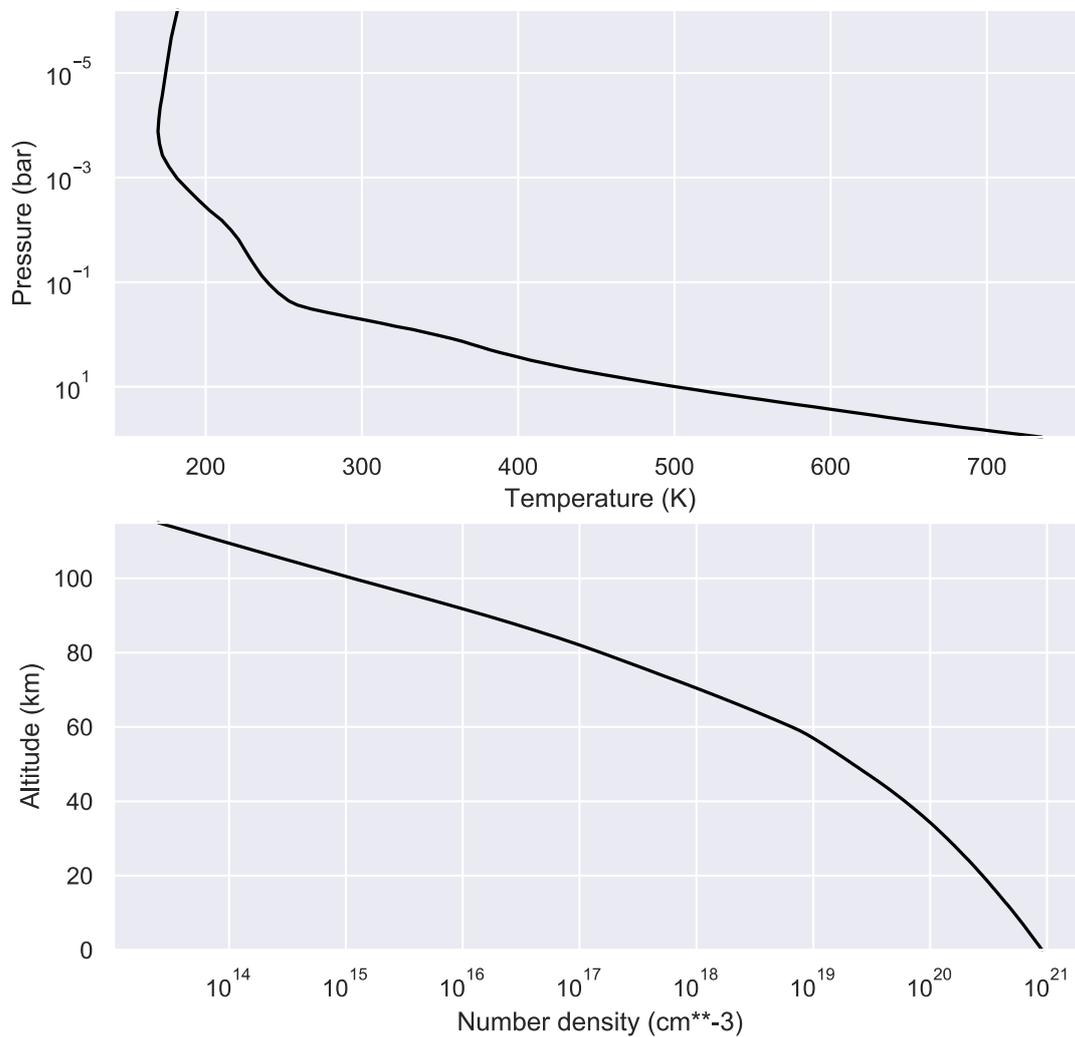

**Fig. S1.** Temperature-pressure profile used in photochemical modeling of the Venusian atmosphere, following (Krasnopolsky 2007; Krasnopolsky 2012). From (Greaves *et al.* 2020c), their Fig. S8.

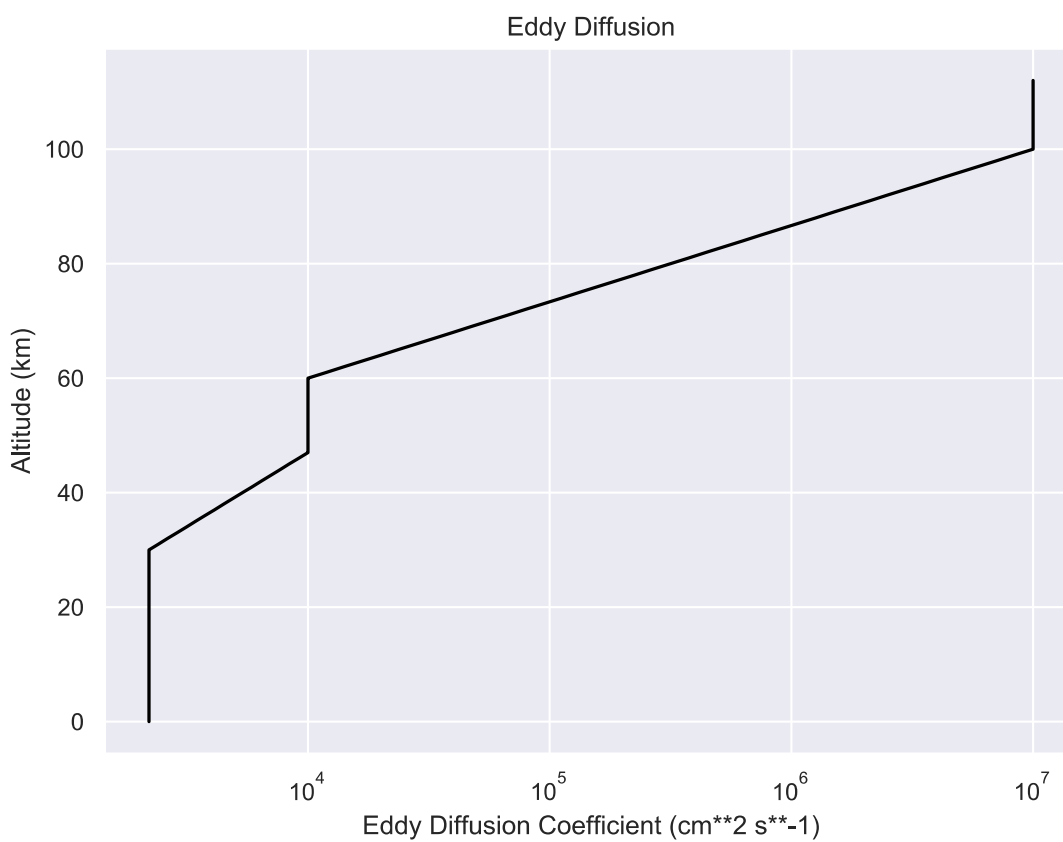

**Fig. S2.** Eddy diffusion profile used in the photochemical modelling of the Venusian atmosphere, following (Krasnopolsky 2007; Krasnopolsky 2012). From (Greaves *et al.* 2020c), their Fig. S8.

### 1. 1. 3. Initial Chemical Boundary Conditions

We use the fixed surface boundary conditions, which are based on the surface abundances of (Krasnopolsky 2007) for the major atmospheric species, and with initial surface boundary conditions from (Krasnopolsky 2007) for minor species, radicals and atoms. The initial surface abundances for our model are shown in Table S1.

| Species | Mixing Ratio |
|---|---|
| $CO_2$ | 0.96 |
| $N_2$ | 0.03 |
| $SO_2$ | $1.5 \times 10^{-4}$ |
| $H_2O$ | $3.0 \times 10^{-5}$ |
| CO | $2.0 \times 10^{-5}$ |
| OCS | $5.0 \times 10^{-6}$ |
| $S_2$ | $7.5 \times 10^{-7}$ |
| HCl | $5.0 \times 10^{-7}$ |
| $S_n$ ($3 \leq n \leq 8$) | $3.3 \times 10^{-7}$ |
| NO | $5.5 \times 10^{-9}$ |
| $H_2$ | $3.0 \times 10^{-9}$ |
| $H_2S$ | $1.0 \times 10^{-9}$ |
| SO | $3.0 \times 10^{-11}$ |
| $ClSO_2$ | $3.0 \times 10^{-11}$ |
| $SO_2Cl_2$ | $1.0 \times 10^{-11}$ |
| HS | $8.0 \times 10^{-13}$ |
| SNO | $1.0 \times 10^{-13}$ |
| SCl | $6.7 \times 10^{-15}$ |
| HSCl | $2.8 \times 10^{-15}$ |
| $Cl_2$ | $1.0 \times 10^{-16}$ |
| S | $7.5 \times 10^{-17}$ |
| H | $7.3 \times 10^{-19}$ |
| OH | $7.3 \times 10^{-19}$ |

**Table S1.** Initial surface conditions for atmospheric chemistry. Table adapted from (Greaves *et al.* 2020c), their Table S2.

We include a source of $PH_3$ in the clouds, with flux:

$$\Phi(z) = 0.5\Phi_0 \left[\tanh\left(\frac{z-45\ km}{2\ km}\right)\tanh\left(\frac{65\ km-z}{2\ km}\right) + 1\right] \quad (23)$$

where $\Phi(z)$ (cm$^{-2}$ s$^{-1}$) is the $PH_3$ flux at height z (km), and $\Phi_0=10^7$ (cm$^{-2}$ s$^{-1}$) is assigned to reproduce a 1 ppb $PH_3$ concentration, which is the lower bound of the values inferred by (Greaves *et al.* 2020b).

### 1. 1. 4. Photochemical Model Validation

Here we compare observations of CO, OCS, $H_2O$, $SO_2$, $H_2S$, HCl, $S_3$, SO, to model predictions (Figure S3 and Table S2). As shown in (Greaves *et al.* 2020c), all species agree with observations to within an order of magnitude in concentration, within +/- 5 km, with the exception of $H_2O$ and $O_2$. Photolysis of water is very efficient for our model, and depletion by reaction with $SO_3$ is significant, so that in our model predicted water vapor drops off rapidly above 70 km, leading to a discrepancy between observed $H_2O$ and model $H_2O$ of several orders of magnitude. This discrepancy is accompanied by higher concentrations of OH and O above 70 km, and so we probably underestimate the lifetime for phosphine above 70 km. However, the lifetime of phosphine at these heights is very short, on the order of days to seconds, for all published models of Venus's middle atmosphere (e.g. by Zhang (Zhang *et al.* 2012) and Bierson & Zhang (Bierson and Zhang 2019)). Our model also predicts too much $O_2$ in the middle atmosphere of Venus, within an order of magnitude of the concentrations predicted by Zhang (Zhang *et al.* 2012) and Bierson & Zhang (Bierson and Zhang 2019).

We consider the possibility that our model contains an idiosyncrasy or error that leads to significant underestimates of $PH_3$ lifetime, and hence overestimates the difficulty of abiotic buildup. To assess this possibility, we repeated our calculations of $PH_3$ lifetime and required production rates using concentration profiles of H, OH, O, Cl, and $SO_2$ drawn from (Bierson and Zhang 2019) (aided by C. Bierson, personal communication, 2-Aug-2019). This model excludes $PH_3$; consequently, it may overestimate lower-atmosphere radical abundances and underestimate $PH_3$ lifetimes. Use of these radical profiles, instead of the profiles drawn from our model, result in $PH_3$ lifetimes becoming short (<$10^3$ s) at an altitude of 71 or 80 km instead of 63 km in our model, depending on which of the scenarios from Bierson & Zhang we adopt (their nominal vs. their low $K_{zz}$+$S_8$ scenarios). However, this change in destruction altitude does not affect the upper limits on lifetime we calculate strongly enough to affect the conclusions of this paper.

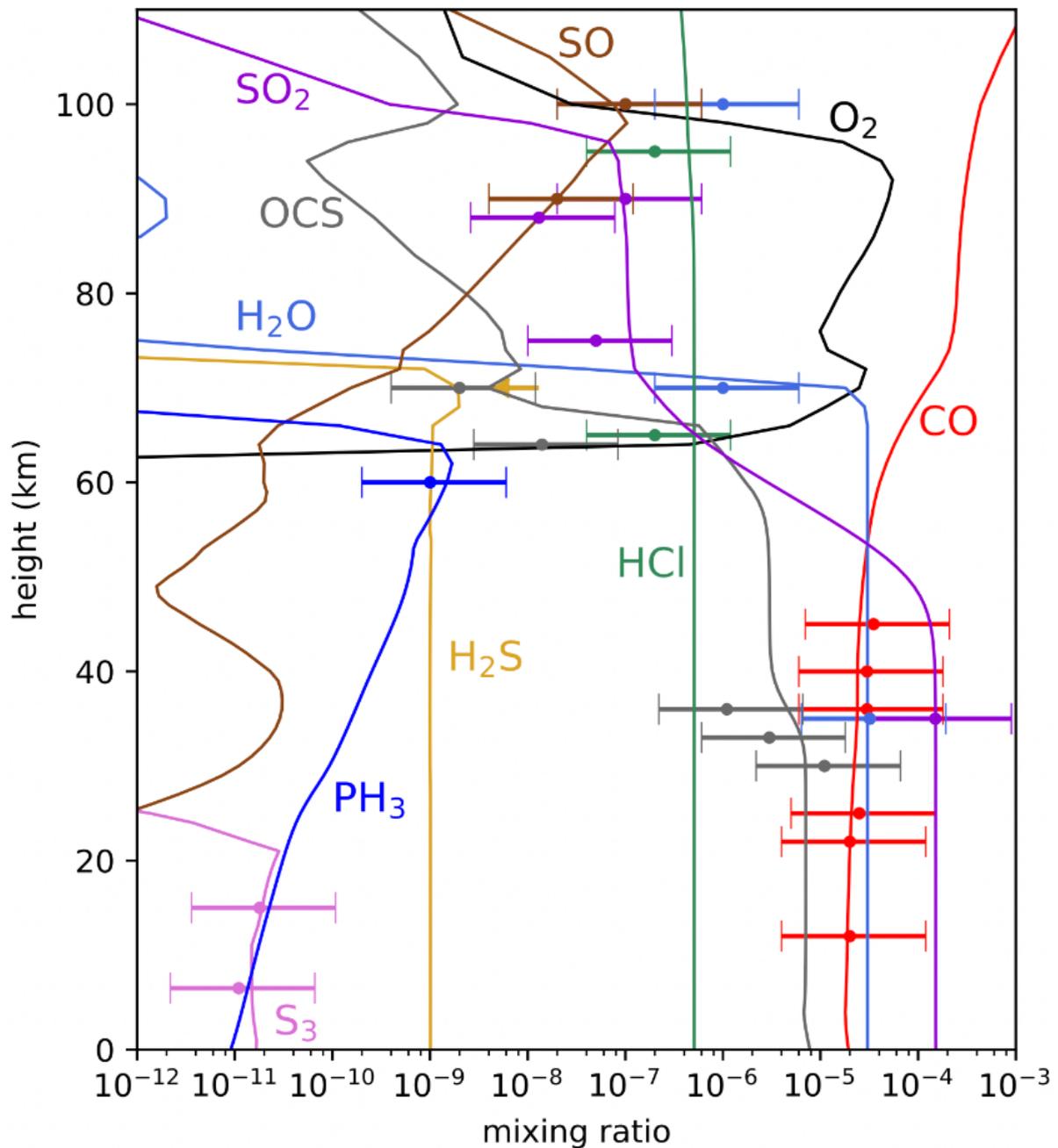

**Fig. S3.** Comparison of Venus model to observations. Mixing ratios (x axis) of various species versus atmospheric height (km) (y axis).

| Species | Atmospheric Height | Mixing Ratio | Error | Reference |
|---|---|---|---|---|
| CO | 12 km | $2 \times 10^{-5}$ | $\pm 5 \times 10^{-6}$ | (Krasnopolsky 2013) |
| | 22 km | $2 \times 10^{-5}$ | $\pm 5 \times 10^{-6}$ | (Krasnopolsky 2013) |
| | 25 km | $2.5 \times 10^{-5}$ | $\pm 5 \times 10^{-6}$ | (Krasnopolsky 2013) |
| | 36 km | $3 \times 10^{-5}$ | $(-7, +9) \times 10^{-6}$ | (Marcq *et al.* 2008) |
| | 40 km | $3 \times 10^{-5}$ | $\pm 5 \times 10^{-6}$ | (Krasnopolsky 2013) |
| | 45 km | $3.5 \times 10^{-5}$ | $\pm 5 \times 10^{-6}$ | (Krasnopolsky 2013) |
| OCS | 30 km | $1.1 \times 10^{-5}$ | $\pm 1 \times 10^{-6}$ | (Krasnopolsky 2013) |
| | 33 km | $3 \times 10^{-6}$ | $\pm 2 \times 10^{-6}$ | (Marcq *et al.* 2008) |
| | 36 km | $1.1 \times 10^{-6}$ | $\pm 1 \times 10^{-7}$ | (Krasnopolsky 2013) |
| | 64 km | $1.4 \times 10^{-8}$ | $(-1.2, +2.8) \times 10^{-8}$ | (Krasnopolsky 2008) |
| | 70 km | $2 \times 10^{-9}$ | $(-1.8, +6) \times 10^{-9}$ | (Krasnopolsky 2008) |
| $H_2O$ | 35 km | $3.2 \times 10^{-5}$ | $\pm 4 \times 10^{-6}$ | (Marcq *et al.* 2008) |
| | 70 km – 100 km | $1 \times 10^{-6}$ | $(-5, +20) \times 10^{-7}$ | (Bertaux *et al.* 2007), Constant between these heights |
| $SO_2$ | 35 km | $1.5 \times 10^{-4}$ | $\pm 1.4 \times 10^{-4}$ | (Marcq *et al.* 2008) |
| | 75 km | $5 \times 10^{-8}$ | $(-4, +45) \times 10^{-8}$ | (Belyaev *et al.* 2012), Average from several observations |
| | 90 km | $1 \times 10^{-7}$ | $(-9, +90) \times 10^{-8}$ | (Belyaev *et al.* 2012), Average from several observations |
| | 100 km | $1 \times 10^{-7}$ | $(-9, +90) \times 10^{-8}$ | (Belyaev *et al.* 2012), Average from several observations |
| $H_2S$ | 70 km | $< 2.3 \times 10^{-8}$ | | (Krasnopolsky 2008), Upper Limit |
| HCl | 65 km – 95 km | $2 \times 10^{-7}$ | $\pm 5 \times 10^{-8}$ at 65 km; $\pm 2 \times 10^{-7}$ at 95 km | (Bertaux *et al.* 2007), Constant between these heights |
| $S_3$ | 6.5 km | $1.1 \times 10^{-11}$ | $\pm 3 \times 10^{-12}$ | (Krasnopolsky 2013), Heights are approximate |
| | 15 km | $1.8 \times 10^{-11}$ | $\pm 3 \times 10^{-12}$ | (Krasnopolsky 2013), Heights are approximate |
| SO | 90 km | $2 \times 10^{-8}$ | $\pm 1 \times 10^{-8}$ | (Belyaev *et al.* 2012), Average from several observations |
| | 100 km | $1 \times 10^{-7}$ | $(-9, +90) \times 10^{-8}$ | (Belyaev *et al.* 2012), Average from several observations |
| $PH_3$ | 60 km | $1 \times 10^{-8}$ | $(-0.5, +1) \times 10^{-8}$ | This work and (Greaves *et al.* 2020c) |

**Table S2.** Observational constraints on atmospheric concentrations. Adapted from (Greaves *et al.* 2020c), their Table S3.

### 1. 1. 5. Details of Estimation of the Lifetime of $PH_3$ in the Venusian Atmosphere

The lifetime of phosphine in the Venusian atmosphere is computed in the photochemical code. Here we break down the destruction rate into its components to gain a better understanding of the chemical sinks of $PH_3$ as a functional of altitude and to enable comparison with other models of the Venusian atmosphere.

The destruction rate components of phosphine are: reactions with O, H, OH, and Cl radicals, direct photolysis by UV radiation, and thermolytic decay, discussed above, in Supplementary Section 1.1.1. The destruction rates due to each of these radicals are shown in Figure 2, in the main text. We also examine the vertical transport of $PH_3$ in the atmosphere of Venus (Supplementary Section 1.1.5.1.) and its effects on the on the final lifetime calculations (Supplementary Section 1.1.5.2.). We close this section with the discussion of the limitations of our lifetime calculations (Supplementary Section 1.1.5.3.).

#### 1. 1. 5. 1. *Vertical Transport Lifetime of $PH_3$*

The photochemical lifetime of $PH_3$ can be long in the deep atmosphere (<50 km), but is always short in the high atmosphere (>78-98 km) where UV-generated radicals efficiently destroy $PH_3$. In the deep atmosphere, transport to the upper atmosphere limits $PH_3$ lifetime. To account for the effects of transport on $PH_3$ lifetime, we calculate the transport timescale for $PH_3$ at altitude $z_1$ due to eddy diffusion, via $t_{transport}=\Delta z^2/(2K_{zz})$ ((Jacob 1999), Eqn. 4.23), where $K_{zz}$ is the eddy diffusion coefficient, and $\Delta z = z_0 - z_1$, where $z_0$ is the vertical altitude at which $PH_3$ lifetimes are short due to photochemistry. The pseudo-first order rate constant of $PH_3$ loss due to eddy diffusion is thus $1/t_{transport} = 2K_{zz}/\Delta z^2$. Out of an abundance of caution, we adopt $K_{zz} = K_{zz}(0)$; since $K_{zz}$ monotonically nondecreases with $z$, this underestimates $K_{zz}$, overestimates $t_{transport}$, and underestimates the destruction rate. We similarly adopt $z_1 = 0$ for conservatism. This means that we assume it is as slow for $PH_3$ to diffuse from the detection altitude, 61 km, to the upper atmosphere, as it is for $PH_3$ to diffuse from the surface to the upper atmosphere. This assumption again overestimates $t_{transport}$, and underestimates the destruction rate.

#### 1. 1. 5. 2. *Overall Lifetime Calculation*

In each altitude bin, we adopt the minimum of the transport timescale to 78 km (where the photochemical lifetimes are, ≤1 s; Figure 2, in the main text) and the photochemical lifetime as our overall lifetime. We invert the lifetime to obtain the pseudo-first-order destruction rate constant. This approach assumes that molecules are lost either to transport or to photochemistry. In reality, both loss mechanisms apply; consequently, this approach underestimates $PH_3$ destruction rates and overestimates the $PH_3$ lifetime.

#### 1. 1. 5. 3. *Chlorine Atom Chemistry in the Lower and Middle Atmosphere of Venus*

In our model Cl peaks at the surface and at 25-35 km. The surficial peak is due to thermal chemistry common to the models of (Zhang *et al.* 2012) and (Krasnopolsky 2007).

The reason for the Cl peak between 25 – 35 km in our model is due to the combination of the networks of K07 (Krasnopolsky 2007) and Z12 (Zhang *et al.* 2012). Within 5 km of the surface of Venus, the temperatures are great enough for the following reactions occur with reasonable efficiency (reaction numbers given from in the form of K# for K07 and Z# for Z12):

$S_2 + CO \rightarrow OCS + S,$      K18;

$H_2S + S \rightarrow 2\,HS,$      K27;

$HS + HCl \rightarrow H_2S + Cl,$      K37;

$Cl + SO_2 + M \rightleftarrows ClSO_2 + M,$      K44;

$S_2 + ClSO_2 \rightarrow SO_2 + ClS_2,$      Z298;

The $ClS_2$ diffuses upward and is photodissociated with increasing efficiency with altitude, since the photodissociation rate depends on a cross-section that is reasonably large and that extends in wavelength to 485 nm (Zhang *et al.* 2012). At heights greater than 30 km, Cl becomes consumed by the products of $H_2SO_4$ dissociation from the bottom of the clouds.

There are differences in lower atmospheric Cl abundances of two or more orders of magnitude between each of the models. In our model, the number density of Cl between 25 and 35 km is between 100 and 702 $cm^{-3}$, which is much greater at that height than the number densities of Bierson (Bierson and Zhang 2019), with a maximum of 128 $cm^{-3}$ at 10 km, or Krasnopolsky (Krasnopolsky 2007), which has a maximum Cl density at the surface of 105 cm-3, dropping rapidly to less than 1 $cm^{-3}$ at >27 km. Consequently, between 25 and 35 km, our predicted lifetime for $PH_3$ is substantially shorter than that inferred from other models, and is shorter even than thermochemical destruction or the diffusion timescale. This explains the peak that is present in our model but not in the models of Krasnopolsky (Krasnopolsky 2007) or Bierson (Bierson and Zhang 2019).

*1. 1. 5. 4. Caveats and Limitations of the Lifetime Calculation*

The main limitations of the calculations presented here is that they likely underestimate the $PH_3$ destruction rate and overestimate the $PH_3$ lifetimes in the deep atmosphere. We have already remarked on our use of room-temperature rate constants for Cl attack, our neglect of N attack, and that our calculation methodology may overestimate diffusion timescale and hence underestimate loss due to vertical transport. For transport, we have further assumed that $PH_3$ must be transported to the high atmosphere to be destroyed, whereas in some cases transport to the bottom of the atmosphere where thermolytic decay is fastest may even more efficiently destroy $PH_3$. We may have neglected other relevant loss processes, due primarily to incompleteness of knowledge of $PH_3$ loss. Nearly all modern phosphine degradation kinetics measurements come from the organometallic vapor phase epitaxy (MOVPE/OMVPE) literature, which is concerned with fractional breakdown of organophosphines over semi-conductor surfaces, in which phosphine is sometimes included as a reference compound. Similarly, theoretical work on thermal decomposition of $PH_3$ is lacking. An early isolated theoretical study that calculates a theoretical rate of thermal decomposition of phosphine gives reaction constant values corresponding to lifetime of 2630 years at 673 K (Buchan and Jasinski 1990), a result that differs significantly from the extrapolated experimental measurements. Only two experimental studies by (Hinshelwood and Topley 1924; Larsen and Stringfellow 1986) give data, on large enough volumes and without catalytic metals, from which free gas kinetics can be extracted, and suggest a half-life of phosphine to thermal breakdown under Venusian surface conditions is 27.2 hours, or 4.2 days as 670 K, which is consistent with the textbook comment that phosphine breaks down 'slowly' at 400 °C (Prescott 1939). For example, the silica surface-catalyzed thermal decomposition of phosphine is well-known from the semi-conductor industry, where surface-catalyzed rates are several orders of magnitude faster than gas phase rates (Hinshelwood and Topley 1924; Larsen and Stringfellow 1986). If similar surface catalysts exist on the surface of Venus, $PH_3$ thermolysis rates may be larger than we have modeled here.

### 1. 2.   Creation of the Forward Chemical Kinetic Network of Phosphorus Species

Many photochemically generated radicals in the Venusian atmosphere (e.g. H) can in principle, very efficiently react with oxidized phosphorus species in the atmosphere leading to their reduction and hence to the potential formation of phosphine. We explore the potential photochemical phosphine production by modeling the kinetics of the chemical reactions between the photochemically generated radicals and the oxidized phosphorus species. We construct a network of possible reactions, and calculate the maximum possible flux through the forward chemical kinetic network (*neglecting* any back reactions), as a function of altitude.

We consider phosphoric acid ($H_3PO_4$) as a starting point of the network because the kinetics of other oxidized phosphorus species, that could serve as alternative starting points of the network (e.g. $P_4O_6$, $P_4O_{10}$) are unknown, and because $H_3PO_4$ is predicted to be the dominant form of phosphorus in the clouds of Venus (See Section 3.2.1.1 of the main paper for the discussion of the dominant phosphorus species in the atmosphere of Venus). We note that choosing $H_3PO_4$ as a starting point is a conservative approach and the "best case scenario" for the production of phosphine. In contrast to other dominant phosphorus species $H_3PO_4$ can serve as a source of both P and H needed for the formation of $PH_3$ in the network. The network contains all reaction rates where their kinetics parameters (which give rate as a function of temperature) are known. Kinetic data for reactions were extracted from the NIST kinetics database (Linstrom and Mallard 2001), supplemented by (Kaye and Strobel 1984) and (Bolshova and Korobeinichev 2006) . The network is shown in Figure S4.

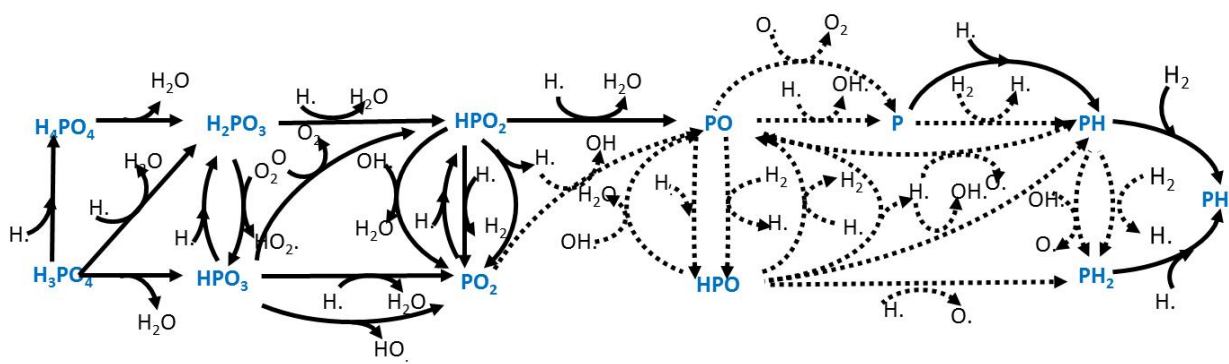

**Fig. S4.** The forward reaction network of phosphorus species to form phosphine. Solid lines represent reactions of phosphorus species for which kinetic data is available in the NIST reaction kinetics database or in (Kaye and Strobel 1984) or (Bolshova and Korobeinichev 2006) are considered. Dotted lines are reactions where phosphorus species kinetics are not known and the analogous nitrogen species reaction kinetics was used instead (see Supplementary Section 1.2.2.). Figure modified from (Greaves *et al.* 2020c).

For each sequence of reactions that lead to $PH_3$, there will be one reaction that is slower than the rest. Such reaction is the "rate-limiting step" and the rate of this reaction accurately represents the rate of the entire sequence of reactions. We next consider whether the rate of this reaction is sufficient to explain the observed amounts of phosphine in Venus' atmosphere. We illustrate the rationale for the approach on a simpler example that considers just six reactions (Figure S5). If *any* of the reactions are too slow to produce the required flux of phosphine, then production of phosphine is not kinetically possible *no matter* how fast the other steps are, or what are the concentration of the other intermediates. For example, if reaction 4 in the simplified network presented on Figure S5 could only proceed at $10^{-4}$ times the rate needed to compensate for the rate of destruction of phosphine, phosphine would not be produced at the required rate to explain the observed abundance of phosphine. This holds even if all the phosphorus in the atmosphere was present as PO, and regardless of the rates of reactions 1, 2, 3, 5 and 6. In the case where reaction 4 is the slowest, most of the non-phosphine phosphorus in the atmosphere would be present as PO. If one reaction in a network of reactions such as that in Figure S4 is the rate limiting step from start to end product, then most of the phosphorus species will 'accumulate' as reactants for that reaction. Therefore, as a limiting case, we calculated the rate of reaction assuming that *all* the phosphorus in the atmosphere was present as the reacting phosphorus species in *each* reaction. We realize that this is not self-consistent. In the simple reaction scheme in Figure S5, all the atmosphere's phosphorus cannot be present as $H_3PO_4$ *and* $H_2PO_3$ *and* $HPO_2$ etc. Such approach will overestimate the rate of reaction through the network and is consistent with our goal of estimating the maximum possible rate at which phosphine could be produced through reaction of oxidized phosphorus species with photochemical intermediates.

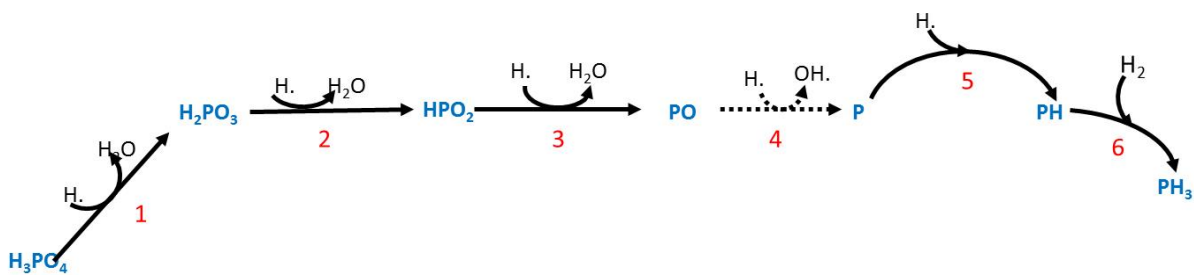

**Fig. S5.** Illustration of the rationale behind the creation of the network of reactions of phosphorus species with Venusian atmospheric components. We assess which reactions in the network are "rate limiting" and are too slow to produce the required flux of phosphine. In the simplified reaction network shown, the reaction 4 is the "rate limiting" step (dotted line).

The rate of reaction for all of the reactions shown in Figure S4 was calculated for 1 km steps in altitude from 0 to 115 km. Reactions were calculated to 115 km because this is the limit of the estimated number densities for H, OH and O species. Some of the reactions between reactive radicals such as H and PO are expected to happen on timescales relevant to this study (i.e. days or hours) even at ~180 K, the temperature of Venus' atmosphere at 115 km. (this contrasts to the thermodynamic calculations discussed below, where reactions are between stable species and hence will happen at negligible rates below ~260 K). The concentration of all species except the phosphorus species was taken from the photochemical model as described above. We emphasize that our network is purely a model of the reductive reactions in the phosphorus species network. It is not an equilibrium model incorporating the back reactions. This therefore represents the maximum possible rate of production of phosphine.

### 1. 2. 1. Estimation of the Gas Phase Concentration of Phosphorus Species.

The concentration of the dominant phosphorus species in the Venusian atmosphere is uncertain. The only measure of atmospheric phosphorus was provided by the Vega descent probe. With an exception of the detection of the trace ~ ppb amounts of phosphine by (Greaves *et al.* 2020a; Greaves *et al.* 2020b; Greaves *et al.* 2020c) and possible identification of $PH_3$ in re-analyzed data from Pioneer 13 LNMS (Mogul *et al.* 2021), no subsequent studies comment on the presence of the phosphorus species in Venus' atmosphere.

The Vega descent probe found fluctuating level of phosphorus in an elemental analysis of materials captured on a filter. In the altitude range of 52 and 47 km the abundance of phosphorus appears to be in the same order as the abundance of sulfur (Andreichikov 1987; Andreichikov 1998b; Surkov *et al.* 1974; Vinogradov *et al.* 1970a), as reviewed in (Titov *et al.* 2018). Above 52 km no phosphorus was detected, and at 47 km the probe appeared to fail. It is therefore plausible that phosphorus is present as a condensed, liquid or solid, phase in the cloud layer. We therefore assume that phosphorus in the gas phase is a saturated vapor over phosphorus in a condensed phase above the base of the clouds. Below the clouds, gaseous phosphorus is assumed to be well-mixed (Figure S6). We estimate the vapor pressure of phosphorus species as follows.

We use the vapor pressure of $P_4O_{10}$ over solid $P_4O_{10}$ as estimate of the vapor pressure over condensed phosphorus species in the clouds, as $P_4O_{10}$ represents the oxidation state of phosphorus expected to be most abundant at the level of the clouds that has a well-defined vapor pressure. We note that phosphorus at the altitude of the clouds is expected to be overwhelmingly present as oxidized phosphorus species. $H_3PO_4$ does not have a well-defined vapor pressure as it decomposes on boiling to mixed anhydrides, of which $P_4O_{10}$ is the end member. We describe the estimation of the vapor pressure over $P_4O_{10}$ below (Supplementary Section 1.3.2.2.).

Our results formally over-estimate the vapor pressure of phosphorus species, as some phosphorus in the clouds will be in the thermodynamically most favored state, $H_3PO_4$. The likely over-estimation of the concentration of the phosphorus species in the gas phase is a conservative approach that overall favors the formation of phosphine in the atmosphere of Venus.

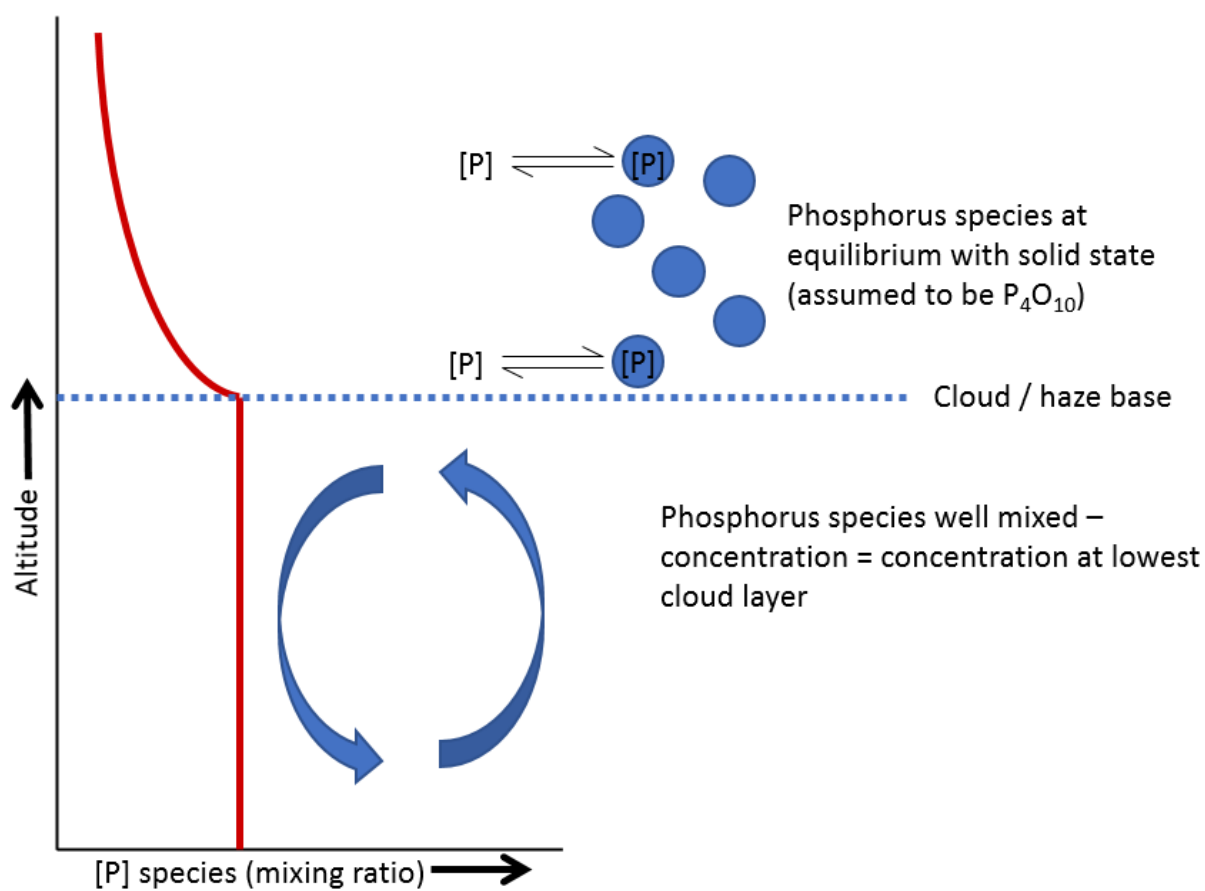

**Fig. S6.** Model of the concentration of phosphorus species in Venus' atmosphere. x axis: gas phase concentration of the phosphorus species in the atmosphere of Venus. y axis: the altitude in the atmosphere of Venus. Gas phase phosphorus is assumed to be saturated over condensed, liquid or solid phase phosphorus in the clouds. The vapor pressure of the gas phase phosphorus species (i.e. the concentration of the gaseous phosphorus species) falls as the altitude rises (as the temperature decreases). The concentration of the gaseous phosphorus species below the cloud decks (~48 km) is unknown and is assumed to be well-mixed (see

Supplementary Section 1.1.5.1 on vertical mixing below the cloud layer). See Supplementary Section 1.3.2.2. for more details on the estimation of the temperature dependent vapor pressure of the phosphorus species.

### 1. 2. 2. Kinetic data for Nitrogen Species as a Substitute for Missing Phosphorus Reactions

The crucial kinetic data for some reactions of phosphorus species are missing. For example, the kinetics of reactions in which a P=O bond is reduced to a P-H bond have not been studied in the gas phase.

To fill in missing kinetic data we construct the kinetic network where reactions of nitrogen species are used as a replacement for analogous reactions of phosphorus species. In particular, we are concerned to model the steps from PO and $PO_2$ to P and PH (by analogy from NO, $NO_2$, N and NH respectively), which are crucial steps in the formation of phosphine (Figure S4).

The likely reason that the P=O -> P-H class of reactions has not been studied is because P=O radical chemistry is investigated exclusively in the context of (terrestrial) combustion (Ballistreri *et al.* 1983; Haraguchi and Fuwa 1976), and especially in relation to phosphorus compounds as flame retardants. In these circumstances, P=O is the most abundant phosphorus-containing species present in the flame (Peters 1979), but in the presence of oxygen gas and its excited states P and PH species are not expected to exist. HP=O and P=O are not as stable as their nitrogen analogues HN=O and N=O, and cannot be isolated as pure gases at Standard Temperature and Pressure (STP) (Dittrich and Townshend 1986). $PO^{2-}$ is the phosphorus analogue of nitrite ($NO^{2-}$), and is also known. P=O double bonds are very common in phosphorus chemistry, being formed in $H_3PO_4$, $H_3PO_3$, and $P_4O_{10}$. Hypophosphite ($H_2PO_2^-$) forms a P=O double bond in preference to a structure with two single P-O bonds (although hypophosphite is only stable in aqueous solution).

The chemistry of phosphorus and nitrogen species is similar in some respects and such analogies between N and P elements are widely validated, e.g. in the theoretical spectroscopy literature (e.g. (Sousa-Silva *et al.* 2014; Sousa-Silva *et al.* 2016)).

We further justify the analogy between N and P in detail below and on Figure S7.

The bond energies of P=O (~588 kJ/mol) (Rao *et al.* 1981; Toy 2016) and N=O (~639 kJ/mol) (Mayer 1969) are similar (Figure S7(a)). The energy of forming the transition state in cleavage of H-N=O and H-P=O is also similar (Figure S7(b)). Similarly, the reaction chemistry of N=O and P=O forming metal complexes is similar across a wide range of metals (Corrigan *et al.* 1994; Herrmann 1991; Johnson *et al.* 1997; Scherer *et al.* 1991). We note that this is consistent with our informal observation of the close similarity in shape and orientation of HOMO and LUMO orbitals in P=O and N=O. Such P=O and N=O metal complexes are relatively stable at STP (Bérces *et al.* 2000; Scoles *et al.* 2001; Yamamoto *et al.* 1998).

Reactions of N=O- and P=O-containing species with H and OH radicals in which the P=O and N=O bonds are broken have similar kinetics overall (data from NIST (Linstrom and Mallard 2001)) (Figure S7(c), (d)). In the four cases where equivalent reactions have kinetic data available for N and P species, presented on (Figure S7(c), (d)), reactions of phosphorus are slower than reactions of nitrogen, so assuming that N reactions are representative of P reactions, the modelling a phosphorus reaction with a nitrogen reaction analogue will slightly over-estimate the rate of that reaction.

By contrast, reactions where the P-H bond is broken are very substantially faster than reactions where the N-H bond is broken, as would be expected from the much lower energy of the P-H bond compared to the N-H bond (as noted above in the discussion of the thermolytic decay of phosphine), therefore further formally overestimating the possibility of the formation of phosphine (Figure S7(e)).

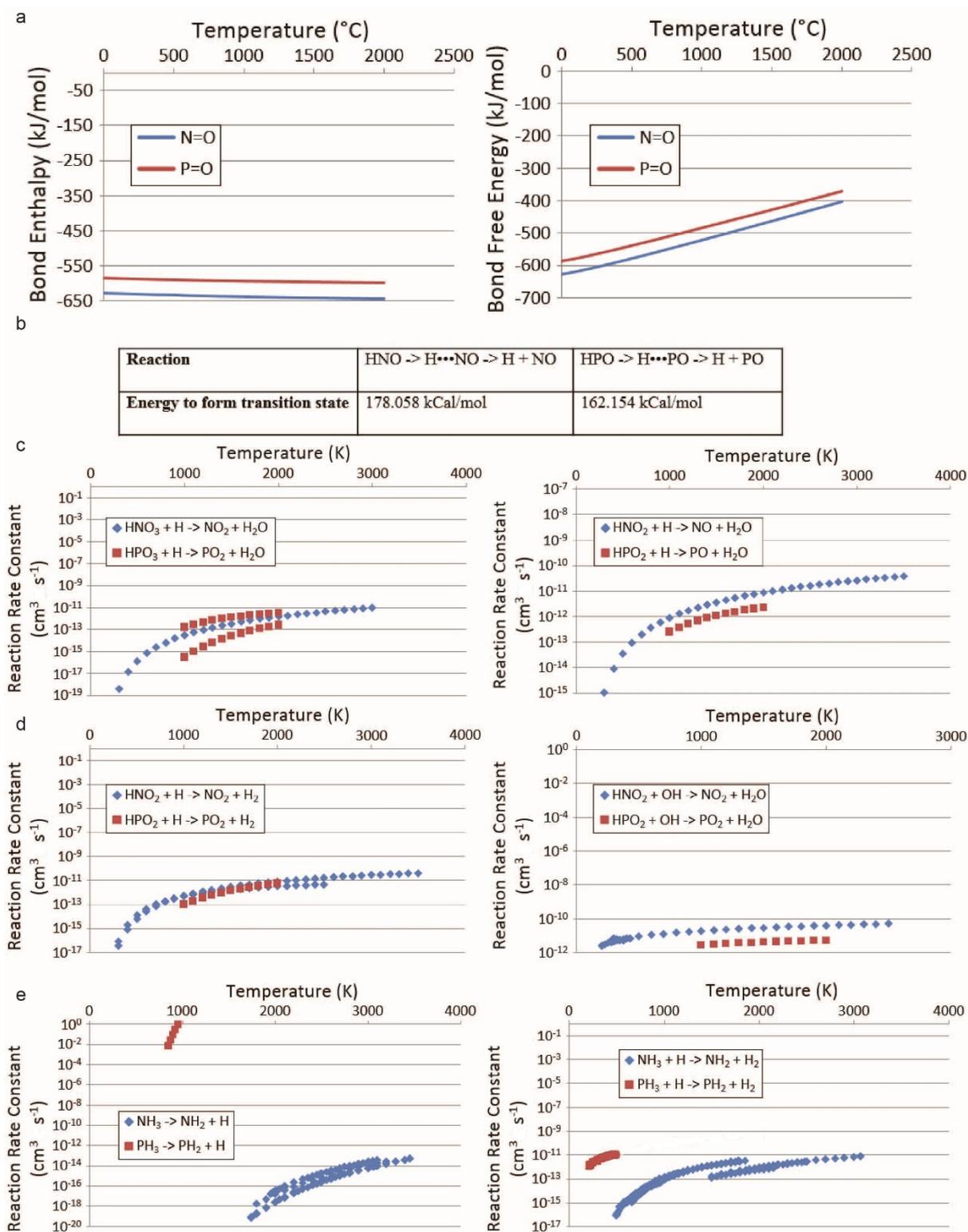

**Fig. S7.** The analogy between N and P chemistry. (a) Bond enthalpy (*left panel*) energy and Gibbs free energy of formation (*right panel*) for N=O and P=O are similar. x axis: Temperature (°C), y axis: Bond Enthalpy (kJ/mol) and Bond Energy (kJ/mol). The bond enthalpy and bond energy values are calculated from JANAF tables of free energy and entropy for reactions N + O -> NO and P + O -> PO (Chase 1998). (b) The energy of forming the transition state in cleavage of H-N=O and H-P=O is similar. The energy values are calculated using *Ab initio* methods using B3LYP approximation to 6-311G level of theory, using GAMESS (Schmidt *et al.* 1993). (c-e) For reactions where temperature-dependent rate information is available and is consistent between experiments, reactions of N=O and P=O-containing species with H and OH radicals have similar kinetics (data from NIST (Linstrom and Mallard 2001)). x axis: Temperature (K), y axis: Reaction Rate Constant ($cm^3 s^{-1}$). (c)

*Left panel*: HNO₃ + H -> NO₂ + H₂O *vs* HPO₃ + H -> PO₂ + H₂O. Note that two sources gave significantly different rates for the phosphorus reaction, which bracket the nitrogen value *Right panel*: HNO₂ + H -> NO + H₂O *vs* HPO₂ + H -> PO + H₂O. N species react ~3-fold faster than P species. (d) *Left panel*: HNO₂ + H -> NO₂ + H₂ vs HPO₂ + H -> PO₂ + H₂. The phosphorus reaction has only been measured at high temperatures, where it has a rate very similar to the nitrogen reaction. *Right panel*: HNO₂ + OH -> NO₂ + H₂O *vs* HPO₂ + OH -> PO₂ + H₂O. Reaction rate constants differ by a factor of ~5 over the range where both are measured. (e) *Left panel*: NH₃ -> NH₂ + H *vs* PH₃ -> PH₂ + H. Note that the reactions have not been measured at the same temperature range, but it is clear that the two sets of points belong to substantially different curves. *Right panel*: NH₃ + H -> NH₂ + H₂ *vs* PH₃ + H -> PH₂ + H₂. Note that the phosphorus species data is only for low temperatures. In both reactions (left and right panels) the P-H bond is broken much faster than the N-H bond. PH₃ is expected to be much more efficiently destroyed than its nitrogen counterpart, NH₃, which leads to the formal overestimation of the formation of phosphine.

### 1. 3. Methods Used in the Thermodynamic Analysis of Potential Phosphine-Producing Reactions

#### 1. 3. 1. Overview of Method for Calculating the Gibbs Free Energy of the Reaction (ΔG)

We calculate the Gibbs Free Energy on the basis of established textbook knowledge (Greiner *et al.* 2012; Perrot 1998) and previously published work (Bains *et al.* 2019). In brief, the free energy of a reaction occurring in non-standard conditions is given by

$$\Delta G = \Delta G^0 + R.T.\ln(Q). \tag{24}$$

Here $\Delta G$ is the free energy of reaction, $\Delta G^0$ is the standard free energy (i.e. the energy where all the reagents are in their standard state), R is the gas constant, T is the absolute temperature and Q is the reaction quotient. The standard free energy of a reaction is the sum of the standard free energy of the products minus the standard free energy of the reactants. The reaction quotient Q is given by

$$Q = \frac{\prod_1^n ap_i^{si}}{\prod_1^m ar_i^{si}} \tag{25}$$

Where *ap*$_i$ is the activity of product number *i*, and *si* is the number of moles of product *i* in the reaction, and *ar*$_i$ is the activity of the reactant *i*, and again *si* is the number of moles of that reactant in the reaction. Thus, for the reaction

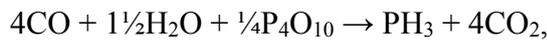
4CO + 1½H₂O + ¼P₄O₁₀ → PH₃ + 4CO₂,

$$Q = \frac{\{PH_3\}\cdot\{CO_2\}^4}{\{CO\}^4\cdot\{H_2O\}^{1.5}\cdot\{P_4O_{10}\}^{0.25}} \tag{26}$$

where {species} is the activity of that species.

Thus, to calculate the free energy of a reaction, and hence estimate whether it will proceed spontaneously, we need to know the standard free energy of the reactants and products and the activities of the reactants and products.

The standard free energy of reaction is itself a function of temperature. The values for the standard free energy ($\Delta G^0$) as a function of temperature between 250 K and 1000 K (where relevant – e.g. there is no free energy of liquid water at temperatures over 673 K because liquid water does not exist above this temperature) were obtained from the sources listed in Table S3.

| Species and phase | Source |
|---|---|
| $P_4O_{10(g)}$, $P_4O_{10(s)}$, $H_3PO_{4(s/l)}$, $H_2SO_{4(g)}$, $PH_{3(g)}$, $H_2O_{(l)}$, $H_2O_{(g)}$, $SO_{2(g)}$, $CO_{(g)}$, $CO_{2(g)}$, $H_2S_{(g)}$, $S_{(g)}$, $OCS_{(g)}$ | (Chase 1998) |
| $H_3PO_{3(aq)}$, $H_3PO_{4(aq)}$ | (Barner and Scheurman 1978) |
| $H_3PO_{3(cr/l)}$ | Calculated from (Barner and Scheurman 1978; Guthrie 1979) |
| $PH_{3(aq)}$ | Calculated from (Chase 1998; Fu *et al.* 2013) |
| $SO_{2(aq)}$, $CO_{2(aq)}$, $CO_{(aq)}$, $H_2S_{(aq)}$ | (Amend and Shock 2001) |
| $CaO_{(s)}$, $Al_2O_{2(s)}$, $MgO_{(s)}$, $CaF_{2(s)}$, $AlPO_{4(s)}$, $Ca_3(PO_4)_{2(s)}$, $Ca_5(PO_4)_3F_{(s)}$, $Mg_3(PO_4)_{2(s)}$, $CaSO_{4(s)}$, $MgSO_{4(s)}$, $FeO_{(s)}$, $FeS_{2(s)}$, $Fe_2O_{3(s)}$, $Fe_3O_{4(s)}$ | (Robie and Hemingway 1995) |
| $NAD_{(aq)}$, $FAD_{(aq)}$, Coenzyme-$Q_{(aq)}$ | (Lehninger 2004; Pratt and Cornley 2014) |
| Ferredoxins$_{(aq)}$ | (Smith and Feinberg 1990) |

**Table S3.** Sources for the values for the standard free energy ($\Delta G^0$) as a function of temperature between 250 K and 1000 K for chemical species used in this study.

At high pressures gas activities differ significantly from their partial pressures. Gas activity was corrected for pressure and temperature according to Berthelot's equation (Rock 1969):

$$a = P \cdot exp\left[\frac{9T_c}{128P_c \cdot T} \cdot \left(1 - \frac{6T_c^2}{T^2}\right) \cdot P\right], \tag{27}$$

where *a* is the activity of the species, *P* is the pressure, *T* is the absolute temperature, $T_c$ is the species' critical temperature and $P_c$ is the species critical pressure. Critical pressures and temperatures were obtained from the sources listed in Table S4.

| Species | $T_c$, $P_c$ source |
|---|---|
| $CO_2$, $H_2S$, $H_2O$, $N_2$ | (Ballesteros *et al.* 2019) |
| $SO_2$ | (Médard 2019) |
| OCS | (Robinson and Senturk 1979) |
| CO | (ToolBox 2003) |
| $H_2$ | (Hoge and Lassiter 1951) |

**Table S4.** Sources for critical pressure ($P_c$) and critical temperature ($T_c$) values of gaseous chemical species used to calculate their gas activities in this study.

The critical temperature of $H_3PO_4$, $P_4O_{10}$ and $P_4O_6$ were assumed to be sufficiently high that these species behaved like a near perfect gas at Venus temperatures.

The activity of solids was assumed to be 1. The standard state of a *pure* solid reagent is 1; mixtures may have different activities, but as the nature of the mixtures are not known all solids were assumed to be single chemical species. At Venus surface pressures, pressure corrections will not introduce material activity changes in solids. Changes in free energy with temperature are included in the sources given above in Table S3.

### 1. 3. 2. Modelling of the Thermodynamics of the Atmospheric and Surface Reactions

Next, we discuss our reasoning behind the choice of chemical reactants (Section 1.3.2.1) and their input concentrations (Section 1.3.2.2). We also present the relevant chemical reactions that could in principle lead to the phosphine formation in the Venusian environment (Section 1.3.2.3).

#### *1. 3. 2. 1. Choice of Reactants*

*Choice of dominant atmospheric phosphorus species.* Phosphorus-containing species have not been modeled for Venus' atmosphere before and our work represents the first attempt to model the dominant phosphorus species in Venusian atmosphere.

We analyzed what phosphorus-containing chemicals are likely to be present in Venus' atmosphere by calculating which species would be most thermodynamically stable under Venusian atmospheric conditions (e.g., concentrations of water and reducing gases).

The equilibrium between the four phosphorus species was calculated as follows by calculating the ln(Q) values for each of the following reactions which would result in a ΔG=0. In all cases, the activities of the other components were calculated as described in the main text, and therefore the calculated concentration of phosphorus species cover a range of values.

There are two classes of reactions involved; dehydration reactions and reduction reactions. For dehydration reactions only one reaction is possible. For reduction reactions five reactions are considered, corresponding to the five reducing gases likely to be present in trace amounts in Venus' atmosphere, and the average free energy of the reactions was used. (Note that if the atmosphere were at equilibrium then each of the reactions would give that same result; however, the atmosphere is not at equilibrium.)

*Dehydration reactions*
$H_3PO_4 \rightarrow \frac{1}{4} P_4O_{10} + 1\frac{1}{2} H_2O$
$H_3PO_3 \rightarrow \frac{1}{4} P_4O_6 + 1\frac{1}{2} H_2O$

*Reduction reactions*
$H_3PO_4 + \frac{1}{3} H_2S \rightarrow H_3PO_3 + \frac{1}{3} SO_2 + \frac{1}{3} H_2O$
$H_3PO_4 + H_2 \rightarrow H_3PO_3 + H_2O$
$H_3PO_4 + CO \rightarrow H_3PO_3 + CO_2$
$H_3PO_4 + \frac{1}{2} S \rightarrow H_3PO_3 + \frac{1}{2} SO_2$
$H_3PO_4 + \frac{1}{3} OCS \rightarrow H_3PO_3 + \frac{1}{3} SO_2 + \frac{1}{3} CO_2$

¼P₄O₁₀ + ⅓H₂S → ¼P₄O₆ + ⅓H₂O + ⅓SO₂

¼P₄O₁₀ + H₂ → ¼P₄O₆ + H₂O

¼P₄O₁₀ + CO → ¼P₄O₆ + CO₂

¼P₄O₁₀ + ½S → ¼P₄O₆ + ½SO₂

¼P₄O₁₀ + ⅓OCS → ¼P₄O₆ + ⅓CO₂ + ⅓SO₂

*Choice of reducing agents.* The conversion of the oxidized variant of phosphorus in the P(+3) or P(+5) oxidation state to phosphine requires a reducing agent and a source of hydrogen atoms.

Two reducing agents – H₂S and H₂ – are themselves sources of hydrogen atoms, a further three reducing gases – CO, OCS and elemental sulfur – contain no hydrogen atoms, and hence require a reaction involving water to provide hydrogen. Gas phase elemental sulfur is taken as the most stable species (S₂ or S₈) at each temperature. In principle, N₂ (which can be oxidized to HNO₃) or HCl (which can be oxidized to perchlorate) could also act as reducing agents. Preliminary calculations suggested that the energy requirements to use the oxidation of N₂ or HCl as a source of electrons to reduce phosphates to phosphine were very high, and so these reactions were not considered further.

The reducing agents on the surface of Venus are unknown, but solid mineral reducing agents are likely to be salts of redox active metals. Iron(II) compounds are potential reductants for phosphorus species (Herschy et al. 2018), and the presence of H₂S and HCl in the Venusian atmosphere suggest that FeS₂ and FeCl₂ should be considered as potential reductants. In the presence of excess liquid water Fe(II) oxidation (serpentenization reactions) have been shown to be capable to reducing phosphate to phosphite at 25°C (Pasek *et al.* 2020), although of course this chemistry could not happen on Venus' surface as there is negligible water there, and water is present in gas phase only. In addition, FeS₂ is unstable under Venus surface conditions (Fegley 1997) and FeCl₂ may be unstable below the cloud deck level on Venus (Figure S8), both spontaneously forming Fe(III) species. We therefore ignored calculations involving FeCl₂ at altitudes below which FeO is thermodynamically favored over FeCl₂.

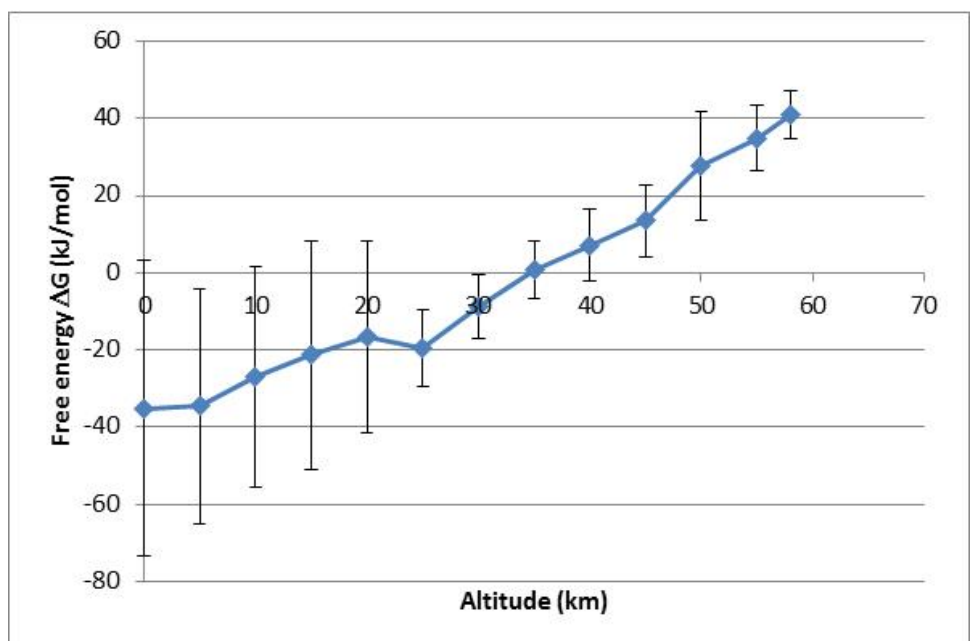

**Figure S8.** FeCl$_2$ may be unstable below the cloud deck level on Venus. x axis: altitude. y axis: free energy of reaction of hydrolysis of FeCl$_2$ by atmospheric water under Venus conditions. Bars show range of ΔG values resulting from different partial pressures of HCl and H$_2$O. The free energy of the reaction FeCl$_2$ + H$_2$O -> FeO + 2HCl was calculated as explained in the main text. We ignored calculations involving FeCl$_2$ at altitudes below that at which the free energy of the reaction forming FeCl$_2$ from FeO and HCl was negative (i.e. at altitudes below which FeO is thermodynamically favored over FeCl$_2$).

*Choice of the surface minerals.* The chemical composition of the surface minerals of Venus is poorly known. The only data on the crustal composition of Venus comes from the X-ray fluorescence measurements of the bulk composition of the crust by Vega (Surkov *et al.* 1986) and Venera (Surkov *et al.* 1984) landers. The measurements suggest that the Venus' crustal composition is extremely similar to terrestrial tholeitic basalts. Terrestrial basalts contain very low amounts of phosphorus (0.08% - 0.45%). If phosphorus is present on the surface of Venus, it is likely to be in the form of phosphate salts. We have considered phosphate salts of Mg, Ca, Al and K, with fluorapatite included as well as HF is probably present in the atmosphere. Phosphate minerals were assumed to be present as differentiated minerals, i.e. as pure solids whose activity is 1. The presence of pure solids is geologically unlikely, but (as with many other assumptions in this paper) presents a 'best case scenario' for making phosphine chemically. Reduction of these minerals by all the reducing atmospheric gases mentioned above was also modeled.

The selection of relevant reactants that build subsurface rocks and minerals of Venus is discussed in Section 1.3.3.

### 1. 3. 2. 2.   *Choice of Vertical Concentration Profiles*

For each of the trace gases in Venus' atmosphere, we use two different vertical concentration profiles, representing a maximum and minimum concentration as reported in the literature (Table S5). Some sources are either directly measured but most gas species are theoretically estimated (Table S5). Gas concentrations were only explored up to 60 km altitude, above which the temperature is below 260 K and reactions would be so slow that thermodynamics would not effectively predict what species would be present. Note that 60 km is a lower height limit than was applied for the kinetics modeling above in Section 1.2. This is because in this section we are concerned with the reactions of stable chemical species with each other. These have extremely slow reaction kinetics below the freezing point of water, and so we can neglect any reaction by these species above 60 km, where the temperature <260 K. By contrast, the reaction network shown in Section 1.2. involved reactive radicals, which have very fast reaction kinetics at temperatures above 150 K. The difference is illustrated from everyday experience on Earth. Chemicals such as isoprene are stable for decades even in the presence of oxygen in the absence of light, because the kinetics of direct reaction of oxygen with isoprene are immeasurably slow. However, the reaction destroys isoprene in the range of hours or minutes under sunlight because photolysis generates reactive radicals which then initiate reaction (Zhan *et al.* 2020).

| Temperature (K) | Pressure (bars) | Altitude (km) | Lowest partial pressure | T (theoretical) vs M (measured) | Reference | Highest partial pressure | T (theoretical) vs M (measured) | Reference |
|---|---|---|---|---|---|---|---|---|
| SO$_2$ | | | | | | | | |
| 735 | 92.1 | 0 | 1.7E-06 | T | A | 2.3E-04 | T | A |
| 697 | 66.5 | 5 | 1.8E-06 | M | F | 2.3E-04 | T | A |
| 658 | 47.4 | 10 | 4.5E-06 | M | F | 5.0E-04 | M | B |
| 621 | 33.0 | 15 | 2.2E-05 | T | G | 1.9E-04 | M | B |
| 579 | 22.5 | 20 | 2.3E-05 | T | G | 1.9E-04 | T | A |
| 537 | 14.9 | 25 | 2.4E-05 | T | A | 1.9E-04 | T | A |
| 495 | 9.85 | 30 | 2.5E-05 | M | F | 1.9E-04 | T | G |
| 453 | 5.92 | 35 | 2.5E-05 | M | F | 2.0E-04 | T | A |
| 416 | 3.50 | 40 | 2.6E-05 | T | G | 2.0E-04 | T | A |
| 383 | 1.98 | 45 | 3.0E-05 | T | G | 2.0E-04 | T | A |
| 348 | 1.07 | 50 | 3.3E-05 | M | F | 2.0E-04 | T | A |
| 300 | 0.53 | 55 | 3.7E-05 | T | G | 2.1E-04 | T | A |
| 263 | 0.24 | 60 | 4.3E-05 | M | F | 2.2E-04 | T | A |
| H$_2$S | | | | | | | | |
| 735 | 92.1 | 0 | 5.0E-09 | T | B | 2.2E-06 | T | A |
| 697 | 66.5 | 5 | 1.4E-08 | T | B | 2.2E-06 | T | A |
| 658 | 47.4 | 10 | 3.7E-08 | T | B | 2.2E-06 | T | A |
| 621 | 33.0 | 15 | 7.7E-08 | T | C | 2.2E-06 | T | A |
| 579 | 22.5 | 20 | 7.8E-08 | T | B | 2.3E-06 | T | A |
| 537 | 14.9 | 25 | 9.0E-08 | T | B | 2.3E-06 | T | B |
| 495 | 9.85 | 30 | 1.0E-07 | T | C | 9.1E-07 | T | A |
| 453 | 5.92 | 35 | 1.2E-07 | T | C | 2.1E-06 | T | A |
| 416 | 3.50 | 40 | 1.2E-07 | T | B | 2.1E-06 | T | A |
| 383 | 1.98 | 45 | 1.4E-07 | T | C | 2.2E-06 | T | A |
| 348 | 1.07 | 50 | 1.6E-07 | T | C | 2.2E-06 | T | A |
| 300 | 0.53 | 55 | 1.7E-07 | T | C | 2.2E-06 | T | A |
| 263 | 0.24 | 60 | 1.7E-07 | T | C | 2.2E-06 | T | A |
| S | | | | | | | | |
| 735 | 92.1 | 0 | 8.8E-10 | T | B | 3.4E-06 | T | C |
| 697 | 66.5 | 5 | 8.8E-10 | T | B | 5.7E-06 | T | C |
| 658 | 47.4 | 10 | 8.8E-10 | T | B | 9.3E-06 | T | C |
| 621 | 33.0 | 15 | 8.8E-10 | T | B | 1.4E-05 | T | C |
| 579 | 22.5 | 20 | 8.8E-10 | T | B | 2.2E-05 | T | C |
| 537 | 14.9 | 25 | 8.8E-10 | T | B | 3.0E-05 | T | C |
| 495 | 9.85 | 30 | 1.1E-09 | T | B | 3.2E-05 | T | C |
| 453 | 5.92 | 35 | 3.7E-09 | T | B | 3.4E-05 | T | C |
| 416 | 3.50 | 40 | 1.0E-08 | T | B | 2.1E-07 | T | B |
| 383 | 1.98 | 45 | 1.3E-08 | T | B | 4.0E-07 | T | B |
| 348 | 1.07 | 50 | 1.5E-08 | M | E | 4.0E-07 | T | B |
| 300 | 0.53 | 55 | 1.5E-08 | M | E | 5.9E-07 | T | B |
| 263 | 0.24 | 60 | 2.0E-08 | M | E | 1.5E-06 | T | C |
| | | | | OCS | | | | |
| 735 | 92.1 | 0 | 7.0E-12 | T | B | 2.6E-05 | T | C |
| 697 | 66.5 | 5 | 4.6E-10 | T | B | 2.7E-05 | T | C |
| 658 | 47.4 | 10 | 6.5E-10 | T | B | 5.6E-05 | T | A |
| 621 | 33.0 | 15 | 6.0E-09 | T | B | 5.9E-05 | T | A |
| 579 | 22.5 | 20 | 1.0E-08 | M | E | 6.0E-05 | T | A |
| 537 | 14.9 | 25 | 1.2E-08 | T | B | 6.0E-05 | T | A |
| 495 | 9.85 | 30 | 1.6E-08 | T | A | 6.1E-05 | T | A |
| 453 | 5.92 | 35 | 2.1E-08 | T | A | 6.2E-05 | T | A |
| 416 | 3.50 | 40 | 2.6E-08 | T | A | 1.3E-05 | T | A |
| 383 | 1.98 | 45 | 4.2E-08 | T | A | 1.5E-05 | M | E |
| 348 | 1.07 | 50 | 1.1E-07 | T | B | 1.6E-05 | T | C |
| 300 | 0.53 | 55 | 1.3E-07 | T | B | 2.0E-05 | T | C |
| 263 | 0.24 | 60 | 2.8E-07 | T | A | 2.3E-05 | T | C |
| | | | | $H_2O$ | | | | |
| 735 | 92.1 | 0 | 5.7E-06 | T | A | 1.4E-03 | T | B |
| 697 | 66.5 | 5 | 1.1E-05 | T | A | 1.4E-03 | T | B |
| 658 | 47.4 | 10 | 1.9E-05 | T | A | 2.2E-03 | M | B |
| 621 | 33.0 | 15 | 2.3E-05 | T | C | 5.2E-03 | M | B |
| 579 | 22.5 | 20 | 2.4E-05 | T | G | 8.0E-03 | M | I |
| 537 | 14.9 | 25 | 2.4E-05 | T | G | 1.4E-03 | T | B |
| 495 | 9.85 | 30 | 2.5E-05 | M | E | 1.4E-03 | T | B |
| 453 | 5.92 | 35 | 2.5E-05 | M | E | 1.4E-03 | T | B |
| 416 | 3.50 | 40 | 2.5E-05 | M | E | 1.4E-03 | T | B |
| 383 | 1.98 | 45 | 2.5E-05 | M | E | 1.4E-03 | T | B |
| 348 | 1.07 | 50 | 2.6E-05 | M | E | 1.4E-03 | T | B |
| 300 | 0.53 | 55 | 2.6E-05 | M | E | 1.4E-03 | T | B |
| 263 | 0.24 | 60 | 2.6E-05 | M | E | 1.4E-03 | T | B |
| | | | | CO | | | | |
| 735 | 92.1 | 0 | 1.6E-12 | T | B | 3.4E-05 | T | A |
| 697 | 66.5 | 5 | 1.8E-12 | T | B | 3.7E-05 | T | C |
| 658 | 47.4 | 10 | 3.3E-11 | T | B | 3.8E-05 | T | C |
| 621 | 33.0 | 15 | 3.7E-11 | T | B | 4.2E-05 | T | A |
| 579 | 22.5 | 20 | 4.9E-10 | T | B | 4.2E-05 | T | A |
| 537 | 14.9 | 25 | 5.4E-10 | T | B | 4.5E-05 | T | A |
| 495 | 9.85 | 30 | 5.3E-09 | T | B | 4.6E-05 | T | A |
| 453 | 5.92 | 35 | 5.7E-09 | T | B | 2.3E-05 | T | C |
| 416 | 3.50 | 40 | 4.1E-08 | T | B | 2.4E-05 | T | A |
| 383 | 1.98 | 45 | 4.4E-08 | T | B | 2.9E-05 | T | C |
| 348 | 1.07 | 50 | 2.2E-07 | T | B | 3.0E-05 | M | E |
| 300 | 0.53 | 55 | 2.4E-07 | T | B | 3.0E-05 | M | B |
| 263 | 0.24 | 60 | 8.8E-07 | T | B | 3.2E-05 | M | B |
| | | | | $H_2$ | | | | |
| 735 | 92.1 | 0 | 3.0E-13 | T | B | 5.8E-08 | T | B |
| 697 | 66.5 | 5 | 2.1E-12 | T | B | 8.2E-08 | T | B |
| 658 | 47.4 | 10 | 1.1E-11 | T | B | 8.8E-08 | T | B |
| 621 | 33.0 | 15 | 3.4E-11 | T | B | 1.1E-07 | T | B |
| 579 | 22.5 | 20 | 4.9E-11 | T | B | 1.3E-07 | T | B |
| 537 | 14.9 | 25 | 1.7E-10 | T | B | 1.4E-07 | T | B |
| 495 | 9.85 | 30 | 4.7E-10 | T | B | 4.5E-09 | T | D |
| 453 | 5.92 | 35 | 7.8E-10 | T | B | 4.5E-09 | T | D |
| 416 | 3.50 | 40 | 1.1E-09 | T | B | 4.6E-09 | T | D |
| 383 | 1.98 | 45 | 1.9E-09 | T | B | 4.7E-09 | T | D |
| 348 | 1.07 | 50 | 2.6E-09 | T | B | 4.9E-09 | T | D |
| 300 | 0.53 | 55 | 3.0E-09 | T | B | 7.5E-09 | T | B |
| 263 | 0.24 | 60 | 3.0E-09 | T | D | 3.5E-08 | T | B |
| HCl | | | | | | | | |
| 735 | 92.1 | 0 | 2.0E-07 | T | D | 5.4E-07 | T | A |
| 697 | 66.5 | 5 | 2.0E-07 | T | D | 5.5E-07 | T | A |
| 658 | 47.4 | 10 | 2.0E-07 | T | D | 5.5E-07 | T | A |
| 621 | 33.0 | 15 | 4.2E-07 | M | E | 5.6E-07 | T | A |
| 579 | 22.5 | 20 | 4.2E-07 | M | E | 3.1E-06 | T | H |
| 537 | 14.9 | 25 | 4.2E-07 | M | E | 8.6E-06 | T | H |
| 495 | 9.85 | 30 | 5.0E-07 | T | A | 1.3E-05 | T | H |
| 453 | 5.92 | 35 | 5.0E-07 | T | A | 1.3E-05 | T | H |
| 416 | 3.50 | 40 | 5.2E-07 | T | A | 1.7E-05 | T | H |
| 383 | 1.98 | 45 | 5.3E-07 | T | A | 3.6E-05 | T | H |
| 348 | 1.07 | 50 | 5.3E-07 | T | A | 5.3E-07 | T | A |
| 300 | 0.53 | 55 | 5.3E-07 | T | A | 5.3E-07 | T | A |
| 263 | 0.24 | 60 | 5.4E-07 | T | A | 5.4E-07 | T | A |

**Table S5.** List of sources for gas concentrations were collected from the available literature and are either directly measured or theoretically estimated. References used: A: (Taylor and Hunten 2014), B: (Oyama *et al.* 1980), C: (Krasnopolsky 2007), D: (Krasnopolsky 2012), E: (Marcq *et al.* 2018), F: (Vandaele *et al.* 2017), G: (Andreichikov 1998a), H: (Hoffman *et al.* 1979), I: (Vinogradov *et al.* 1970b).

We estimated the total concentration of phosphorus species in the gas phase as a saturated vapor over phosphorus species in a condensed phase above the base of the clouds.

The vapor pressure of $P_4O_{10}$ over solid $P_4O_{10}$ as a function of temperature is predicted from equations that were developed to empirically predict that vapor pressure. Three such equations are available; the geometric mean was taken as the value for this work.

$$P_v = \sqrt[3]{\alpha \cdot \beta \cdot \gamma}, \qquad (28)$$

where Pv=vapor pressure and α, β, γ are the results of the predictive equations from:

α from (DIPPR (https://dippr.aiche.org/))

$$LN(V_p) = 79.33 - \frac{12766}{K} - 7.3289 * LN(K) + 1.1 \cdot 10^{-18} \cdot K^6 \text{ (over solid, in pascals)} \qquad (29)$$

β from (DIPPR (https://dippr.aiche.org/))

$$LN(V_p) = -10.768 - \frac{9004.2}{K} - 5.8118 \cdot LN(K) - 2.5 \cdot 10^{-6} \cdot K^2 \text{ (over liquid, in pascals)} \qquad (30)$$

γ from ((Yaws and Knovel 1999))

$$LOG_{10}(P_v) = -55.9316 - \frac{2852.9}{K} + 27.0 \cdot LOG_{10}(K) - 0.029138 \cdot K + 9.47 \cdot 10^{-6} \cdot K^2 \text{ (over solid, in mm Hg) } (31)$$

Some reactions only occur in the clouds, and so the cloud top altitude also affects our calculations. For completeness, we varied cloud bases from 35 km to 60 km, and cloud tops from 40 km to 60 km, with the caveat that a 5 km thick cloud layer was always assumed. In the droplet phase in the clouds, phosphorus species concentration is assumed to be 1 molal. This is an arbitrary amount, chosen purely for convenience, as the concentration of phosphorus species in the cloud droplet phase is unknown. We note that this is likely an overestimation of the concentration of phosphorus, and a conservative approach aimed to make phosphine production more favorable. If phosphorus species are less abundant then the formation of phosphine is much less likely.

*Concentration of sulfuric acid*. It is widely assumed that cloud droplets are composed primarily of sulfuric acid. (Titov *et al.* 2018). Given the water vapor profile presented in the Table S5, the concentration of sulfuric acid can be calculated from the relationship between the vapor pressure of water over sulfuric acid with temperature and concentration of the acid (Greenewalt 1925). Notably, at lower cloud levels this calculation gives concentrations of slightly over 100%, which is consistent with the suggestion that there is $SO_3$ in the lower atmosphere of Venus, likely dissolved in the sulfuric acid droplets (Craig *et al.* 1983) to give 'oleum' (Greenwood and Earnshaw 2012). Given the vapor pressure of gaseous sulfuric acid over liquid sulfuric acid from (Ayers *et al.* 1980), the partial pressure of gaseous sulfuric acid at the base of the clouds can be calculated, and this is assumed to be well mixed throughout the atmosphere below the clouds. This means that the partial pressure of sulfuric acid is dependent on the altitude of the base of the clouds.

*Concentration of gases dissolved in cloud droplets*. For completeness, we assume that all species present in the gas phase are also present in droplets. The solubility of atmospheric gases in cloud and haze droplets is unknown. Of the components of the Venusian atmosphere, only the solubility of carbon dioxide and sulfur dioxide in concentrated sulfuric acid have been studied (Hayduk *et al.* 1988; Markham and Kobe 1941; Zhang *et al.* 1998). However, assuming that the gaseous species are chemically stable in cloud droplets, then their solubility is not important for calculating the thermodynamics of the reaction with species originally in the gas phase. If trace gases in the atmosphere are at equilibrium with solvated species in droplets then, by definition, any energy released by their solution must be compensated by their greater concentration in the liquid phase, so the net free energy of reaction is not affected.

### *1. 3. 2. 3. Choice of Reactions*

To probe the source of phosphine in Venusian atmosphere we have assembled a representative list of all possible chemical reactions involving the species (summarized in Table S5) that could theoretically lead to the reduction of oxidized phosphorus species in Venusian atmosphere, surface, and subsurface and the formation of phosphine.

We identified 75 potential reactions that could involve various oxidized phosphorus species and reducing agents present in the Venusian environment (see Table S6 for the list of all chemical reactions considered). We calculated the Gibbs Free Energy for each reaction, for a total of 256 partial pressure and 15 cloud altitude combinations. A total of 3840 conditions were tested for each of the 75 reactions. Below we describe particulars of both the choice of phosphorus species and their reactions with reductants.

Because phosphorus could be present in gas phase as $P_4O_6$, $P_4O_{10}$ and $H_3PO_4$ (albeit at very different partial pressures), the reactions of all three species were modelled (see Supplementary Section 1.3.2.1 and Section 3.2.1.1 in the main text for details on the dominant phosphorus species on Venus).

Some reactions can only take place in liquid or solid phases. The examples of such reactions include reduction of phosphorus species to phosphite ($H_3PO_3$), reduction by solid phase sulfur in hazes or reduction of solid phosphate minerals.

Reduction of phosphite to phosphine can only occur in liquid phase as phosphites disproportionate before they evaporate. Phosphite could be made in cloud or haze droplets at high altitude, therefore the reactions of reduction of $H_3PO_3$ to phosphine by atmospheric gases or by minerals transported to the atmosphere as dust are only considered in the clouds.

Reduction by solid phase sulfur in hazes in the atmosphere of Venus is considered to take place only in solid (or liquid/melt) phase, i.e. where the activity of sulfur ~1.

Solid mineral phosphates could be reduced by atmospheric gases to phosphine. Such reactions are most likely to occur on the surface. In principle, surface dust could be carried into the atmosphere either by air movement, or volcanic eruptions. While there is no evidence for either process transporting significant mineral mass into Venus' atmosphere, we have considered the reactions of reduction of minerals at all altitudes for completeness.

We divide the reactions into reduction of P(V) species, reduction of P(III) species, and disproportionation of P(III) species, as summarized in the main text and in Table S6.

| **Reduction of P(V) species** |
|---|
| *Reduction of $H_3PO_4$ to $PH_3$* |
| 1) $H_2S + H_3PO_4 \rightarrow PH_3 + H_2SO_4$ |
| 2) $^4/_3H_2S + H_3PO_4 \rightarrow PH_3 + ^4/_3SO_2 + ^4/_3H_2O$ |
| 3) $4H_2 + H_3PO_4 \rightarrow PH_3 + 4H_2O$ |
| 4) $4CO + H_3PO_4 \rightarrow PH_3 + 4CO_2$ |
| 5) $2S + H_3PO_4 \rightarrow PH_3 + 2SO_2$ |
| 6) $^4/_3OCS + H_3PO_4 \rightarrow PH_3 + ^4/_3SO_2 + ^4/_3CO_2$ |
| 7) $^4/_5FeS_2 + H_3PO_4 \rightarrow PH_3 + ^4/_5FeO + ^8/_5SO_2$ |
| 8) $^8/_{11}FeS_2 + H_3PO_4 \rightarrow PH_3 + ^4/_{11}Fe_2O_3 + ^{16}/_{11}SO_2$ |
| 9) $8FeO + H_3PO_4 \rightarrow PH_3 + 4Fe_2O_3$ |
| 10) $8FeCl_2 + 8H_2O + H_3PO_4 \rightarrow PH_3 + 4Fe_2O_3 + 16HCl$ |
| *Reduction of $P_4O_{10}$ to $PH_3$* |
| 11) $H_2S + 1½H_2O + ¼P_4O_{10} \rightarrow PH_3 + H_2SO_4$ |
| 12) $^4/_3H_2S + ¼P_4O_{10} + ^1/_6H_2O \rightarrow PH_3 + ^4/_3SO_2$ |
| 13) $4H_2 + ¼P_4O_{10} \rightarrow PH_3 + 2½H_2O$ |
| 14) $4CO + 1½H_2O + ¼P_4O_{10} \rightarrow PH_3 + 4CO_2$ |
| 15) $2S + 1½H_2O + ¼P_4O_{10} \rightarrow PH_3 + 2SO_2$ |
| 16) $^4/_3OCS + 1½H_2O + ¼P_4O_{10} \rightarrow PH_3 + ^4/_3SO_2 + ^4/_3CO_2$ |
| 17) $^4/_5FeS_2 + 1½H_2O + ¼P_4O_{10} \rightarrow PH_3 + ^4/_5FeO + ^8/_5SO_2$ |
| 18) $^8/_{11}FeS_2 + 1½H_2O + ¼P_4O_{10} \rightarrow PH_3 + ^4/_{11}Fe_2O_3 + ^{16}/_{11}SO_2$ |
| 19) $8FeO + 1½H_2O + ¼P_4O_{10} \rightarrow PH_3 + 4Fe_2O_3$ |
| 20) $8FeCl_2 + 9½H_2O + ¼P_4O_{10} \rightarrow PH_3 + 4Fe_3O_4 + 16HCl$ |
| **Reactions with sulfur haze** |
| 21) $2S_{(s)} + ^1/_4P_4O_{10} + 1½H_2O \rightarrow PH_3 + 2SO_2$ [solid sulfur in haze particles] |
| **Reduction of phosphate minerals at the surface of the planet or as dust in atmosphere** |
| *Whitlockite ( $Ca_3(PO_4)_2$ )* |
| 22) $1½H_2S + ½Ca_3(PO_4)_2 \rightarrow PH_3 + ½CaSO_4 + CaO + ½S + ½SO_2$ |
| 23) $4H_2S + ½ Ca_3(PO_4)_2 \rightarrow PH_3 + ^3/_2CaO + 2½H_2O + 4S$ |
| 24) $4H_2 + ½ Ca_3(PO_4)_2 \rightarrow PH_3 + ^3/_2CaO + 2½H_2O$ |
| 25) $4CO + ½ Ca_3(PO_4)_2 + 1½ H_2O \rightarrow PH_3 + 1½ CaO + 4CO_2$ |
| 26) $1½S + ½Ca_3(PO_4)_2 + 1½H_2O \rightarrow PH_3 + CaSO_4 + ½CaO + ½SO_2$ |
| 27) $OCS + ½Ca_3(PO_4)_2 + 1½H_2O \rightarrow PH_3 + CaSO_4 + CO_2 + ½CaO$ |
| *Fluorapatite ( $Ca_5(PO_4)_3F$ )* |
| 28) $1½H_2S + ^1/_3Ca_5(PO_4)_3F \rightarrow PH_3 + ^1/_6CaF_2 + ^1/_3CaSO_4 + 1^1/_6CaO + ^3/_4SO_2 + ^5/_{12}S$ |
| 29) $4H_2 + ^1/_3Ca_5(PO_4)_3F \rightarrow PH_3 + ^1/_6CaF_2 + 1½CaO + 2½H_2O$ |
| 30) $4CO + 1½H_2O + ^1/_3Ca_5(PO_4)_3F \rightarrow PH_3 + ^1/_6CaF_2 + 1½CaO + 4CO_2$ |
| 31) $2S + 1½H_2O + ^1/_3Ca_5(PO_4)_3F \rightarrow PH_3 + ^1/_6CaF_2 + 1½CaO + 2SO_2$ |
| 32) $1^1/_3OCS + 1½H_2O + ^1/_3Ca_5(PO_4)_3F \rightarrow PH_3 + ^1/_6CaF_2 + 1½CaO + 1^1/_3CO_2 + 1^1/_3SO_2$ |
| *Magnesium phosphate ( $Mg_3(PO_4)_2$ )* |
| 33) $1½H_2S + ½Mg_3(PO_4)_2 \rightarrow PH_3 + ½MgSO_4 + MgO + ½S + ½SO_2$ |
| 34) $4H_2S + ½ Mg_3(PO_4)_2 \rightarrow PH_3 + ^3/_2MgO + 2½H_2O + 4S$ |
| 35) $4H_2 + ½ Mg_3(PO_4)_2 \rightarrow PH_3 + ^3/_2MgO + 2½H_2O$ |
| 36) $4CO + ½ Mg_3(PO_4)_2 + 1½ H_2O \rightarrow PH_3 + 1½ MgO + 4CO_2$ |
| 37) $1½ S + ½Mg_3(PO_4)_2 + 1½H_2O \rightarrow PH_3 + MgSO_4 + ½MgO + ½SO_2$ |
| 38) $OCS + ½Mg_3(PO_4)_2 + 1½H_2O \rightarrow PH_3 + MgSO_4 + CO_2 + ½MgO$ |
| *Potassium phosphate ( $K_3PO_4$ )\** |
| 39) $K_3PO_4 + 4H_2S \rightarrow PH_3 + 1½K_2O + 2½H_2O + 4S$ |
| 40) $K_3PO_4 + 2H_2S \rightarrow PH_3 + 1½K_2O + ½H_2O + S + SO_2$ |
| 41) $K_3PO_4 + 4H_2 \rightarrow PH_3 + 1½K_2O + 2½H_2O$ |
| 42) $K_3PO_4 + 4CO + 1½H_2O \rightarrow PH_3 + 1½K_2O + 4CO_2$ |
| 43) $K_3PO_4 + 2S + 1½H_2O \rightarrow PH_3 + 1½K_2O + 2SO_2$ |
| 44) $K_3PO_4 + ^4/_3OCS + 1½H_2O \rightarrow PH_3 + 1½K_2O + ^4/_3SO_2 + ^4/_3CO_2$ |

**Reduction of P(III) species**

| |
|---|
| *Reduction of $P_4O_6$ to $PH_3$* |
| 45) $¾H_2S + 1½H_2O + ¼P_4O_6 \rightarrow PH_3 + ¾H_2SO_4$ |
| 46) $H_2S + ¼P_4O_6 + ½H_2O \rightarrow PH_3 + SO_2$ |
| 47) $3H_2 + ¼P_4O_6 \rightarrow PH_3 + 1½H_2O$ |
| 48) $3CO + 1\frac{1}{2}H_2O + \frac{1}{4}P_4O_6 \rightarrow PH_3 + 3CO_2$ |
| 49) $1\frac{1}{2}S + 1\frac{1}{2}H_2O + \frac{1}{4}P_4O_6 \rightarrow PH_3 + 1\frac{1}{2}SO_2$ |
| 50) $OCS + 1\frac{1}{2}H_2O + \frac{1}{4}P_4O_6 \rightarrow PH_3 + SO_2 + CO_2$ |
| 51) $^3/_5FeS_2 + 1\frac{1}{2}H_2O + \frac{1}{4}P_4O_6 \rightarrow PH_3 + {}^3/_5FeO + {}^6/_5SO_2$ |
| 52) $^6/_{11}FeS_2 + 1\frac{1}{2}H_2O + \frac{1}{4}P_4O_6 \rightarrow PH_3 + {}^3/_{11}Fe_2O_3 + {}^{12}/_{11}SO_2$ |
| 53) $6FeO + 1\frac{1}{2}H_2O + \frac{1}{4}P_4O_6 \rightarrow PH_3 + 3Fe_2O_3$ |
| 54) $6FeCl_2 + 7\frac{1}{2}H_2O + \frac{1}{4}P_4O_6 \rightarrow PH_3 + 3Fe_2O_3 + 12HCl$ |
| 55) *Reduction of $H_3PO_3$ (in droplets) to $PH_3$* |
| *56)* $\frac{3}{4}H_2S + H_3PO_3 \rightarrow PH_3 + \frac{3}{4}H_2SO_4$ |
| 57) $H_2S + H_3PO_3 \rightarrow PH_3 + SO_2 + H_2O$ |
| 58) $3H_2 + H_3PO_3 \rightarrow PH_3 + 3H_2O$ |
| 59) $3CO + H_3PO_3 \rightarrow PH_3 + 3CO_2$ |
| 60) $1\frac{1}{2}S + H_3PO_3 \rightarrow PH_3 + 1\frac{1}{2}SO_2$ |
| 61) $OCS + H_3PO_3 \rightarrow PH_3 + SO_2 + CO_2$ |
| 62) $^3/_5FeS_2 + H_3PO_3 \rightarrow PH_3 + {}^3/_5FeO + {}^6/_5SO_2$ |
| 63) $^8/_{11}FeS_2 + H_3PO_4 \rightarrow PH_3 + {}^4/_{11}Fe_2O_3 + {}^{16}/_{11}SO_2$ |
| 64) $6FeO + H_3PO_3 \rightarrow PH_3 + 3Fe_2O_3$ |
| 65) $6FeCl_2 + 6H_2O + H_3PO_3 \rightarrow PH_3 + 3Fe_2O_3 + 12HCl$ |

**Disproportionation of P(III) species**

| |
|---|
| *Disproportionation of $H_3PO_3$ (in droplets)* |
| 66) $4H_3PO_3 \rightarrow PH_3 + 3H_3PO_4$ |
| $4H_3PO_3 \rightarrow PH_3 + \frac{3}{4}P_4O_{10} + 3H_2O$ |
| 67) *Disproportionation of $P_4O_6$ (via notional $H_3PO_3$) in gas phase* |
| 68) $P_4O_6 + 1\frac{1}{2}H_2O \rightarrow PH_3 + \frac{3}{4}P_4O_{10}$ |
| 69) $P_4O_6 + 6H_2O \rightarrow 4H_3PO_3 \rightarrow 3H_3PO_4 + PH_3$ |
| 70) $P_4O_6 + 3HCl + 3H_2O \rightarrow 3H_3PO_3 + PCl_3 \rightarrow 2\frac{1}{4} H_3PO_4 + \frac{3}{4} PH_3 + PCl_3$ |
| 71) $P_4O_6 + 6HCl \rightarrow 2H_3PO_3 + 2PCl_3 \rightarrow 1\frac{1}{2} H_3PO_4 + \frac{1}{2} PH_3 + 2PCl_3$ |
| 72) $P_4O_6 + 9HCl \rightarrow H_3PO_3 + 3PCl_3 + 3H_2O \rightarrow \frac{3}{4}H_3PO_4 + \frac{1}{4}PH_3 + 3PCl_3 + 3H_2O$ |
| 73) $P_4O_6 + 9HCl \rightarrow 3POCl_3 + PH_3 + 3H_2O$ |
| 74) $P_4O_6 + 3H_2S \rightarrow 2\ H_3PO_3 + \frac{1}{2}P_4S_3 + 1\frac{1}{2}S \rightarrow \frac{1}{2}P_4S_3 + 1\frac{1}{2}S + \frac{1}{2}PH_3 + 1\frac{1}{2}H_3PO_4$ |

**Excluded reactions**

| |
|---|
| 75) $^4/_5N_2 + {}^{23}/_{10}H_2O + \frac{1}{4}P_4O_{10} \rightarrow PH_3 + {}^8/_5HNO_3$ |
| 76) $HCl + \frac{1}{4}P_4O_{10} + 1\frac{1}{2}H_2O \rightarrow PH_3 + HClO_4$ |

**Table S6.** A list of reactions used to calculate the thermodynamics 'heat maps' (Figure 6 and Figure 7 in the main text) in this paper, for potential $PH_3$ production pathways. *Aluminum sulfate decomposes at Venus surface temperatures (Truex *et al.* 1977) and so was not considered as a product. Two reactions were considered but excluded because initial evaluation suggested that their free energy would always be >200 kJ mol$^{-1}$.

### 1. 3. 3. Calculation of Subsurface Thermodynamics of Phosphine Production

Oxygen fugacity (fO$_2$) is a geochemically relevant, quantitative method to calculate the redox state of a mineral, and hence whether that mineral could drive a redox reaction such as the reduction of phosphate to phosphine (Frost 1991). Fugacities are often referred to by reference to standard 'buffers'. Like the more familiar pH buffer, which provides a stable reference for the concentration of hydrogen ions in solution, an fO$_2$ buffer provides a stable reference for the chemical activity of molecular oxygen in a rock system, and hence how reduced or oxidized that system is.

For example, the Quartz-Iron-Fayalite (QIF) buffer is based on a mixture of iron, silicon dioxide and iron(II) silicate. The buffer uses the following reaction to buffer O$_2$:

Fe$_2$SiO$_4$ ↔ 2Fe + SiO$_2$ + O$_2$

QIF buffer's maximum buffering capacity is when:

$$\frac{\{Fe\}^2 \cdot \{SiO_2\}}{\{Fe_2SiO_4\}} = 1 \qquad (32)$$

At equilibrium, at this maximum buffering point, the log of oxygen fugacity is directly related to the Gibbs free energy for the reaction described above, as shown by the following formula:

$$\Delta G = 0 = \Delta G^0 + R \cdot T. ln\left[\frac{\{Fe\}^2 \cdot \{SiO_2\} \cdot \{O_2\}}{\{Fe_2SiO_4\}}\right] \Rightarrow \Delta G^0 = -R \cdot T. ln[\{O_2\}] \qquad (33)$$

Where ΔG is the free energy of reaction and is by definition 0 when the reaction is at equilibrium, and other symbols have meanings given previously. Oxygen fugacity is usually expressed on a log scale, and the more negative it is, the more reducing the rock is. See (Frost 1991) for more detail on the measurement, calculation and application of mineral oxygen fugacity buffers.

A number of standard fO$_2$ buffers are used in geology as references for the redox states of rock. As iron is the major redox-active metal in the crust by mass, most use the redox states of iron. The four standard fO$_2$ buffers used as exemplars in this study are shown in Table S7.

| Abbreviation | Name | Reaction | Buffered species |
|---|---|---|---|
| QIF | Quartz-Iron-Fayalite | FeSiO$_4$ <-> Fe + SiO$_2$ + O$_2$ | Fe(0) / Fe(+2) |
| IW | Iron-Wustite | 2Fe$_x$O <-> xFe + O$_2$ | Fe(0) / Fe(+1 – +1.9) |
| FMQ | Fayalite-magnetite-quartz | 2Fe$_3$O$_4$ + 3SiO$_2$ <-> 3Fe$_2$SiO$_4$ + O$_2$ | Fe(+2) / Fe(+2$^1$/$_3$) |
| MH | Magnetite-hematite | 6Fe$_2$O$_3$ <-> 4Fe$_3$O$_4$ + O$_2$ | Fe(+2$^1$/$_3$) / Fe(+3) |

**Table S7.** Four standard fO$_2$ buffers used in geology as references for the redox states of rock.

We can compare these standard fO$_2$ buffers with the redox state under which phosphorus present in crustal rocks could be reduced to elemental phosphorus (discussed in Supplementary Section 2.5.3.3.) or to phosphine. Two reactions were modelled to plot the reduction of phosphorus on an oxygen fugacity scale as shown in Table S8, together with three to model the balance between H₂S and SO₂ in the rocks.

| Process | Reaction |
|---|---|
| Reduction of P(+5) to phosphine | Mg₃(PO₄)₂ + 1½SiO₂ + 3H₂O → 1½Mg₂SiO₄ + 2PH₃ + 4O₂ |
| Production of elemental phosphorus | Mg₃(PO₄)₂ + 1½SiO₂ → 1½Mg₂SiO₄ + ½P₄ + 2½O₂ |
| Production of H₂S from sulfate | 3MgSO₄ + 1½SiO₂ + 3H₂O → 1½Mg₂SiO₄ + 3H₂S + 6O₂ |
| Production of SO₂ from sulfate | 3MgSO₄ + 1½SiO₂ → 1½Mg₂SiO₄ + 3SO₂ + 1½O₂ |
| Production of H₂S from SO₂ | SO₂ + H₂O → H₂S + 1½O₂ |

**Table S8**. Reactions modelled to plot the reduction of phosphorus in the oxygen fugacity scale. Sulfur is assumed to be present in rocks as magnesium sulfate. The ratio of $H_2S:SO_2$ is calculated from the ratio of the energy of production of $H_2S$ from magnesium sulfate rock compared to the energy of production of $SO_2$ from the same rock.

Following from the equation above (eq. (33)), the oxygen fugacity needed to allow 50% of the phosphorus in a rock to be present as phosphine is given by

$$\Delta G = 0 = \Delta G^0 + RT\ln(Q) \qquad (34)$$

and therefore, for production of $PH_3$, for example (following the reaction shown in Table S8),

$$\Delta G^0 + R \cdot T \cdot \ln\left(\frac{\{Mg_2SiO_4\}^{1.5} \cdot \{PH_3\}^2 \cdot \{O_2\}^4}{\{Mg_3(PO_4)_2\} \cdot \{SiO_2\}^{1.5} \cdot \{H_2O\}^3}\right) = 0 \qquad (35)$$

If we assume $\{SiO_2\}$ is ~ $\{Mg_2SiO_4\}$ (i.e. to within a factor of two or three, the amount of magnesium and the amount of silicon in the rocks is the same), and as we have defined this to be at the reaction half-point, so that by definition $\{PH_3\}^2 = \{Mg_3(PO_4)_2\}$, then

$$\Delta G^0 = -R \cdot T \cdot 2.5 \cdot [\ln(\{O_2\})] \qquad (36)$$

and hence

$$\log(\{O_2\}) = \frac{-\frac{\Delta G^0}{R \cdot T}}{2.5} \cdot 0.4343 \qquad (37)$$

for the reaction shown in Table S8. Because $fO_2$ values typically span tens of orders of magnitudes, they are usually plotted on a $\log_{10}$ scale, hence multiplying the natural log value by 0.4343. $fO_2$ is sensitive to temperature, but relatively insensitive to pressure.

We have also included the $SO_2/H_2S$ couple in our fugacity calculations to validate that the calculation method gave results consistent with real geochemistry that was identified on Earth.

## 2. Supplementary Results

### 2. 1. Example of Thermodynamics Calculation

We provide a detailed exposition of one reaction thermodynamics, as a worked example of the method. For simplicity, we chose a reaction with relatively few components:

$4H_2 + ¼P_4O_{10} \rightarrow PH_3 + 2½H_2O$

At an altitude of 20 km. The conditions at 20 km in our model are:

Pressure = p = 22.5 bar
Temperature = t = 579 K

The standard free energy of formation ($\Delta G^0$) of $H_2$ is 0 by definition (see Table S3 for sources for the values of the standard free energy ($\Delta G^0$) as a function of temperature). The standard free energy of formation of the other reagents at 579 K is:

$\Delta G°$ ($P_4O_{10(g)}$) = d = -2442.79 kJ/mol
$\Delta G°$ ($PH_{3(g)}$) = e = 10.22 kJ/mol
$\Delta G°$ ($H_2O_{(g)}$) = f = -215.026 kJ/mol

and therefore the free energy of the reaction at 579 K is

$\Delta G°$ ($4H_2 + ¼P_4O_{10} \rightarrow PH_3 + 2½H_2O$) = R = (e+2.5f)-(0.25d) = 83.36 kJ/mol

The mixing ratio (m.r.) of the gases we will take as their maximum values:

m.r.($H_2$) = g = 1.30E-07
m.r.($PH_3$) = h = 1.00E-09
m.r.($H_2O$) = i = 8.00E-03

The partial pressures (p.p.) of the reagents are therefore (following the data shown in Table S5):

p.p.($H_2$) = g*p = 2.93E-06
p.p.($PH_3$) = h*p = 2.25E-08
p.p.($H_2O$) = i*p = 1.80E-1

Because 20 km is below the base of the clouds, we take the partial pressure of $P_4O_{10}$ to be the partial pressure at the base of the clouds, which is determined by the vapour pressure at the base of the clouds, which for this calculation we assume is at 40 km. Here

p.p.($P_4O_{10}$) = k = 2.34E-9

This is corrected to activity (a) as described in Supplementary Section 1.3.1 (the full calculation is not given here, and for almost all conditions correcting for non-ideality of the gas makes little difference to the result, as illustrated by the small difference between activity (below) and partial pressure (above).

a($H_2$) = u = 2.95E-6
a($PH_3$) = v = 2.236E-08
a($H_2O$) = w = 1.71E-1
a($P_4O_{10}$) = x = 2.34E-9

We can then calculate LN(Q)

$$LN(Q) = LN\left(\frac{v \cdot w^{2.5}}{u^4 \cdot x^{0.25}}\right) = 33.872$$

and so the overall free energy of reaction

ΔG = ΔG° + RTLN(Q) = 83.36+0.008314*579*33.872 = 246.123 (note that, as energy units are kJ/mol, the gas constant R is in kJ/mol/K)

This positive free energy shows that this reaction is not at equilibrium with $H_2$, $P_4O_{10}$ and $H_2O$ at Venus atmosphere conditions with phosphine present at 1 ppb, and equilibrium will lie is the reverse direction of the reaction as stated, i.e. <$PH_3$, <$H_2O$ or both

## 2. 2. Individual Reaction Results

Below we show the individual curves of free energy of reaction as a function of altitude for the reactions listed in Table S6. For each reaction a maximum and minimum free energy is calculated for each altitude (different free energies result from different assumptions about the gas concentrations in the atmosphere, as discussed above); the overall maximum and minimum values for each set of reactions is show by the two dashed lines on each graph (Figure S9 and Figure S10, Figure S11, Figure S12).

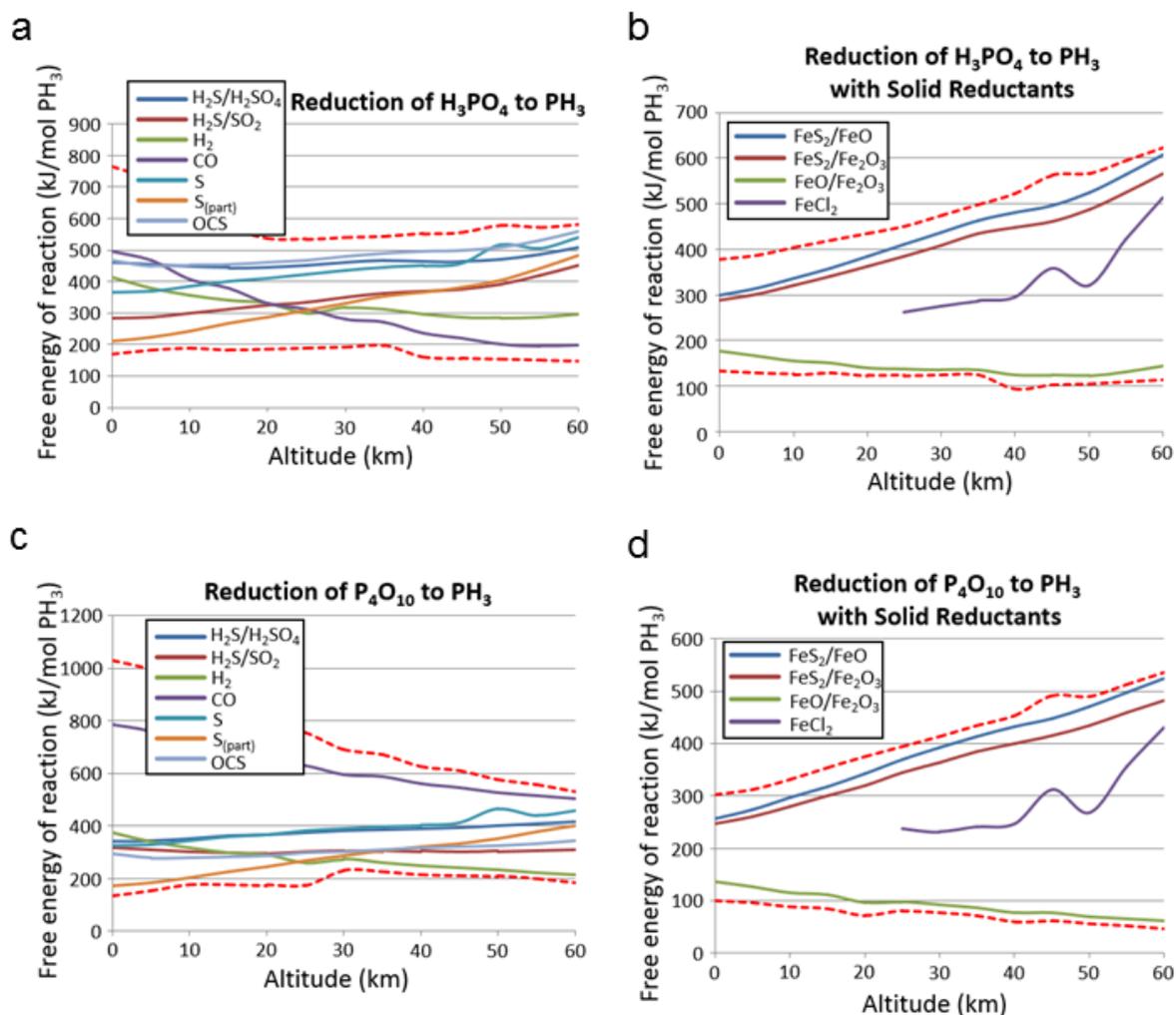

**Fig. S9.** Thermodynamics of phosphine production, by reduction of P(V) species, in the Venusian atmosphere-surface environment. x axis: altitude (km), y axis: Gibbs free energy of reaction (ΔG) (kJ/mol). Dashed lines show the limits of the free energy found for any combination of gas partial pressures, for any altitude, for any reaction in a set of reactions. Each solid line represents a different reductant, and in the case of $H_2S$ as a reductant, a different oxidized product. 'S' is elemental sulfur in gas phase, '$S_{(part.)}$' is elemental sulfur in solid (particle) phase. (a) Free energy of reduction of orthophosphoric acid by gaseous reductants under Venus atmosphere conditions. (b) Free energy of reduction of orthophosphoric acid by mineral reductants under Venus atmosphere conditions. Note that the line for $FeCl_2$ only covers altitudes from 35 km upwards. Below 35 km $FeCl_2$ is unstable to hydrolysis to HCl and FeO under Venus atmosphere conditions in nearly all scenarios (Figure S8). Calculations are done for altitudes up to 60 km because, in principle, minerals could be carried to the cloud tops as dust. (c) Reduction of $P_4O_{10}$ to $PH_3$ by atmospheric components. (d) Reduction of $P_4O_{10}$ to $PH_3$ by mineral / dust. In conclusion (a-d), the formation of phosphine in the Venusian atmosphere-surface environment cannot proceed spontaneously (i.e. none of the conditions considered result in a negative free energy).

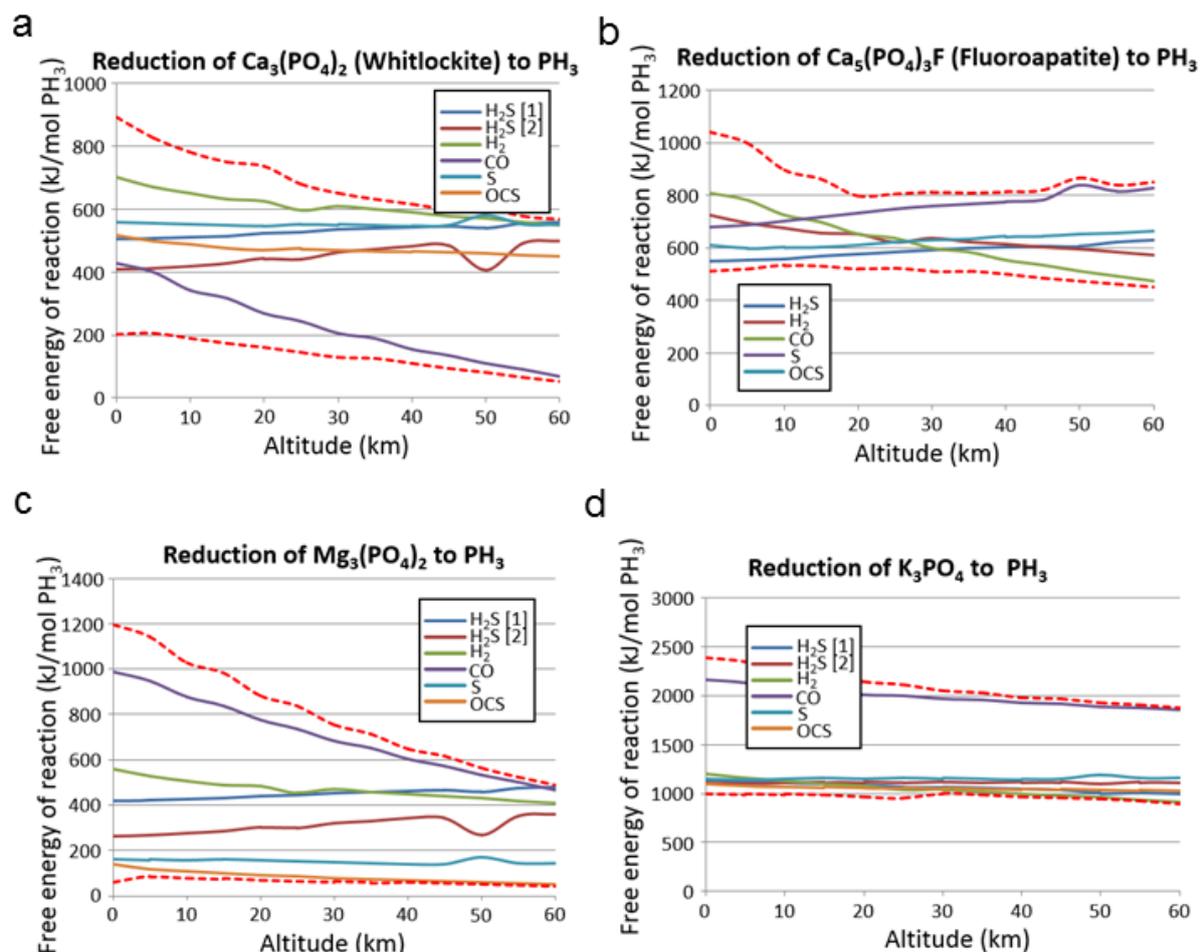

**Fig. S10.** Thermodynamics of the reduction of mineral phosphates by atmospheric gases. x axis: altitude (km), y axis: Gibbs free energy of reaction (ΔG) (kJ/mol). Dashed lines show the limits of the free energy found for any combination of gas partial pressures, for any altitude, for any reaction in a set of reactions. Each solid line represents a different reductant. (a) Reduction of calcium phosphate ($(Ca_3(PO_4)_2)$; whitlockite) by Venusian trace atmospheric gases. Reactions are calculated to 60 km altitude to cover the possibility that dust could be carried into the cloud layer. $H_2S$ [1]: reaction $1½H_2S + ½Ca_3(PO_4)_2 \rightarrow PH_3 + ½CaSO_4 + CaO + ½S + ½SO_2$ ; $H_2S$ [2]: reaction $4H_2S + ½ Ca_3(PO_4)_2 \rightarrow PH_3 + ^3/_2CaO + 2½H_2O + 4S$ (b) Reduction of calcium fluorophosphate (fluorapatite) to $PH_3$ by atmospheric gases. (c) Reduction of magnesium phosphate ($Mg_3(PO_4)_2$) to phosphine by atmospheric gases. (d) Reduction of potassium phosphate ($K_3PO_4$) to phosphine by atmospheric gases. In conclusion (a-d), formation of phosphine by reduction of surface mineral phosphates in the Venusian atmosphere-surface environment cannot proceed spontaneously (i.e. none of the conditions considered result in a negative free energy).

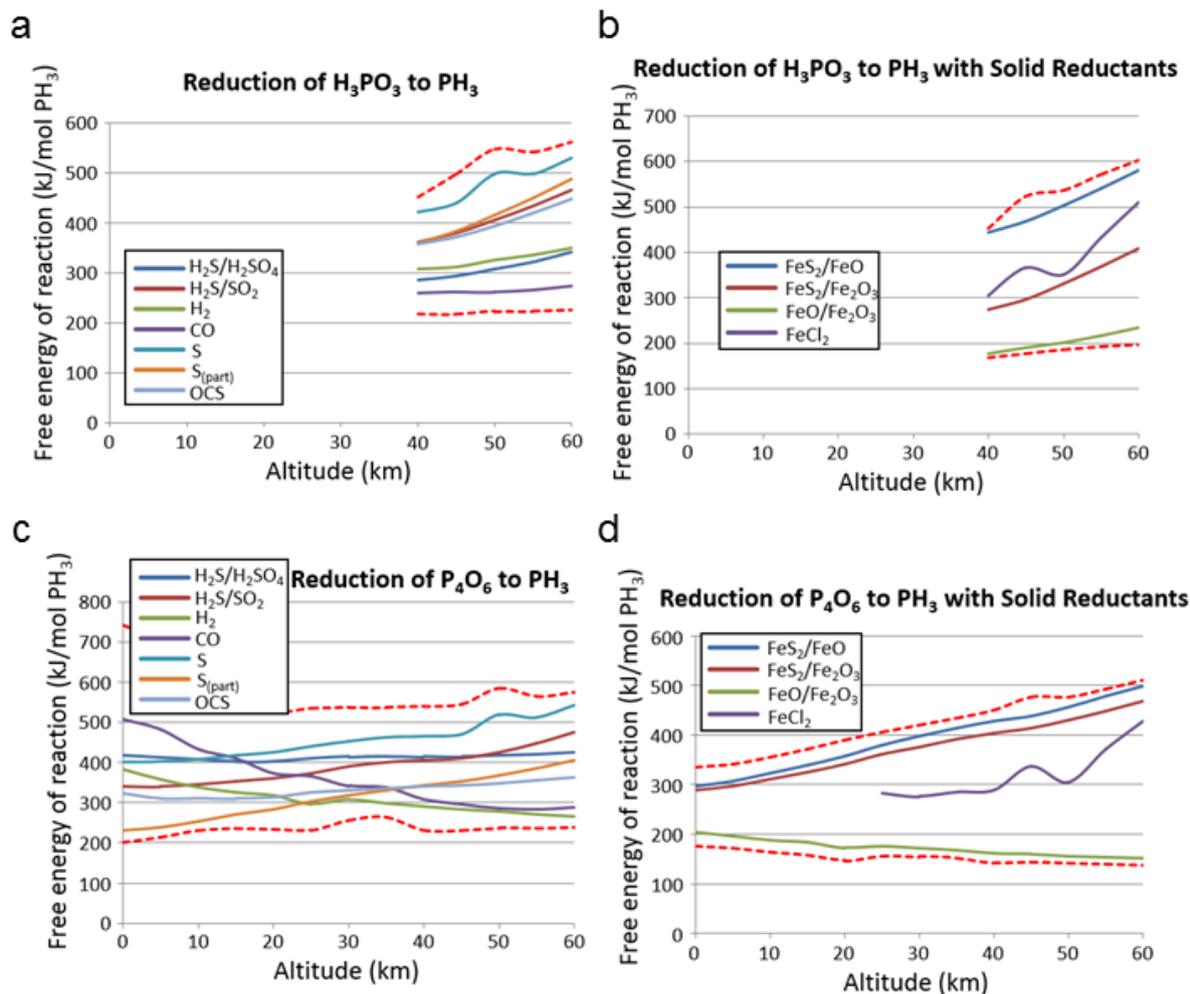

**Fig. S11.** Thermodynamics of phosphine production, by reduction of P(III) species, in the Venusian atmosphere-surface environment. x axis: altitude (km), y axis: Gibbs free energy of reaction (ΔG) (kJ/mol). Dashed lines show the limits of the free energy found for any combination of gas partial pressures, for any altitude, for any reaction in a set of reactions. Each solid line represents a different reductant, and in the case of $H_2S$ as a reductant, a different oxidized product. 'S' is elemental sulfur in gas phase, '$S_{(part.)}$' is elemental sulfur in solid (particle) phase. (a) Reduction of $H_3PO_3$ to phosphine by atmospheric reductants. Note that $H_3PO_3$ cannot exist in gas phase, outside a liquid droplet under Venus' clouds temperatures, and so these calculations are only performed for altitudes at which cloud droplets could exist. (b) Reduction of $H_3PO_3$ to $PH_3$ by mineral reductants. (c) Reduction of $P_4O_6$ by atmospheric components. (d) Reduction of $P_4O_6$ by mineral / dust. In conclusion (a-d), the formation of phosphine in the Venusian atmosphere-surface environment cannot proceed spontaneously (i.e. none of the conditions considered result in a negative free energy).

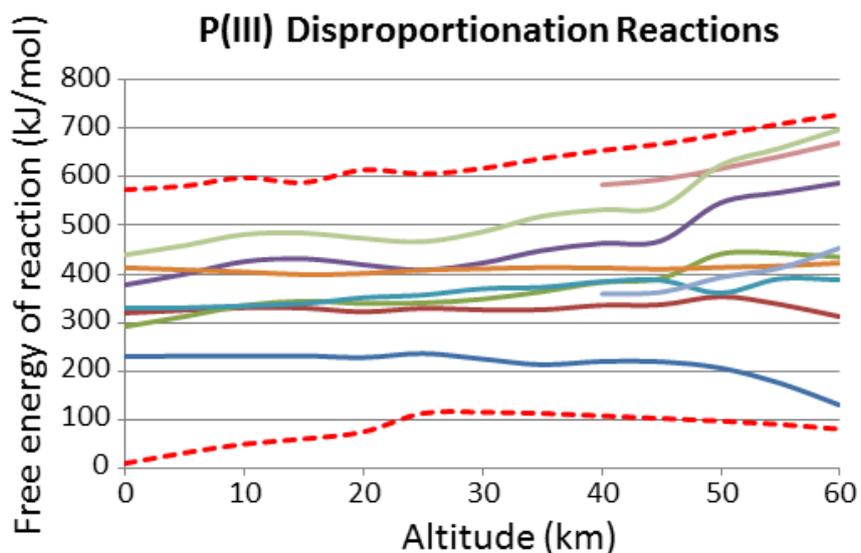
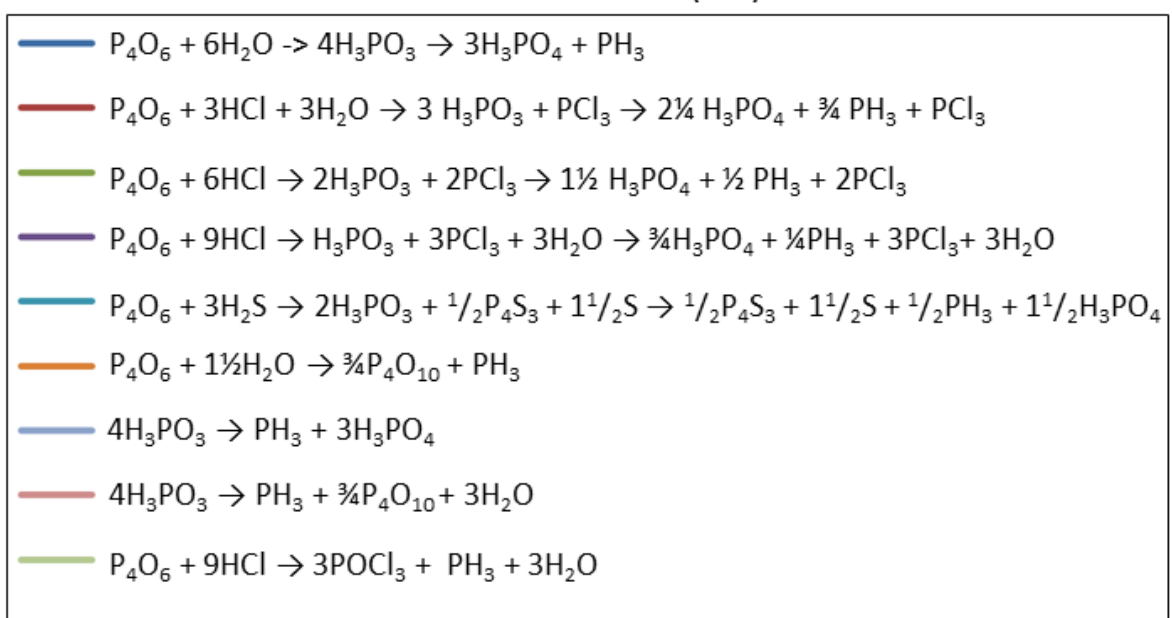

**Fig. S12.** Thermodynamics of phosphine production, by disproportionation of P(III) species, in the Venusian atmosphere-surface environment. x axis: altitude (km), y axis: Gibbs free energy of reaction (ΔG) (kJ/mol). Dashed lines show the limits of the free energy found for any combination of gas partial pressures, for any altitude, for any reaction in a set of reactions. Each solid line (*top panel*) represents a different reaction (*bottom panel*).

### 2. 3. Sensitivity Analysis to Variations of Venus Atmospheric Gas Concentrations

To test the sensitivity of our results to the assumptions about gas concentrations, we asked how much each gas concentration listed in Table S5 would have to be changed for *any* of the reactions listed in Table S6 to be exergonic for phosphine production at *any* altitude. The results are shown in Figure S13.

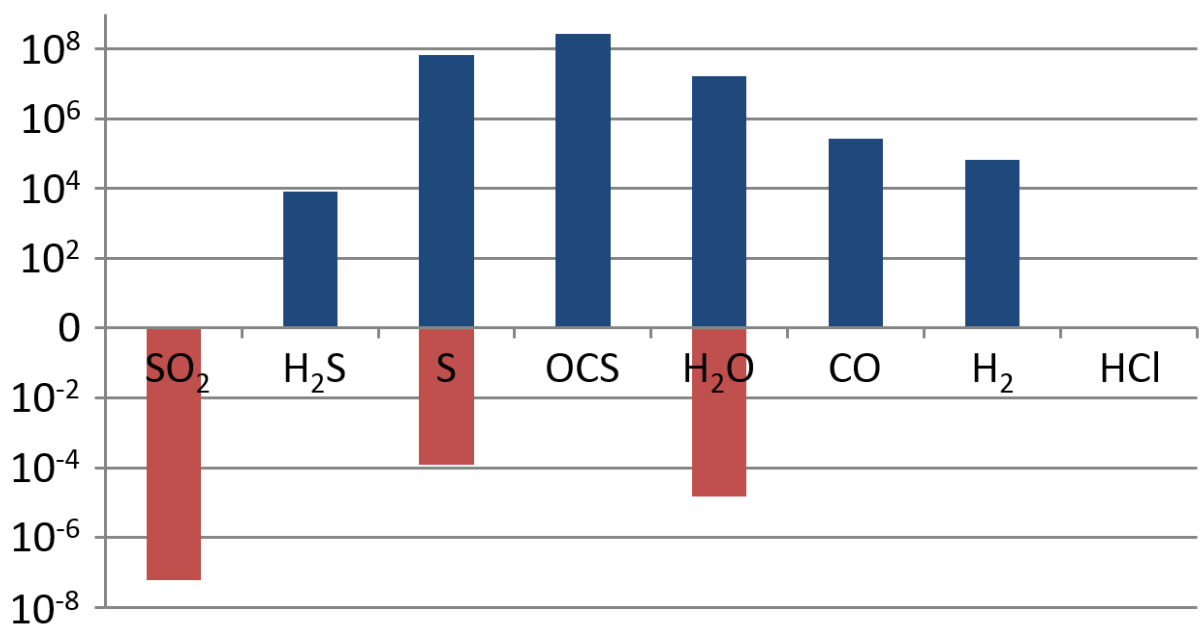

**Fig. S13.** Fractional change in partial pressure of trace gases needed to produce a negative ΔG value for phosphine production through any chemistry at any altitude. x axis: gaseous reductant. y axis: fold increase (y>1) or decrease (y<1) in partial pressure needed for thermodynamically favorable conditions for phosphine production. Gases were set to the geometric mean of the maximum and minimum values in Table S5, and then each gas was increased in steps to a maximum of $10^9$ of its mean value, or decreased to a minimum of $10^{-9}$ of its mean value. Bars represent the smallest change that would give a negative ΔG for phosphine production at *any* altitude using *any* reaction: bars above y=1 imply an increase in partial pressure is favorable, bars below y=1 imply a decrease in partial pressure is favorable. If there is no bar above the axis, then no increase in gas concentration can drive phosphine production. Similarly, no bar below the axis implies that no reduction in gas can drive phosphine production. $CO_2$ was not varied, as its partial pressure is well known to within a few percent. In summary, the estimations of gas concentrations would have to be incorrect by more than four orders of magnitude for our conclusions to change, i.e. for the formation of phosphine to be exergonic and likely to occur spontaneously.

For example, no tested change in HCl partial pressure resulted in phosphine production. Only $H_2S$ and CO have values which suggest that very substantial systematic errors in measurements or modelling could account for the production of phosphine. If the maximum concentration of $H_2S$ was ~$10^4$-fold higher than the highest level reported in the literature, or that of CO was $3.10^5$-fold higher, then under some conditions they could drive phosphine production. Such an unlikely scenario would be equivalent to 0.1% $H_2S$ or 1% CO in Venus' atmosphere. All other gases would require physically unrealistic changes in their partial pressures to drive phosphine production. For example, a reduction of $SO_2$ by a factor of $5.8 \times 10^{-8}$, necessary to allow phosphine production, implies a partial pressure of $10^{-13}$, which is at least 4 orders of magnitude below the detection limits of the instruments that have detected $SO_2$ on Venus.

### 2. 4. Validation of the Fugacity Calculations

#### 2. 4. 1. $H_2S/SO_2$ as a Qualitative Validation of the Fugacity Calculations

As a qualitative validation of the fugacity calculation we calculate the fugacity of the terrestrial $H_2S/SO_2$ equilibrium (Figure S14 - blue line). For example, at 1000 K (the temperature of the vertical black line, on Figure S14), in a rock with the oxygen fugacity of QIF, at low temperatures, sulfur will predominantly be reduced (yellow QIF line is *below* blue $H_2S/SO_2$ line), whereas at high temperatures sulfur will predominantly be oxidized (yellow QIF line is *above* blue $H_2S/SO_2$ line).

The results from the $SO_2/H_2S$ line are qualitatively consistent with field observations on Earth and modelling on Mars. Specifically, Terrestrial and Martian mantle rocks typically have $fO_2$ values between FMQ-4 and FMQ+3 (Ballhaus *et al.* 1990), a region shaded in grey on the graphs on Figure S14 and Figure S15. The $SO_2/H_2S$ $fO_2$ curve falls largely within this zone, and indeed terrestrial volcanoes can emit $SO_2$, $H_2S$ or a mixture from primary degassing. Consistent with this observation, gases evolved from rocks at higher temperature, or rocks containing less water or with higher oxygen fugacity (smaller negative log number) have a lower $H_2S/SO_2$ ratio (Gerlach 1982; Hoshyaripour *et al.* 2012; Whitney 1984) on Earth and on Mars (Gaillard and Scaillet 2009).

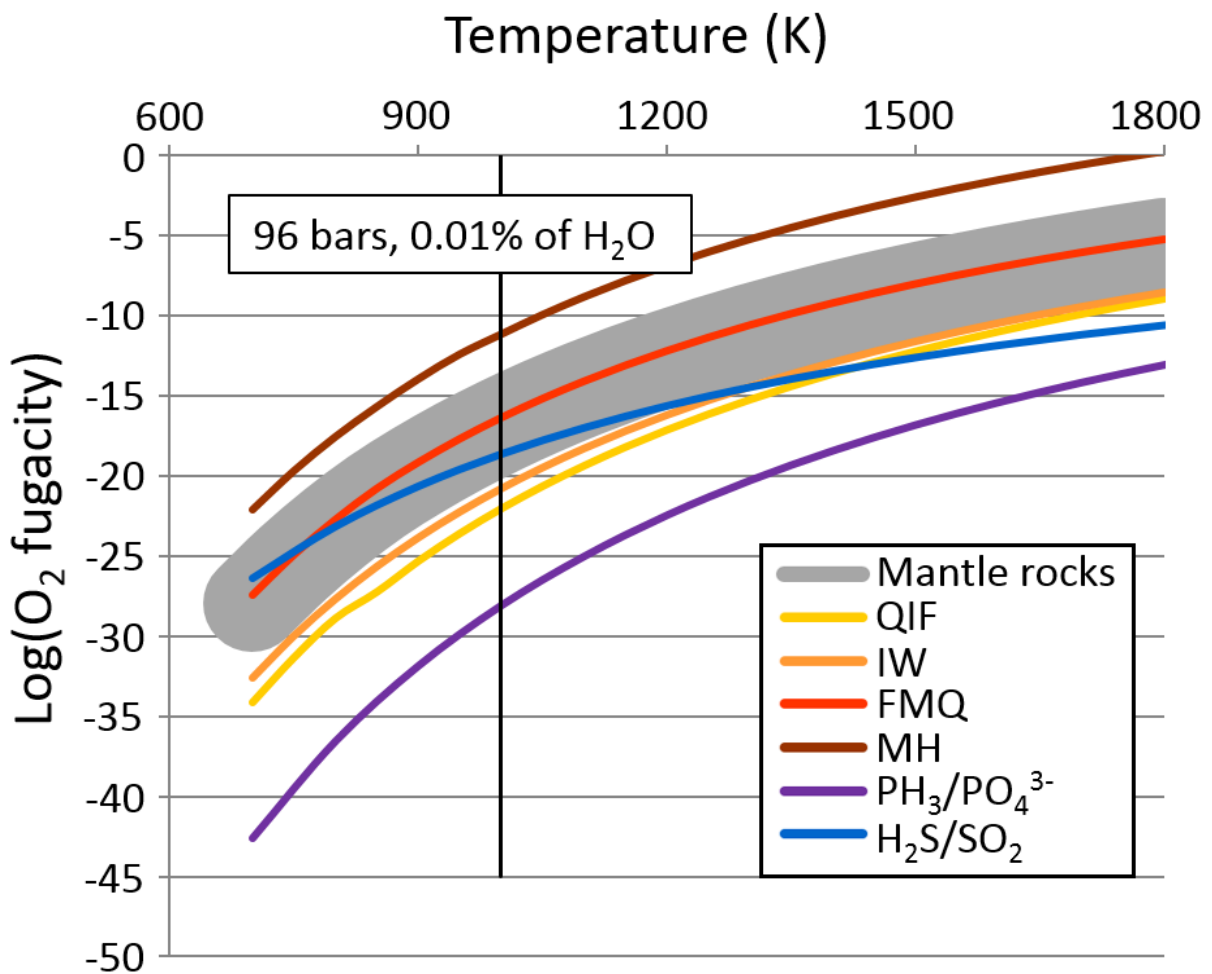

**Fig. S14.** Comparison of the fugacity of the phosphate/phosphine equilibrium to the fugacity of the standard mineral buffers of terrestrial rocks and the fugacity of the terrestrial $H_2S/SO_2$ equilibrium (blue line). x axis: log $O_2$ fugacity, y axis: Temperature (K). Fugacity of the production of phosphine from phosphate minerals is calculated for 96 bars and 0.01% water in the rocks. The fugacity of the phosphate/phosphine equilibrium is shown as a purple line. The other curves are $O_2$ fugacities of standard rock buffers. The phosphate/phosphine $fO_2$ curve lies below the QIF buffer line (the most reduced rock of the buffers shown) which falls below the typical $fO_2$ of terrestrial mantle or crustal rocks (grey band region). Therefore, typical terrestrial rocks are too

oxidized to produce PH₃ from phosphates and the formation of phosphine is highly unlikely under Venusian subsurface conditions.

### 2. 4. 2. Sensitivity Analysis on Subsurface Fugacity Calculations.

We modelled combinations of f(O₂) of the phosphate/phosphine equilibrium in the plausible Venusian pressure range, and for water content of the rocks of 0.01-5% (unrealistically high for modern Venus, but found in some recently subducted rocks on Earth). We note that the mineral redox buffers are also pressure sensitive (Frost 1991), but this effect is trivial at crustal pressures. Phosphine production is not favored under any plausible crustal conditions (Figure S15).

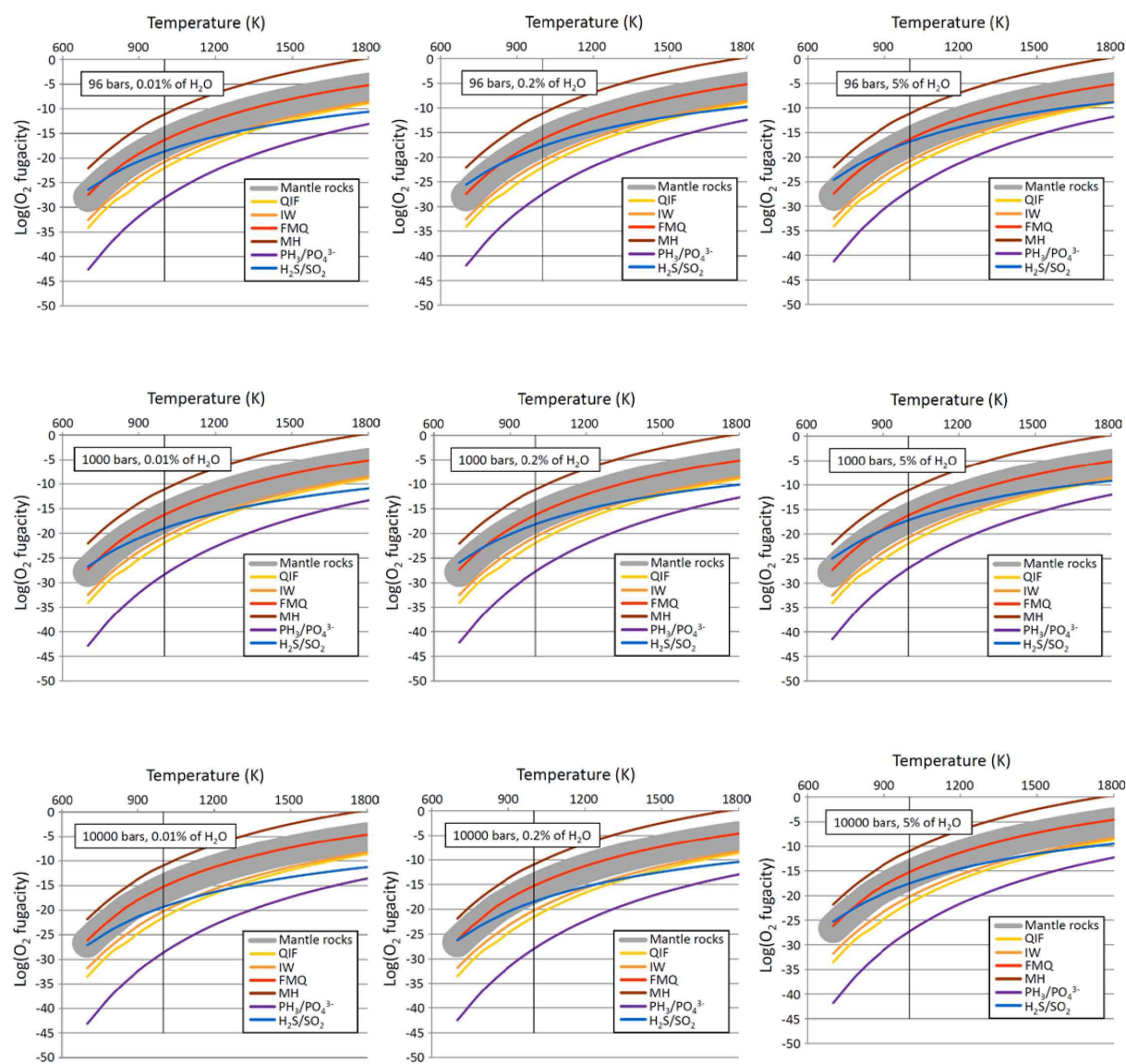

**Fig. S15.** Oxygen fugacity of the phosphate/phosphine equilibrium under variable pressures and water content values. x axis: logO₂ fugacity, y axis: Temperature (K). Fugacity of the production of phosphine from phosphate minerals is calculated for range of pressure values (96-10000 bars) and subsurface water abundances ranging from very low (0.01%) to very high (5%) water content. Phosphine production is not favored under any plausible crustal conditions (the phosphate/phosphine fO₂ curve lies way below the most rock, the QIF buffer line).

### 2. 4. 3. Amount of Phosphine Produced by Volcanism

The amount of volcanism required to produce a given flux of phosphine was calculated as follows. We calculated the ratio of P(+5):P(-3) based on the f($O_2$) values of six redox buffers with redox states between IW (Iron/Wustit: Fe/FeO) and MH (Magnetite/Haematite: $Fe_3O_4$/$Fe_2O_3$) buffers, including the IW and MH buffers themselves. IW and MH buffers represent the limits of oxygen fugacity commonly found in terrestrial mantle rocks. The P(+5):P(-3) ratio calculations were done for a temperature range of 700 K to 1600 K (representing the extremes of temperatures seen in outgassing in terrestrial volcanoes), 100 to 10000 bar and 0.00015 to 0.015 rock water content. The results are shown in Figure S16A.

From this we can estimate the total amount of phosphorus that has to be outgassed in order to provide a flux of ~26 kg/second across the planet (~8.0x$10^{11}$ grams per (terrestrial) year) that is needed to maintain an atmospheric concentration of ~1 ppb by the following equation:

$$P_T = 26000 \cdot \frac{P_O}{P_R},$$

where $P_T$ = the total phosphorus outgassing needed in grams/second/planet, and $P_O$/$P_R$ is the ratio of oxidized to reduced phosphorus in the outgassed phosphorus species, assuming that almost all the phosphorus is present in an oxidized form. This flux is plotted in Figure 9 of the main paper.

We relate the flux of phosphorus outgassing necessary to maintain a 1 ppb atmospheric level to terrestrial volcanic outgassing rates as follows. The rate at which phosphorus is outgassed from terrestrial volcanoes is not known, as phosphorus is produced in volcanoes as non-volatile phosphate species or as $P_4O_{10}$ which rapidly condenses with atmospheric water to form phosphoric acid. We therefore assume that the ratio of phosphorus to sulfur production by volcanoes is the same as the ratio of phosphorus to sulfur in metamorphic rock, and estimate phosphorus 'outgassing' by reference to sulfur outgassing. The ratio of sulfur to phosphorus in metamorphic rock varies widely with the rock, but averages ~1.9 (Figure S16B). Data on the phosphorus and sulfur content of igneous rocks were downloaded from the PetDB Database (www.earthchem.org/petdb) on 12th December 2020, using the following parameters: Chemistry = Chemistry    P EXISTS, S EXISTS, Materials = Igneous. Sulfur is outgassed at a rate of approximately 285 kg/second (Halmer *et al.* 2002) on Earth, suggesting ~143 kg/second of phosphorus is outgassed on Earth.

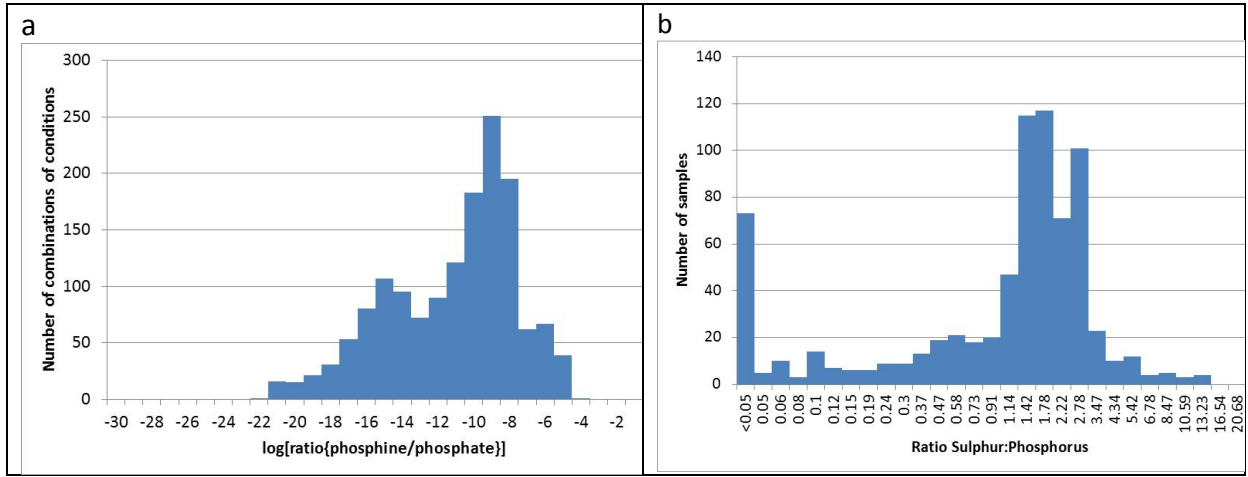

**Fig. S16.** Calculation of the flux of phosphine from volcanic eruption. a) Log of ratio of P(-3)/P(+5) based on fugacity calculation for a total of 1500 combinations of conditions, as described in the text. x axis: $\log_{10}(P^{(-3)}/P^{(+5)})$, in bins of 1 log unit. y axis: number of condition combinations producing this ratio. b) Ratio of sulfur to phosphorus in 745 igneous rock samples where both S and P concentrations were measured. x axis: sulfur/phosphorus ratio, y axis: number of samples with that ratio of S/P. Note that some samples are volcanic material from which nearly all the sulfur has been lost as sulfur gases. The average ratio S/P is 1.91, i.e. sulfur is approximately twice as abundant in these samples as phosphorus. The data were downloaded from the PetDB Database (www.earthchem.org/petdb; (Lehnert *et al.* 2000)).

### 2. 4. 4. Lower Mantle Phosphorus Abundance

Lower mantle rocks are believed to be reduced and volatile-poor. There are two methods for probing the likely phosphorus content of plume volcanism originating from the lower mantle. The most reliable is to ask whether volcanic basalts whose isotope ratio indicate a lower mantle origin have lower phosphorus. Low $^{87}Sr/^{86}Sr$ and high $^{143}Nd/^{144}Nd$, $^{206}Pb/^{204}Pb$ are associated with lower mantle source rocks (Hart *et al.* 1992). On Figure S17 we plot the abundance of sulfur and phosphorus as a function of $^{87}Sr/^{86}Sr$, $^{143}Nd/^{144}Nd$ and $^{206}Pb/^{204}Pb$ in igneous rocks in the PetDB database (data downloaded as above, in Supplementary Section 2.4.3.). The results show that igneous rocks that originate from the lower mantle do not have significantly higher phosphorus content than rocks originating nearer the surface (Figure S17).

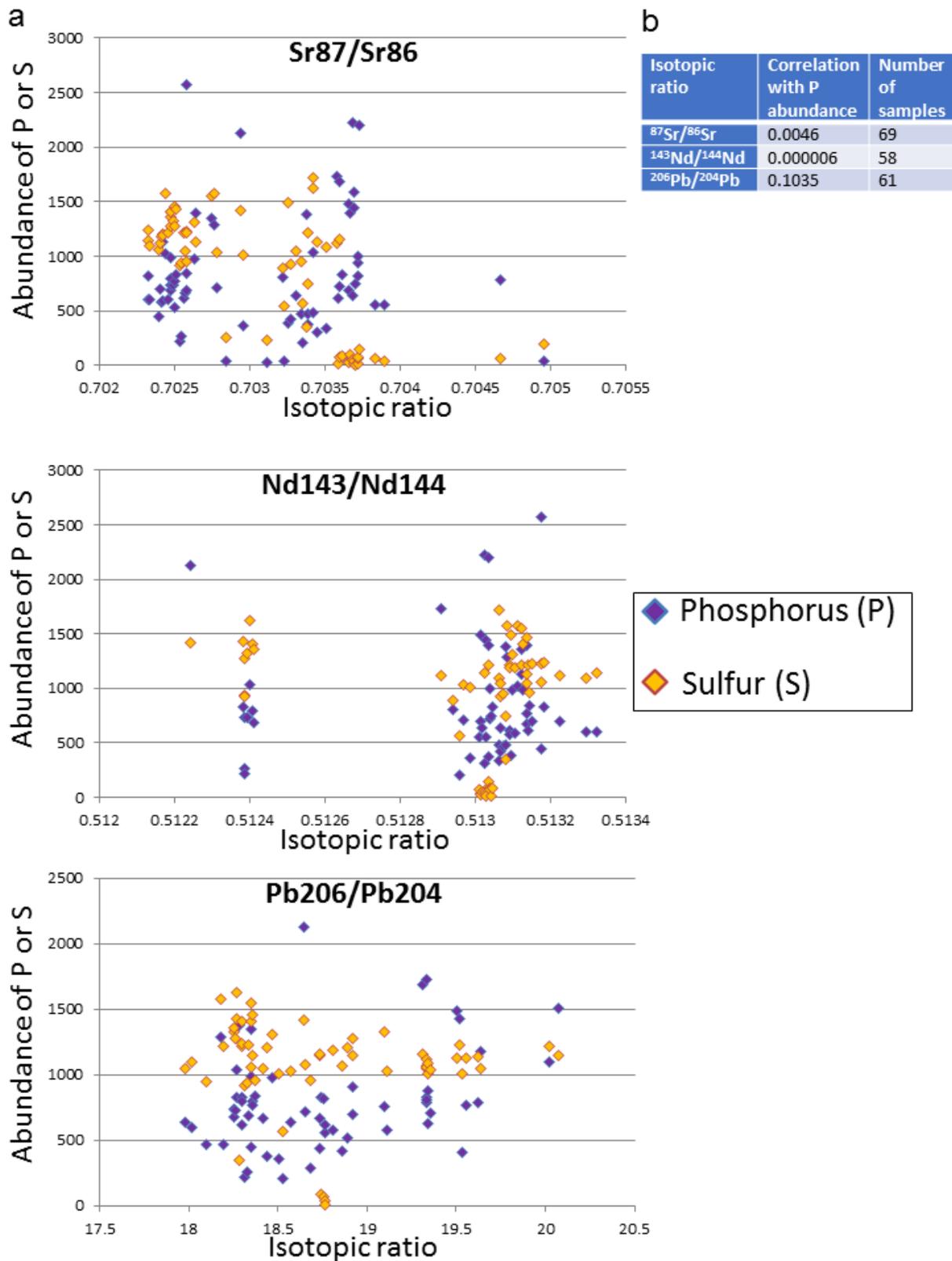

**Fig. S17.** Igneous rocks that originate from the lower mantle do not have significantly higher phosphorus content than rocks originating nearer the surface. (a) Abundance of phosphorus (purple) and sulfur (yellow) in terrestrial igneous rocks, as a function of isotope ratios that are characteristic of lower mantle rocks. x axis: isotopic ratio, y axis: abundance of P or S (b) The correlation coefficients of the isotope ratios and phosphorus abundances. None of the correlations are significant at the p=0.1 level.

The second approach is to examine the phosphorus content in diamond inclusions originating from the lower mantle. There is some evidence that diamonds originating in the lower mantle originate in an environment with higher abundance of volatiles, and therefore potentially in an environment with higher abundance of P than other mantle rock (Kaminsky *et al.* 2013). This however is not surprising, and should not be taken as evidence of higher abundance of P in lower mantle rocks. Diamonds are derived from carbon, which is predominantly hosted in metals in the lower mantle (Dasgupta and Hirschmann 2010). To form diamonds carbon is processed out of this phase into a volatile-enriched phase by subducted hydrated lithosphere (Regier *et al.* 2020). Thus, diamonds represent a very specific processing end-member, and are not representative of the lower mantle as a whole.

## 2. 5.  Details on Other Potential Processes of Phosphine Formation

### 2. 5. 1. Formation of Phosphine by Lightning

Our assumptions are as follows. If the Vega data represents genuine atmospheric phosphorus, then the column density of phosphorus is (order of magnitude) similar to sulfuric acid (Titov *et al.* 2018). Density of cloud and haze materials in the cloud layer is ~0.2 μg/m$^3$ (assuming droplet density of 2 g/ml) (Knollenberg and Hunten 1979), of which maybe ¼ is phosphate (i.e. $^1/_{12}$ is phosphorus atoms). We assumed that the overall dimensions and frequency of lightning strikes are similar to those found on Earth; the average lightning bolt is 25 mm wide and 8 km long[1], i.e. ~4 m$^3$; with 100 lightning strikes per second; noting that this is a matter of debate (Lorenz 2018).

The production of materials in lightning is the result of complex chemistry in a cooling plasma, which it is impractical to model. We have therefore simplified it as follows.
- We assume that the gas in the atmosphere and the liquid in the droplets is atomized to individual atoms. We consider only P, H, S, and O atoms, as other atoms are likely to form species that hydrolyze to phosphorous oxyacids.
- We add atoms one at a time to a phosphorus atom until it's valences are filled – three valency for phosphine, five valency for any compound with an oxygen attached to a phosphorus. We consider only stable compounds as end products. Phosphine oxide itself (PH$_3$O) is not a stable compound.
- We assume that the probability of an atom being added to a P atom is solely determined by its abundance in the plasma.
- We assume that once an atom is added it cannot be taken off again. We note that this is unrealistic – at the temperature of a lightning bolt plasma, the chance that a P-H bond will be attacked by an O atom to break the P-H bond is significant; however to include such reactions would require a full kinetic network which is not practical.
- We assume that any compounds other than H$_3$PO$_4$ could form phosphine; either it is phosphine (PH$_3$) or it could disproportionate to phosphine (H$_3$PO$_3$, H$_3$PO$_2$) or hydrolyze to give phosphorous or hypophosphorous acid which could then disproportionate to phosphine (S and C compounds).

---

[1] We note that there are several reports available describing lightning discharges on Earth that are unusually long (up to hundreds of kilometres) but such discharges are extremely rare (Lyons *et al.* 2020; Peterson 2019).

The network of possible reactions is very large; Figure S18 shows only a subset of O and H reactions.

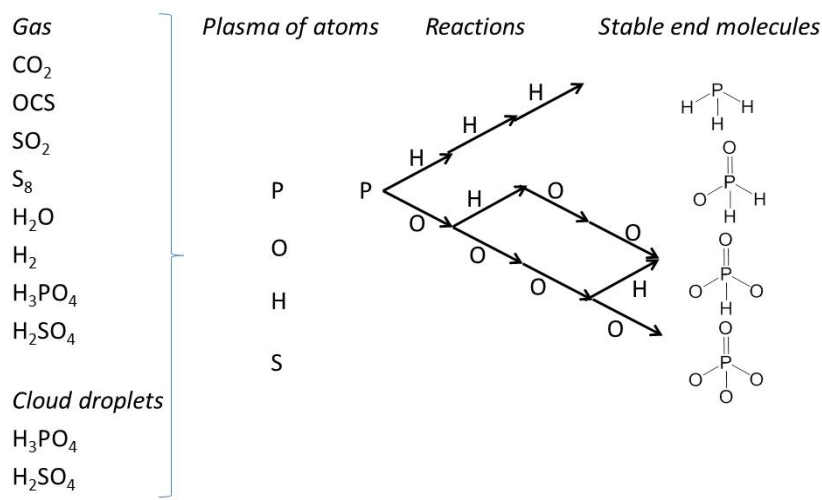

**Fig. S18.** Cartoon of reaction mechanism modelled for lightning. A plasma of atoms is assumed to recombine at random, with a rate unaffected by order. Thus $H_3PO_3$ can be generated by adding an H to a P atom and then three O atoms, or 3 O atoms and then an H atom, or one O, one H and then two more O atoms etc.. P, O, S and H atoms originate from any molecule listed on the left hand side. Their abundance is therefore the abundance of those atoms in those molecules in gas or liquid phase.

However this makes clear that only one path – adding four sequential oxygens to P – ends in $H_3PO_4$, Any species $H_3PO_3$, $H_3PO_2$, $H_3PO$ can be a source of $PH_3$, and so any path *other* than one ending in $H_3PO_4$ end up in potentially phosphine-generating species (It is assumed that once an atom is added to a P atom it cannot be removed, so to form $H_3PO_3$ you need to add one H and three O atoms to a P: to create the final molecule $H_3PO_3$ you then need to add two more H atoms to O atoms). The probability that a phosphorus atom will end up as phosphate is therefore

$$O^4$$

Where $O$ is the abundance of oxygen in the plasma. The probability that a phosphorus atom ends up in a species that is or can generate phosphine is therefore

$$1-O^4$$

(we note that this is not exactly the number of phosphine molecules, as one molecule of $H_3PO_3$ makes ¼ molecule of $PH_3$ on disproportionation, one molecule of $H_3PO_2$ make ½ a $PH_3$ molecule. Therefore, using $1-O^4$ as an estimate of the fraction of P that ends up as $PH_3$ will over-estimate $PH_3$ by up to 4-fold.)

The abundance of atoms in the plasma is calculated as follows. For gas, the abundance of atoms from a gas is given by

$$pp \cdot N \cdot R \cdot A \cdot (298/T) \cdot 0.0224 \quad \text{atoms/m}^3$$

where pp is the partial pressure of the gas, N is the number of atoms of that element in the gas (1 oxygen in CO, 2 in $CO_2$ etc.), R is the pressure, A is Avogadro's number (~$6.023 \times 10^{23}$), T is the absolute temperature and 0.0224 is because one mole of a gas at STP occupies 22.4 litres. For cloud droplets, the abundance is given by

D.C.A

where D is the mass of droplets per cubic meter, and C is the concentration of the target species. Gas partial pressures are the same as for other calculations in this paper. Liquid concentrations are assumed to be 18 M for sulfuric acid and 1 M for phosphoric acid in droplets.

| Altitude (km) | Temp (K) | Pressure (bar) | Gas phase | | | | Droplet phase | | | |
|---|---|---|---|---|---|---|---|---|---|---|
| | | | H atoms | O atoms | P atoms | S atoms | H atoms | O atoms | P atoms | S atoms |
| 0 | 735 | 92.10 | 5.13E+23 | 1.94E+27 | 2.44E+22 | 2.19E+23 | | | | |
| 5 | 697 | 66.50 | 3.91E+23 | 1.48E+27 | 1.86E+22 | 1.67E+23 | | | | |
| 10 | 658 | 47.39 | 2.95E+23 | 1.11E+27 | 1.40E+22 | 1.27E+23 | | | | |
| 15 | 621 | 33.04 | 2.25E+23 | 8.23E+26 | 1.04E+22 | 9.41E+22 | | | | |
| 20 | 579 | 22.52 | 1.64E+23 | 6.02E+26 | 7.58E+21 | 7.16E+22 | | | | |
| 25 | 537 | 14.93 | 1.17E+23 | 4.30E+26 | 5.42E+21 | 5.35E+22 | | | | |
| 30 | 495 | 9.85 | 8.06E+22 | 3.08E+26 | 3.88E+21 | 4.02E+22 | | | | |
| 35 | 453 | 5.92 | 5.29E+22 | 2.02E+26 | 2.54E+21 | 2.91E+22 | | | | |
| 40 | 416 | 3.50 | 3.41E+22 | 1.30E+26 | 1.64E+21 | 2.13E+22 | | | | |
| 45 | 383 | 1.98 | 2.09E+22 | 8.00E+25 | 1.01E+21 | 1.29E+22 | 9.20E+14 | 1.09E+14 | 3.03E+12 | 2.42E+13 |
| 50 | 348 | 1.07 | 1.55E+21 | 4.74E+25 | 7.40E+19 | 3.40E+21 | 9.20E+14 | 1.09E+14 | 3.03E+12 | 2.42E+13 |
| 55 | 300 | 0.53 | 2.91E+20 | 2.74E+25 | 7.24E+17 | 1.39E+21 | 3.83E+15 | 4.54E+14 | 1.26E+13 | 1.01E+14 |
| 60 | 263 | 0.24 | 7.30E+19 | 1.39E+25 | 6.18E+15 | 1.40E+20 | 7.77E+13 | 9.20E+12 | 2.55E+11 | 2.04E+12 |
| 65 | 243 | 0.10 | 3.27E+19 | 6.21E+24 | 2.54E+14 | 6.26E+19 | 7.77E+13 | 9.20E+12 | 2.55E+11 | 2.04E+12 |
| 70 | 230 | 0.04 | 1.31E+19 | 2.48E+24 | 2.37E+13 | 2.50E+19 | 7.77E+13 | 9.20E+12 | 2.55E+11 | 2.04E+12 |

**Table S9.** An example of the calculation of the abundance of atoms, for the lower partial pressure of all input gases (Table S5).

| Altitude band | Mode | Alt max (km) | Alt min (km) | Number density/cm³ | Mean diameter (μm) | volume of droplets (m³) per m³ volume of atmosphere | Altitude (km) | Mass of droplet/m³ |
|---|---|---|---|---|---|---|---|---|
| High | 1 | 70 | 56.5 | 1500 | 0.4 | 5.03E-11 | 45 | 8.04E-11 |
| High | 2 | 70 | 56.5 | 50 | 2 | 2.09E-10 | 50 | 8.04E-11 |
| Middle | 1 | 56.5 | 50.5 | 300 | 0.3 | 4.24E-12 | 55 | 3.35E-10 |
| Middle | 2 | 56.5 | 50.5 | 50 | 2.5 | 4.09E-10 | 60 | 6.79E-12 |
| Middle | 3 | 56.5 | 50.5 | 10 | 7 | 1.80E-09 | 65 | 6.79E-12 |
| Low | 1 | 50.5 | 47.5 | 1200 | 0.4 | 4.02E-11 | 70 | 6.79E-12 |
| Low | 2 | 50.5 | 47.5 | 50 | 2 | 2.09E-10 | | |
| Low | 3 | 50.5 | 47.5 | 50 | 8 | 1.34E-08 | | |

**Table S10.** The contribution of droplets is calculated from the observed abundance of droplets of difference sizes ("modes") at different altitude bands in the cloud decks

Lightning could in principle go from cloud to cloud or cloud to ground. We assume both, with 100-fold more cloud-to-cloud lightning strikes as cloud-to-ground, and average path length of 10 and 45 km respectively, 100 strikes per seconds, and an average diameter of the lightning path of 100 mm. (the actual lightning path on Earth is smaller than this, but we reason that heating effects from the bolt will create plasma several centimeters from the electrical path itself).

With these assumptions, the tables above (Table S9 and Table S10) give the following calculated amounts of $PH_3$ produced (Table S11). Which gives $9.09 \times 10^5$ moles or $3.09 \times 10^6$ grams of $PH_3$/Venusian year. If this is recalculated for all combinations of high and low partial pressures of gases (Table S5), the mean production is 3.524 tones/Venusian year, +/- 0.326 tones (standard deviation), even with these highly optimistic assumptions. This is more than $10^5$ times too little to explain the proposed 1 ppb phosphine concentration in the Venusian atmosphere (Greaves *et al.* 2020c).

| Fraction of P as $PH_3$ | moles of $PH_3$ per cubic meter | Volume of cloud-cloud lightning | Volume cloud-ground lightning | Moles of P per second |
|---|---|---|---|---|
| 1.56E-03 | 3.49E-06 | 0 | 3.53E+02 | 1.23E-03 |
| 1.56E-03 | 2.52E-06 | 0 | 3.53E+02 | 8.92E-04 |
| 1.56E-03 | 1.80E-06 | 0 | 3.53E+02 | 6.36E-04 |
| 1.60E-03 | 1.28E-06 | 0 | 3.53E+02 | 4.54E-04 |
| 1.62E-03 | 8.85E-07 | 0 | 3.53E+02 | 3.13E-04 |
| 1.64E-03 | 5.95E-07 | 0 | 3.53E+02 | 2.10E-04 |
| 1.62E-03 | 3.87E-07 | 0 | 3.53E+02 | 1.37E-04 |
| 1.67E-03 | 2.41E-07 | 0 | 3.53E+02 | 8.50E-05 |
| 1.75E-03 | 1.49E-07 | 0 | 3.53E+02 | 5.26E-05 |
| 1.74E-03 | 8.37E-08 | 7.85E+01 | 0 | 6.57E-04 |
| 4.24E-04 | 1.36E-09 | 7.85E+01 | 0 | 1.07E-05 |
| 2.45E-04 | 6.65E-12 | 7.85E+01 | 0 | 5.22E-08 |
| 6.14E-05 | 1.25E-14 | 7.85E+01 | 0 | 9.78E-11 |
| 6.14E-05 | 4.73E-16 | 7.85E+01 | 0 | 3.72E-12 |
| 6.14E-05 | 4.17E-17 | 7.85E+01 | 0 | 3.27E-13 |

**Table S11.** Predicted amounts of $PH_3$ produced by lightning in the Venusian atmosphere.

### 2. 5. 2. Disproportionation of Phosphorous Acid Within the Clouds

The calculations above can also be used to estimate the amount of phosphorous acid in the droplets in the clouds. The fraction of phosphorus present as $H_3PO_3$ as a function of altitude is derived from Figure 5 of the main paper. Assuming 1 molal concentration of phosphorus species, the mass of cloud particles per $m^3$ as described above (Table S10), and the volume of a 5 km slice of the atmosphere in meters (5000 meters height x $4.602 \cdot 10^{14}$ square meter area of Venus, we calculate the total mass of the available $H_3PO_3$ in the entire Venusian atmosphere (Table S12). This gives a mass of $2.05 \cdot 10^{-2}$ grams of phosphorous acid in the entire atmosphere.

| Altitude | Mass of droplet/m^3 | Moles of P per m^3 | Fraction of P that is $H_3PO_3$ | Volume of atmospheric altitude slice (m^3) | Mols of $H_3PO_3$ |
|---|---|---|---|---|---|
| 40 | 8.04E-11 | 1.37E-05 | 7.796E-18 | 2.30E+18 | 2.45E-04 |
| 45 | 8.04E-11 | 1.37E-05 | 1.450E-19 | 2.30E+18 | 4.56E-06 |
| 50 | 8.04E-11 | 1.37E-05 | 1.129E-20 | 2.30E+18 | 3.55E-07 |
| 55 | 3.35E-10 | 2.21E-06 | 1.290E-23 | 2.30E+18 | 6.56E-11 |
| 60 | 6.79E-12 | 2.60E-07 | 1.149E-27 | 2.30E+18 | 6.87E-16 |
|  |  |  |  | Total | *2.50E-04* |

**Table S12.** Available $H_3PO_3$ in the Venusian atmosphere.

### 2. 5. 3. Unknown Chemistry as an Explanation of Phosphine on Venus

#### *2. 5. 3. 1.    Phosphine Chemistry in Concentrated Sulfuric Acid*

Phosphine is readily oxidized on passing through concentrated sulfuric acid at Earth ambient temperatures. The chemistry has been known for over 140 years, as it was used as a method to remove phosphine from acetylene. Acetylene was widely used as a gas for lighting in the late 19$^{th}$ and early 20$^{th}$ century before the advent of electrification, and the gas was manufactured by the acid hydrolysis of calcium carbide. Trace phosphide and sulfide in the carbide lead to phosphine and hydrogen sulfide in the acetylene, which caused undesirable smell in the gas and 'haze' of $H_3PO_4$ and $H_2SO_4$ produced on burning (Doman 1902). Passing the gas through $H_2SO_4$ efficiently cleared out both gases (reviewed in (Doman 1902; Leeds and Butterfield 1910)). The process fell into disuse in the West in the 1900s, replaced by purification of acetylene over chromic acid (Leeds and Butterfield 1910), and became obsolete when carbides as a source of acetylene were replaced by synthesis by partial oxidation of methane (Sachsse 1954); however the method remains in use in China (Cai *et al.* 2010; Xiao-yong 2009). Reaction temperatures are typically cited as between 0 ºC and 15 ºC.

There is very limited data on the kinetics of the oxidation of phosphine with sulfuric acid reaction. The citations above state that acetylene is passed up a tower down which >95% $H_2SO_4$ is sprayed, implying efficient removal (by rapid oxidation of phosphine to phosphoric acid) in 10s of seconds. (Dorfman *et al.* 1991) state that the reaction occurs at negligible rate at concentrations of acid below ~90% acid by weight. Such reaction behavior suggests attack on $PH_3$ by $SO_3$, which is consistent with the electrophilic attack by $SO_3$ on the lone pair on $PH_3$, and with B3LYP *ab initio* calculations to 6-311-G level of the energy of the reaction:

$PH_3 + SO_3 \rightarrow PH_3:SO_3$              $\Delta H$=-8.25 kCal/mol.

(Lorenz *et al.* 1963) report that the reaction between 99% $H_2SO_4$ and $PH_3$ is at least 99% complete in 40 seconds at 60 °C , and (Perraudin 1961) reports that $PH_3$ is effectively cleared by bubbling through a thin layer of $H_2SO_4$ <~1 cm deep. (Leeds and Butterfield 1910) claim that the reaction is efficient down to -20 °C. It is therefore likely that phosphine will be oxidized efficiently by the sulfuric acid in Venus' lower clouds. Oxidation in the upper clouds, where the concentration of sulfuric acid is below 90% and temperature below 270 K is unknown, but it is very unlikely that any process would *synthesize* phosphine under these conditions.

*2. 5. 3. 2.    Production of Phosphine from Elemental Phosphorus*

Elemental phosphorus is most stable as $P_4$ ("White" phosphorus") at Venus surface conditions. The standard state for elemental phosphorus - "Red" phosphorus - which is more stable at temperatures <540 K (at 1bar) is not volatile, and so would not be present in the atmosphere. Thermodynamic calculations are therefore done for $P_4$. However, the free energy difference between reference P and $P_4$ is <7 kJ per mole of phosphorus atoms at the temperatures considered in this model, and so the difference between the two allotropes will be small.

We modelled the production of phosphine from elemental phosphorus using reducing agents and hydrogen sources available in Venus' atmosphere, using the same approach as described in Supplementary Section 1.3.2. (Figure S19).

1) $1½H_2S + ¼P_4 \rightarrow PH_3 + 1½S$
2) $1½H_2 + ¼P_4 \rightarrow PH_3$
3) $1½CO + 1½H_2O + ¼P_4 \rightarrow PH_3 + 1½CO_2$
4) $^3/_4S + 1½H_2O + ¼P_4 \rightarrow PH_3 + ^3/_4SO_2$
5) $½OCS + 1½H_2O + ¼P_4 \rightarrow PH_3 + ½SO_2 + ½CO_2$

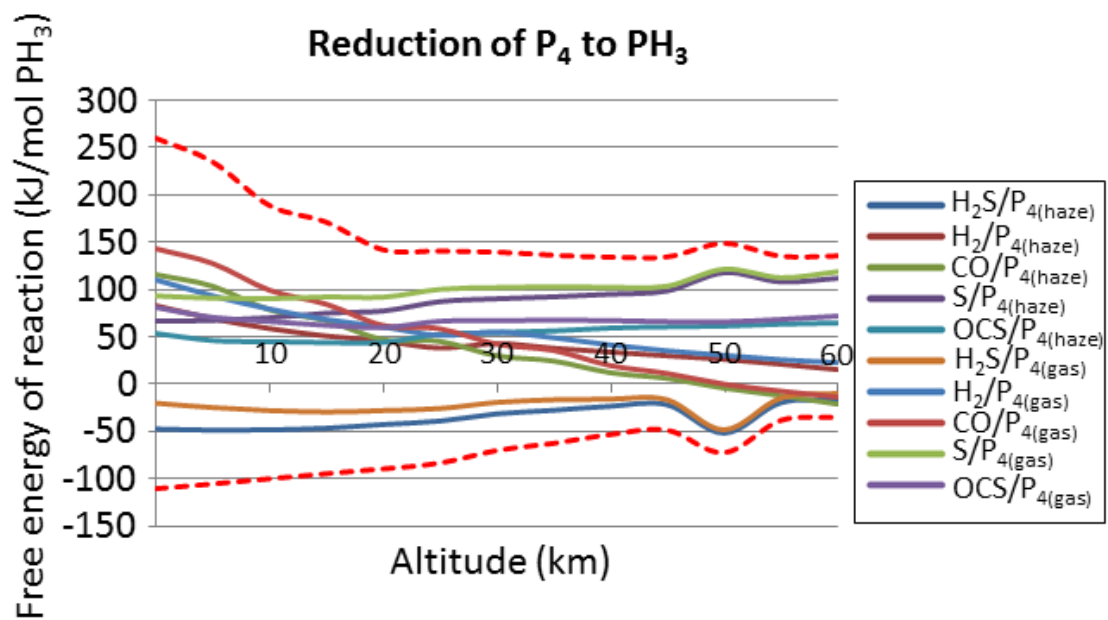

**Fig. S19.** Thermodynamics of phosphine production from reaction of elemental phosphorus in gas phase or as a solid ('haze') in Venus atmosphere. x axis: altitude (km), y axis: Gibbs free energy of reaction (ΔG) (kJ/mol). Dashed lines show the limits of the free energy found for any combination of gas partial pressures, for any altitude, for any reaction in a set of reactions. Only the reactions of elemental phosphorus with hydrogen sulfide may provide a source of phosphine.

The Figure S19 shows the reaction of elemental phosphorus with hydrogen sulfide may provide a source of phosphine. Reduction of elemental phosphorus by CO and water (water to provide hydrogen atoms) is also potentially favourable at cloud level. Even with elemental phosphorus as a source of phosphorus atoms, no other reaction is favourable for making phosphine.

However, the production of elemental phosphorus it itself extremely unlikely under any plausible Venus' surface and subsurface conditions. In brief, fugacity calculations show that elemental phosphorus is no more likely to be outgassed on Venus than phosphine. Therefore, in suggesting elemental phosphorus as a source of phosphine, we have just exchanged the difficulty of making phosphine for the equal difficulty of making elemental phosphorus.

Industrial production of elemental phosphorus is achieved by heating phosphate rock with coke (carbon) in a blast furnace (Greenwood and Earnshaw 2012), and it has been suggested that the 'unknown UV absorber' could be graphite (Shimizu 1977) (but see (Hapke and Nelson 1975)). Could elemental phosphorus be formed by reduction of P(V) or P(III) species by graphite? This idea also does not explain the presence of phosphine, as it itself displaces the improbability onto two other factors; firstly, where does the graphite come from, and secondly how is graphite at cloud temperatures heated to >1000 ºC required to make this reaction thermodynamically favorable (the actual industrial reaction uses mineral phosphate and silicon dioxide, but it is implausible that silicon dioxide would be in the clouds.) (Figure S20).

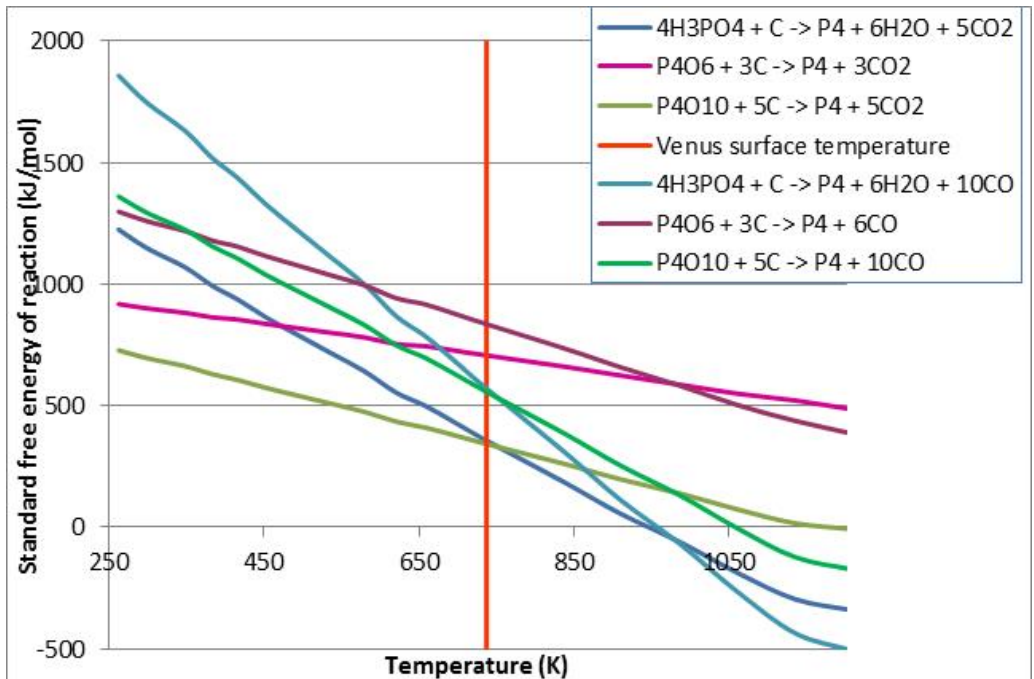

**Fig. S20.** Standard free energy of reduction of phosphoric acid by carbon is positive at temperatures on or above Venus' surface. (Note that this is standard free energy, as the activity of $P_4$ cannot be estimated realistically).

### 2. 5. 3. 3.    Crustal Production of Elemental Phosphorus

We addressed whether crustal chemistry could produce elemental phosphorus as a source of reduced phosphorus species that could subsequently be reduced to phosphine. We replicated the fugacity model shown in Supplementary Section 1.3.3., by comparing mineral oxygen fugacity buffers to the oxygen 'fugacity' of the following reaction:

$Mg_3(PO_4)_2 + 1½SiO_2 \rightarrow 1½Mg_2SiO_4 + ½P_4 + 2½O_2$

The results are shown on Figure S21.

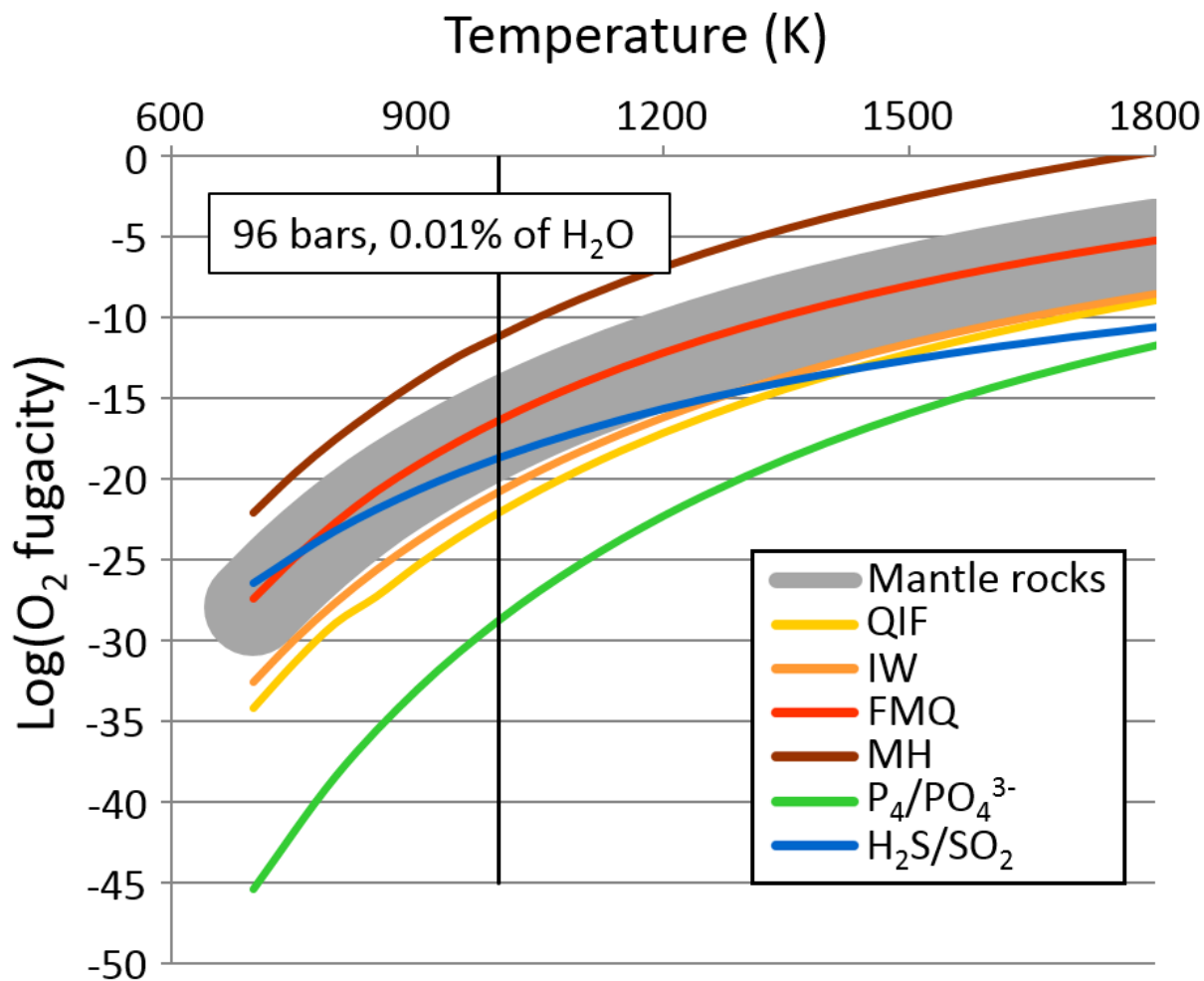

**Fig. S21.** Comparison of the fugacity of the phosphate/$P_4$ equilibrium to the fugacity of the standard mineral buffers of terrestrial rocks. x axis: log $O_2$ fugacity, y axis: Temperature (K). Fugacity of the production of $P_4$ from phosphate minerals is calculated for 96 bars and 0.01% water in the rocks. The fugacity of the phosphate/$P_4$ equilibrium is shown as a green line. The other curves are $O_2$ fugacities of standard rock buffers. The formation of elemental phosphorus is highly unlikely under Venusian conditions. Fugacity of the production of elemental phosphorus from phosphate minerals calculated for 96 bars, 0.01% water.

The results obtained for the formation of elemental phosphorus in subsurface rocks are similar to the results obtained for the possibility of the formation of phosphine (Figure S14 and Figure S15). We conclude that it is extremely unlikely that crustal rocks could produce elemental phosphorus, and as a result it is very unlikely that the observed atmospheric phosphine comes from the reduction of subsurface fraction of the elemental phosphorus.

### 2. 5. 3. 4. *High Altitude Reduction of Calcium Phosphate*

Because the free energy of calcium phosphate reduction to phosphine showed a trend that suggested further steep decline of the ΔG of formation of phosphine with altitude (Figure S22), we have calculated the theoretical free energy of phosphine formation by the reduction of calcium phosphate up to the altitude of 120 km.

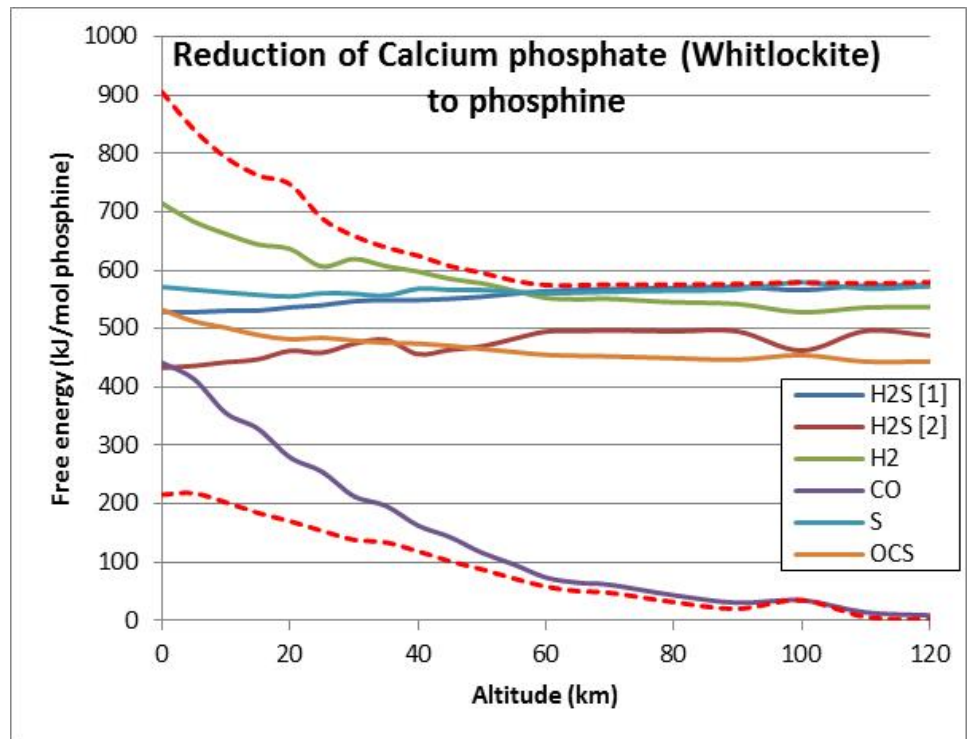

**Fig. S22.** The Thermodynamics of phosphine production by reduction of calcium phosphate (Whitlockite) to phosphine up to the altitude of 120 km. x axis: altitude (km), y axis: Gibbs free energy of reaction ($\Delta G$) (kJ/mol). Dashed lines show the limits of the free energy found for any combination of gas partial pressures, for any altitude, for any reaction in a set of reactions. At 120 km the free energy of reduction of calcium phosphate to phosphine by carbon monoxide is approximately 0, however such scenario for production of phosphine is highly unlikely.

At 120 km the free energy of reduction of calcium phosphate by carbon monoxide is approximately 0, i.e. if the reduction reaction reached thermodynamic equilibrium, the atmospheric loading of phosphine would be comparable to the atmospheric loading of whitlockite. However, for this to be a source of phosphine, two implausible events have to happen:

- Whitlockite, a mineral, has to be transported to an altitude of 120 km, despite very limited vertical air flow on Venus, including being transported through the clouds without being absorbed onto cloud particles
- It must react with CO on a timescale comparable to that of the lifetime of phosphine, in an environment where the temperature is -100 ºC and hence where almost all non-photochemical reactions will have a negligible rate over geological timescales, and where phosphine itself has an extremely short lifetime due to rapid photolysis by unshielded solar UV.

One could hypothesize scenarios under which this could happen (for example nanoparticles of whitlockite mixed with metallic iron/nickel could both have a long enough residence time and a high enough surface:volume ratio to reach 120 km, where the iron/nickel could catalyse reactions with carbon monoxide). However, these are *ad hoc* scenarios that are unjustified by any physical observations of the atmosphere of Venus.

### 2. 5. 4. Formation of Phosphine from Berlinite

The free energy of reaction of $H_2S$ with Berlinite (aluminium phosphate) is negative at Venus surface, as calculated above for the following reactions (Table S13) on Figure S23. However, Berlinite itself requires substantial energy to create from other crustal rocks under Venus conditions: the reaction of Berlinite with calcium and/or magnesium silicate to form aluminium silicate and calcium or magnesium phosphate

$1½CaMgSiO_4 + 2AlPO_4 \rightarrow Al_2SiO_5 + ½Ca_2(PO_4)_3 + ½Mg_2(PO_4)_3 + ½SiO_2$

is highly exothermic (Figure S23). Berlinite would therefore not be expected to form under Venus surface conditions if calcium, magnesium and silica were present, as it is known that they are from Vega lander data (Smrekar *et al.* 2014). Note that on Earth Berlinite is formed in high temperature hydrothermal systems, a mechanism that does not apply to Venus.

| *Berlinite ( AlPO$_4$ )\** |
|---|
| 77) $2H_2S + AlPO_4 \rightarrow PH_3 + ½ Al_2O_3 + ½ H_2O + SO_2 + S$ |
| 78) $4H_2S + AlPO_4 \rightarrow PH_3 + ½ Al_2O_3 + 2 ½ H_2O + 4S$ |
| 79) $4H_2 + AlPO_4 \rightarrow PH_3 + ½ Al_2O_3 + 2 ½ H_2O$ |
| 80) $4CO + 1½H_2O + AlPO_4 \rightarrow PH_3 + ½Al_2O_3 + 4CO_2$ |
| 81) $2S + 1½H_2O + AlPO_4 \rightarrow PH_3 + ½ Al_2O_3 + 2SO_2$ |
| 82) $1^1/_3OCS + 1½H_2O + AlPO_4 \rightarrow PH_3 + ½ Al_2O_3 + 1^1/_3CO_2 + 1^1/_3SO_2$ |

**Table S13.** Reactions of formation of phosphine from Berlinite.

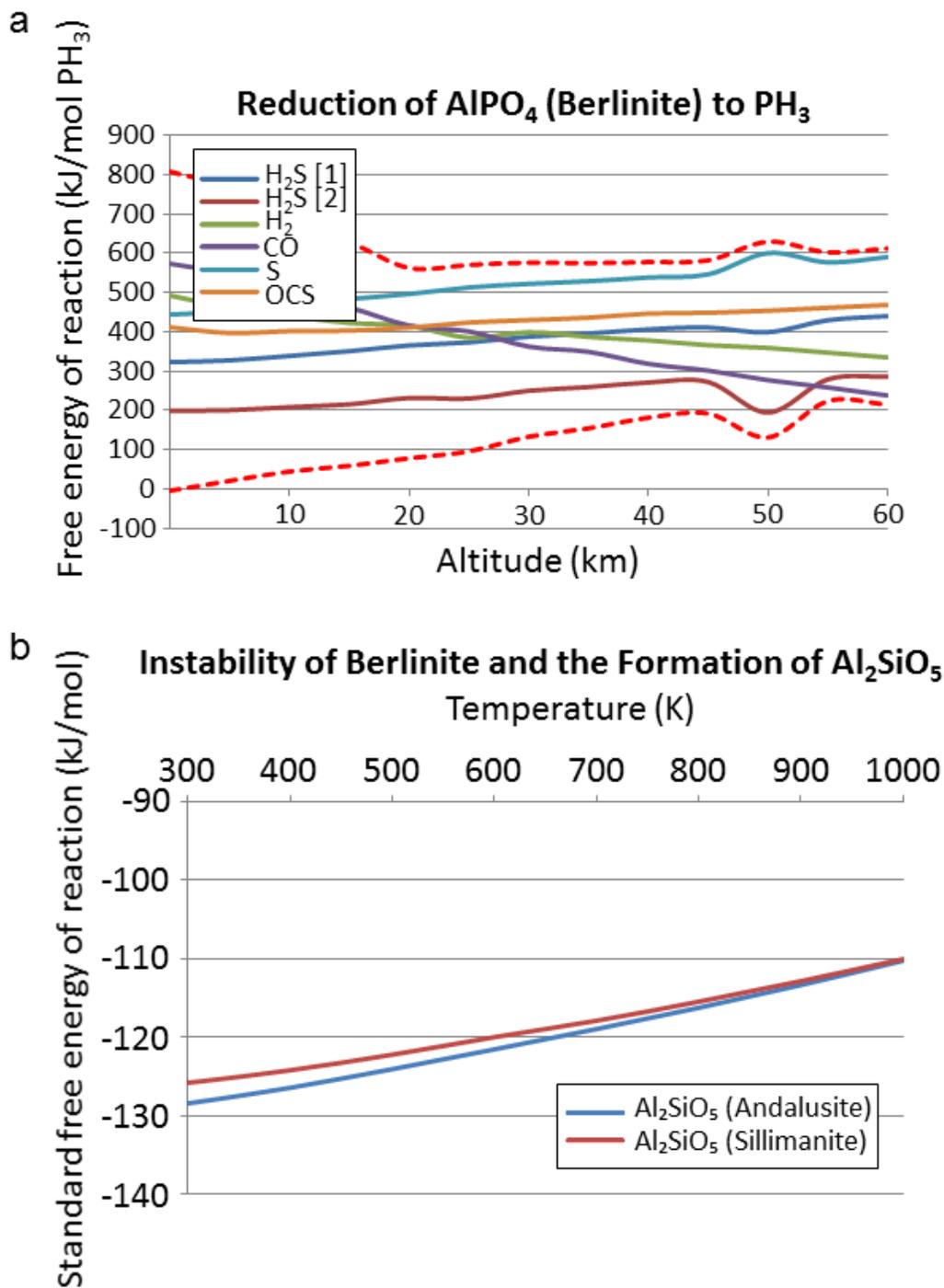

**Fig. S23.** Formation of phosphine from berlinite. (a) Reduction of aluminium phosphate (berlinite) ($AlPO_4$) to phosphine by atmospheric gases. x axis: altitude (km), y axis: Gibbs free energy of reaction ($\Delta G$) (kJ/mol). Dashed lines show the limits of the free energy found for any combination of gas partial pressures, for any altitude, for any reaction in a set of reactions. Each solid line represents a different reductant. We note that some combinations of extreme values of the partial pressure of both $H_2S$ and elemental sulfur ($S_2$) come close to predicting a negative $\Delta G$ value for phosphine production at the surface. $\Delta G$ values are affected by the substantial uncertainty in the partial pressures of $H_2S$, $H_2O$ and gas phase elemental sulfur. (b) conversion of Berlinite to Aluminium silicate is highly expothermic under Venus surface conditions. x axis: temperature (K), y axis: Standard Gibbs free energy of reaction

(ΔG) (kJ/mol). Two free energy curves are shown for two crystal forms of Aluminium silicate: Andalucite and Silmanite.

### 2. 5. 5. Formation of Phosphine by Tribochemical Processes

An intriguing possibility for the production of phosphine from rock phosphorus is coupling of mechanical energy to phosphorus reduction in the presence of fluids, termed tribochemical synthesis. Glindemann et al (Glindemann *et al.* 2005) have explored this, and report variable conversion of rock phosphorus to $PH_3$, the highest values being for quartz and calcium carbonate (limestone and marble). Other rocks reported $10^{-6}$ to $10^{-9}$ conversion of phosphate to phosphine, with the exception of the pulverization of one quartz pebble. Calcium carbonate will not exist on the surface of Venus, as the $CO_2$ will be baked out into the atmosphere (Catling and Kasting 2017; Rasool and de Bergh 1970).

Terrestrial industry produces a large amount of crushed rock – 1.3 bln tonnes in the USA alone (Wilburn 2020), over half of it limestones and marble. However, this is probably dwarfed by the volume of rock fractured or ground up as a result of earthquakes. Marc et al have estimated the volume of landslides caused by earthquakes (Marc *et al.* 2016), which approximates exponential function of earthquake magnitude (as would be expected as magnitude is itself a log scale):

$$V = 3.81 \cdot 10^{-15} \cdot e^{4.14M} \tag{38}$$

where V is the total landslide volume in $km^3$ and M is the Richter magnitude of the earthquake. If we apply this to all Earthquakes in 2019 that were shallower than 20 km, we estimate that those earthquakes caused $\sim 3.10^{11}$ tonnes of rock to shift in landslides (note this is a notional figure, as many of these earthquakes were under water; however the distinction between land and water does not apply to Venus). If there were 650 ppb phosphorus in that rock (the average for metamorphic rock in PetDB Database (www.earthchem.org/petdb; (Lehnert *et al.* 2000) as described in the main text), if all of that rock was efficiently pulverized and if all underwent conversion of phosphate to phosphine with an efficiency of $10^{-6}$ (mid-range for Glindemann's paper, excluding limestones) then that produce 660 tonnes of phosphine per year. The flux needed to explain the phosphine on Venus is 800,000 tonnes/year. Even under these optimistic assumptions, therefore, Venus has to be >1000 times as tectonically active as Earth to sustain the observed phosphine levels.

In practice, the mechanism postulated by Glindemann et al requires very specific types of rock to be rubbed together, and requires fluid inclusions in the rock to provide hydrogen atoms for phosphine production. The former will substantially reduce the estimated production rate, and fluid inclusions will be entirely absent from the Venusian surface. The only relevant hydrogen-containing fluids that could be form on Venus would be supercritical $H_2O$ or HCl; as these are present in the atmosphere at $3.10^{-5}$ and $10^{-7}$ mole fraction respectively, their forming dense supercritical fluid phases in rocks seems unlikely. Sulfuric acid could in principle form liquid if the pressure was high enough, but as sulfuric acid efficiently and rapidly oxidizes $PH_3$ to phosphate, it is an unlikely fluid to participate in $PH_3$'s formation.

This is not to say that tribochemical production of phosphine is not significant *on Earth*, where water is abundant in surface rocks. However, we conclude that tribological phosphine production, while interesting and important chemistry, cannot explain the presence of 1 ppb phosphine in Venus' atmosphere. In this regard the tribochemical production of reduced phosphorus species is similar to the potential reduction of phosphate by serpentenization reactions (Pasek *et al.* 2020); both are potentially important on Earth, but cannot be significant in the highly desiccated surface environment of Venus.

## 2. 6. Model for Phosphate Ion Species Calculation and the Thermodynamics of the Biological Reduction of Phosphorus Species

The free energy needed to transport phosphorus species outside the cell into the cell is calculated as follows. Energy has to be input to drive the balance of ions present in the exterior milieu into that found inside the cell. The energy needed to change the phosphate ions from the equilibrium concentration found outside the cell to their concentration at pH=7 was subtracted from the free energy of reaction. When products are allowed to return to the external environment, energy is released as they relax to their thermodynamic minimum: this energy was added to the final energy of reaction. Thus, the overall Gibbs free energy available for the disproportionation reaction is given by:

$$\Delta G_x = \Delta G + \sum_{N=0}^{3} R.T.\ln\left(\frac{P.HPn}{HP.Pn}\right) - \sum_{n=0}^{2} R.T.\ln\left(\frac{Q.HQn}{HQ.Qn}\right), \quad (39)$$

where: $\Delta G$ is the free energy as calculated in Supplementary Section 1.3.1, P is the concentration of each of the phosphate ions at equilibrium at pH=7, HP is the concentration of the protonated ion at pH=7, for each 3 pairs of phosphate species corresponding to the three $pK_a$s of phosphoric acid ($H_3PO_4/H_2PO_4^-$, $H_2PO_4^-/HPO_4^{2-}$ and $HPO_4^{2-}/PO_4^{3-}$), and $P_n$ and $HP_n$ are the relative concentrations of the same ions at the pH assumed to be outside the cell. Q and HQ are the equivalent terms for phosphite. Equivalent calculations apply to the import and export of phosphite species. See Section 5.3. and Figure 10 of the main text for the overall process modelling.

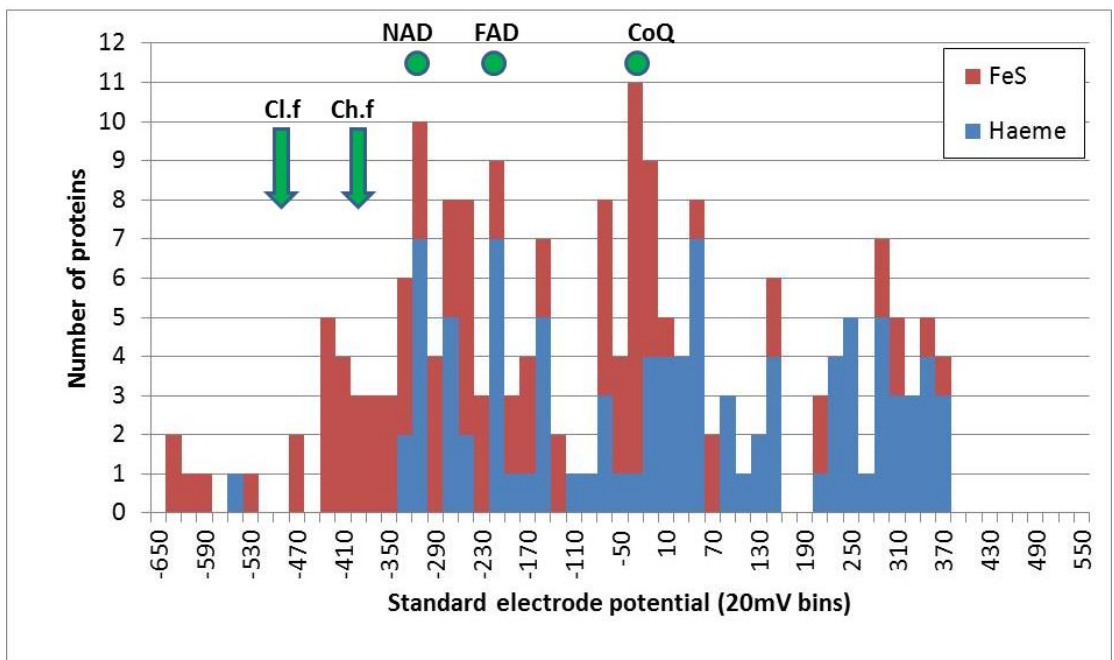

**Fig. S24.** A. Standard electrode potential of Iron/Sulfur proteins (red) and Heme proteins (blue). x axis: standard electrode potential and pH=7. y axis: number of proteins with that potential. Data from (Reedy *et al.* 2007; Zanello 2017; Zanello 2018a; Zanello 2018b; Zanello 2019a; Zanello 2019b). Green circles: NADH/NAD+ couple, FADH/FAD+ couple and Co-enzyme-Q (Yudkin and Offord 1973). Green arrows: standard electrode potential of Iron/Sulfur proteins from *Chromatium vinosum* (Ch.v) and; *Clostridium thermoaceticum* (Cl.f) (Harder *et al.* 1989; Mizrahi *et al.* 1976).

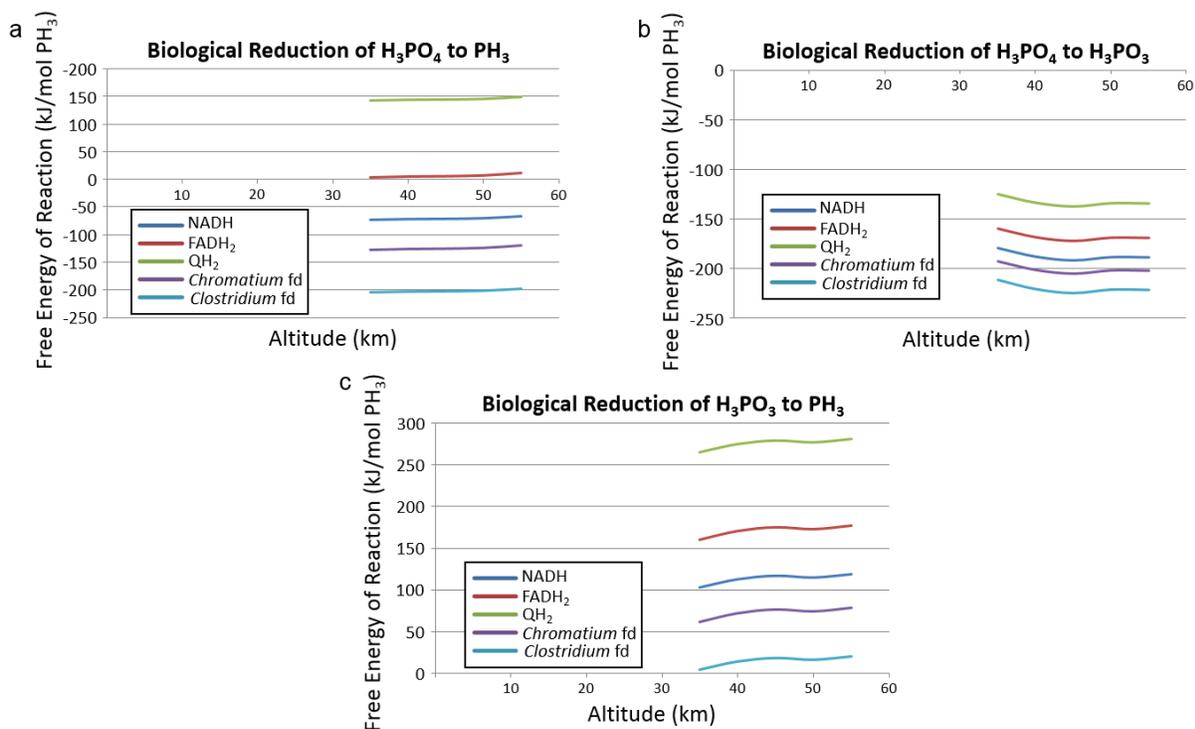

**Fig. S25.** Terrestrial redox proteins and biochemicals could reduce phosphate to phosphine under Venus conditions. The y axes show free energy of reduction of phosphate to phosphine while the x axes show altitude in the Venus atmosphere. Biological reduction of oxidized phosphorus species is assumed to only take place in the clouds at altitudes 35-55 km as any hypothetical organisms are presumed to only live in the cloud, and hence no calculations are performed below 35 km altitude. The different color curves represent the free energy for reduction by an example different biological reducing agents. The curves show that three (NADH and two Fe-/S proteins) out of the five biological agents are thermodynamically favored for the reduction of phosphate to phosphine (i.e., have negative values of free energy). Reduction of phosphite to $PH_3$ is disfavored under all conditions (unless the concentration of phosphite is allowed to rise in the cell – as shown in Figure 10). (a) Biological reduction of phosphate to phosphine (b) Biological reduction of phosphate to phosphite (c) Biological reduction of phosphite to phosphine. The biological reducing agents assumed are redox equivalents of: NADH: nicotinamide adenine dinucleotide; $FADH_2$: Flavine adenine dinucleotide; $QH_2$: ubiquinone (Co-enzyme Q); *Chromatium* ferredoxin (fd): Iron-sulfur protein from *Chromatium vinosum*; *Clostridium* ferredoxin (fd): Iron-sulfur protein from *Clostridium thermoaceticum*.

## 3. Supplementary References: